\documentclass[11pt,a4paper]{article}
\pdfoutput=1

\usepackage{jheppub}
 \usepackage[utf8]{inputenc} 
 \usepackage{bm}                                           % bold in math mode
 \usepackage{bbm}					%double letter eg for Id
 \usepackage{subfigure}
 \usepackage{tensor}					% tensor notation with proper index placement
\usepackage{amsmath,amstext,amssymb}
\addtocontents{toc}{\protect\setcounter{tocdepth}{2}}

\def\AdS#1{AdS$_{#1}$}

\title{Hydrodynamic Regimes of Spinning Black D3-Branes}
\author[a]{Johanna Erdmenger,}
\author[b]{Mukund Rangamani,}
\author[a]{Stephan Steinfurt}
\author[a]{and Hansj\"org Zeller}
\affiliation[a]{Max-Planck-Institut f\"ur Physik (Werner-Heisenberg-Institut),\\
F\"ohringer Ring 6, 80805 M\"unchen, Germany.}
\affiliation[b]{Centre for Particle Theory \& Department of Mathematical Sciences,\\
Durham University, South Road, Durham DH1 3LE, UK.}
\emailAdd{jke@mppmu.mpg.de}
\emailAdd{mukund.rangamani@durham.ac.uk}
\emailAdd{steinfur@mppmu.mpg.de}
\emailAdd{zeller@mppmu.mpg.de}

\abstract{We present the long-wavelength effective description of non-extremal spinning black D3-branes in flat space. Our setup is motivated by recent explorations of low energy dynamics on black brane world-volumes within the blackfold approach and its connections to the fluid/gravity correspondence. The spinning D3-branes with a rigid radial Dirichlet cut-off give rise to an effective field theory. This theory describes a charged plasma which is driven by external forces, given by one vector and two scalar operators. Furthermore, the flavour charge of this plasma is anomalous, allowing us to examine features of anomaly-induced transport in the blackfold context. We calculate the hydrodynamic transport coefficients to first order and show that in the near-horizon limit, they reproduce the conformal charged fluid dynamics of ${\cal N}=4$ Super Yang-Mills theory. More generally the system interpolates smoothly between the blackfold, fluid/gravity, and Rindler fluid dynamics. }

\date{\today}

\preprint{MPP-2014-561, DCPT-14/59}

\begin{document}
\maketitle
%-------------------------------------------------------------------------------------------------------------

% Part 1 - Introduction
%!TEX root=main.tex

%~~~~~~~~~~~~~~~~~~~~~~~~~~~~~~~~~~~~~~~~~~~~~~~
\section{Introduction}
\label{sec:intro}
%~~~~~~~~~~~~~~~~~~~~~~~~~~~~~~~~~~~~~~~~~~~~~~

The thermodynamic nature of black holes was uncovered more than four decades ago. It prompts the natural question whether there exists a connection between black hole dynamics and hydrodynamics. 
An attempt to address this led to the development of the black hole membrane paradigm, which displayed a suggestive analogy between dynamics of black hole horizons and the hydrodynamic equations of Navier-Stokes \cite{Damour:1978cg,Price:1986yy,Thorne:1986iy}. This paradigm has proven immensely useful in building intuition about black hole dynamics, for it reduces the complex gravitational dynamics into an electromechanical system with viscosities \cite{damour1979quelques,damour1982surface} and conductivities \cite{Damour:1978cg,znajek1978electric}. However, given the fact that asymptotically flat black holes have a gapped spectrum of excitations implies that the analogy is at best approximate.

The situation changes dramatically if one considers instead (large) black holes in asymptotically AdS spacetimes, or black branes in Minkowski spacetime, for now the black hole excitations, characterized by the quasinormal modes is gapless. Indeed, this observation was the cornerstone for the development of the connections between linearized hydrodynamics of strongly coupled plasmas and black hole dynamics in AdS in \cite{Policastro:2001yc,Policastro:2002se}. Building on these works and many others, the fluid/gravity correspondence \cite{Bhattacharyya:2008jc} (cf., \cite{Rangamani:2009xk, Hubeny:2011hd} for reviews) demonstrates a precise relation between the dynamics of Einstein's equations and those of the (relativistic) non-linear hydrodynamical equations, using the standard logic of effective field theory. Likewise an attempt to understand higher dimensional black holes led to the construction of the blackfold effective field theory approach to gravity~\cite{Emparan:2009cs, Emparan:2009at,Camps:2012hw} (
see~\cite{Emparan:2011br} for a review), wherein it was shown that the world-volume dynamics of extended black objects is modeled at the leading order in a gradient expansion as hydrodynamics~\cite{Camps:2010br,Camps:2012hw}. Furthermore, in a related development \cite{Bredberg:2011jq} (see also \cite{Compere:2011dx,Compere:2012mt,Eling:2012ni}) argued that the near horizon Rindler geometry of black holes endowed with a suitable cut-off surface admits an effective description in terms of non-relativistic Navier-Stokes equations.\footnote{ The derivation of the non-relativistic Navier-Stokes equations from the fluid/gravity correspondence was outlined in \cite{Bhattacharyya:2008kq}.}

While these connections between fluid dynamics and gravity a-priori appear to operate in different environments, it was argued in \cite{Emparan:2013ila}, building on earlier work of \cite{Brattan:2011my,Emparan:2012be} that these disparate realizations can be brought into a single universal framework. This was achieved by implementing the blackfold construction for a stack of $N$ D3-branes in Type IIB supergravity; the world-volume dynamics of the branes at finite temperature in the precise of a fixed Dirichlet cut-off surface (at finite radial distance from the branes) was argued to be a hydrodynamic theory, which interpolates between the three constructions alluded to above. One finds that the blackfold hydrodynamics holds sway when the Dirichlet cut-off surface is in the asymptotic flat region, while moving this to the AdS throat leads to the standard hydrodynamics of ${\cal N}=4$ SYM plasma; further pushing the cut-off surface towards the branes leads to the Rindler hydrodynamics.

At a technical level the analysis of \cite{Emparan:2013ila} employed a Kaluza-Klein reduction of the D3-brane down to five-dimensions and considered an effective black brane in an Einstein-dilaton theory. The low-energy dynamics of this system was studied with the imposition of a Dirichlet boundary condition for all fields at a cut-off surface located at a fixed radial position from the branes. Having solved the bulk equations of motion subject to regularity at the black hole horizon, it was found that the quasi-local Brown-York tensor induced on the cut-off surface (which is conserved courtesy the momentum constraints of gravity in radial decomposition) captures the low-energy intrinsic dynamics of the original D3-branes. This stress-tensor depends on the number $N$ of D3-branes (held fixed), the temperature $T$ (or the energy above extremality) and the location $r=R$ of the cut-off surface. By scanning through the various regions in the spacetime, using the location of the cut-off surface $R$, the stress tensor was found to interpolate between the different hydrodynamic forms appropriate for blackfolds, fluid/gravity and the (Rindler) membrane paradigm, respectively.

Given the intrinsic connection between the different hydrodynamic regimes, in this paper we would like to upgrade the discussion to include finite charge density. Apart from the obvious motivation of examining how the dynamics of charge interplays with the location of the cut-off, this provides an opportunity to understand geometrically certain aspects of anomaly induced transport in hydrodynamics. It is by now well established that quantum anomalies leave a definite and predictable imprint in hydrodynamics. This was first discovered in the context of the fluid/gravity correspondence in \cite{Banerjee:2008th,Erdmenger:2008rm} by examining the dynamics of R-charged ${\cal N}=4$ SYM plasma. Subsequently, it was understood that the anomalous contributions are necessitated by the second law of thermodynamics \cite{Son:2009tf}. In the AdS/CFT discussion the gravitational setup analyzed in~\cite{Banerjee:2008th,Erdmenger:2008rm} was the bosonic truncation of $SO(6)$ gauged supergravity to an Einstein-Maxwell-Chern-Simons theory; the gauge field in the bulk corresponds by the usual AdS/CFT rules to a boundary conserved current and the bulk Chern-Simons term captures the boundary R-current anomaly. 

The $SO(6)$ R-symmetry of  ${\cal N}=4$ SYM  is geometrically captured by rotations in the space transverse to the branes. One consequence of this is that imparting a spin to the D3-branes in the transverse ${\mathbb R}^6$ induces an effective charge density on the branes~\mbox{\cite{Chamblin:1999tk,Cvetic:1999ne}}.\footnote{ This suggests a geometric origin for the R-current anomaly, which as far as we are aware has never been made precise. The prime difficulty can be traced to the absence of a covariant action for Type IIB supergravity including the self-dual five form (whose presence is responsible for the anomaly). The analogous story for M5-brane dynamics is much cleaner \cite{Harvey:1998bx}.} Inspired by the rich physics of anomaly induced transport, we undertake the task of examining the effective hydrodynamics on the world-volume of spinning D3-branes. A natural outcome of our analysis is the introduction of anomalous currents into the blackfold paradigm (charged blackfolds have been studied in \cite{
Emparan:2011hg}).

Specifically, we consider a stack of $N$ spinning D3-branes in flat space, focussing on the case when all the spins are equal for simplicity. The long-wavelength perturbations of an effective five-dimensional description of these D3-branes is examined subject to Dirichlet boundary conditions on a cut-off surface at constant $r=R$ and regularity on the horizon. The induced quasi-local stress-energy tensor and charge currents characterize the low energy dynamics on the cut-off surface. We determine the constitutive relations and the transport coefficients as a function of the cut-off radius $R$ explicitly.\footnote{ We unfortunately leave the bulk viscosity  undetermined owing to the complexity of the bulk equations.}

The analysis of the effective dynamics is immensely facilitated as in \cite{Erdmenger:2008rm,Banerjee:2008th,Emparan:2013ila} by performing a Kaluza-Klein reduction of 
type IIB supergravity on the $\mathbf{{\bf S}^5}$ transverse to the branes. For the case of interest of equal spins in the three planes of ${\mathbb R}^6$, the rotation of the D3-branes is in the diagonal $U(1)$ subgroup of the $U(1)^3 \subset SO(6)$ Cartan. Regarding ${\bf S}^5$ as a Hopf fibration  of ${\bf S}^1$ over $\mathbb{CP}^2$, the rotation is along the circle fibre.\footnote{ An obvious generalization would be to extend the consideration to ${\cal N}=1$ superconformal field theory plasmas obtained by placing D3-branes at the tip of a cone over a Sasaki-Einstein space. The rotation is then aligned with the Reeb vector of the manifold.} One can then use the consistent truncation discovered in \cite{Maldacena:2008wh} as our effective five dimensional theory. The five-dimensional background solution obtained by reducing the spinning D3-brane metric~\cite{Cvetic:1999xp} has all the fields in the effective action of~\cite{Maldacena:2008wh} non-vanishing. In particular, additional to the gauge field which 
corresponds to the fibre isometry -- the one used in~\cite{Erdmenger:2008rm,Banerjee:2008th} -- one has three additional matter fields. The first is a massive vector field originating from the vector modes of the five-form (obtaining such massive vectors was one of the motivations of \cite{Maldacena:2008wh}). In addition, there are two scalars (dilatons); these correspond to the breathing modes associated with the volume of the ${\bf S}^1$ fibre and $\mathbb{CP}^2$ base space, and lead to further complications (somewhat fatally in the scalar sector).

Starting with this seed solution, we follow the logic of \cite{Bhattacharyya:2008jc} making the physical black hole parameters vary along the world-volume. The resulting perturbation equations for long-wavelength fluctuations of the dynamic fields are decomposed into irreducible representations (scalars, vectors and tensors) under the $SO(3)\subset SO(3,1)$, which decouple from each other. The radial evolution of these away from the cut-off surface is embodied into (i) constraint equations, which describe the conservation of the leading (zeroth) order currents, and (ii) dynamical equations which have to be solved along with suitable boundary conditions (regularity at the horizon and Dirichlet conditions at $r=R$). We note in passing that the presence of additional non-vanishing matter fields leads to source terms in the conservation equations beyond leading order. In particular, the system we consider should be viewed as a charged generalization of the forced fluid set-up of \cite{Bhattacharyya:2008ji}.

We are able to solve the dynamical equations in the tensor and vector sector analytically;
owing to the extra fields the scalar sector has proved intractable thus far. The upshot of the analysis is a first order stress tensor and charge current which depends explicitly on the cut-off radius $R$. From here we read off the transport coefficients in the various regimes.\footnote{ For reasons described in \cite{Bhattacharyya:2008kq}, in the non-relativistic limit forced upon us by the near-horizon Rindler geometry, the charge dynamics freezes out.} In particular, working in the fluid/gravity regime, by placing the cut-off surface in the AdS throat region recovers the  well-known results from the perturbations of the AdS Reissner-Nordstr\"om solutions~\cite{Erdmenger:2008rm,Banerjee:2008th}.

The ratio of shear viscosity and entropy density takes the universal value $1/4\pi$. This was to be expected, as the arguments for the universality of this ratio in two derivative gravity in \cite{Buchel:2003tz, Kovtun:2004de} continue to hold in the presence of the extra matter fields. In the vector sector, we encounter the parity-even conductivity in addition to the parity odd contributions proportional to the vorticity (and magnetic field). The parity-even vector contributions which correspond to conductivity (or charge diffusion) ends up being more intricate. The complications of the Kaluza-Klein reduction make their presence felt in this analysis for not only does the physical (anomalously) conserved current have a source (the chemical potential) and an expectation value, it appears that with the na\"ive choice of counter-terms, so does the non-conserved vector operator. As a result we encounter 
(a) a modification of the Smarr relation which is a statement about equilibrium thermodynamics, involving contributions from the vector operator and (b) the physical charge diffusion constant ends up being contaminated by the vector operator acquiring a vacuum expectation value. Furthermore, in the canonical basis of fields we encounter, structures reminiscent of Weyl invariant fluids (the na\"ive diffusion constant appears to multiply the Weyl covariant derivative of the charge densities), despite conformal symmetry being explicitly broken by the background (and by the cut-off surface).

The outline of the paper is as follows: In  \S\ref{sec:consistent_truncations}, we start off with a quick summary of the Kaluza-Klein reduction and consistent truncation of \cite{Maldacena:2008wh} which will be convenient for our analysis. We also obtain the equilibrium seed solution from the  spinning D3-branes geometry \cite{Cvetic:1999xp} and explicitly map out the connections between our parameterization and those in the earlier analysis of \cite{Emparan:2013ila} and \cite{Banerjee:2008th,Erdmenger:2008rm} in appropriate limits for later comparisons. In  \S\ref{sec:black_brane_background_and_currents} and \S\ref{sec:blackfold_setup} we display this geometry in a useful coordinate chart which is regular through the future horizon and thence proceed to allow fluctuations along the world-volume. In  \S\ref{sec:tensor_sector} and \S\ref{sec:vector_sector} we find the leading order correction to the background geometry with the prescribed boundary conditions,  briefly commenting on the scalar sector in  \S\ref{sec:scalar_sector}. Finally, \S\ref{sec:physical_results} is devoted to the physical results from our analysis and we examine the features of the transport coefficients as a function of $R$. We conclude with a brief discussion in  \S\ref{sec:discuss}. The appendices contain some more technical details of the computation:
Appendix \ref{sec:rotating_D3_branes} elaborates on the Kaluza-Klein truncation, while details of the vector sector computations such as source terms and fixing integration constants are described in 
Appendices \ref{ap:source_terms} and \ref{sec:fix_int_const} respectively.

% Part 2 - Consistent Truncation
%!TEX root=main.tex

%~~~~~~~~~~~~~~~~~~~~~~~~~~~~~~~~~~~~~~~~~~~~~~~
\section{Effective description of spinning D3-branes}
\label{sec:consistent_truncations}
%~~~~~~~~~~~~~~~~~~~~~~~~~~~~~~~~~~~~~~~~~~~~~~

We aim for a minimal generalization of the interpolating blackfold fluid presented in \cite{Emparan:2013ila} to incorporate non-vanishing charge density. We therefore take a stack of D3-branes in flat space and let them spin along their transverse directions. This naturally incorporates the description of a charged fluid from the world-volume point of view \cite{Cvetic:1999ne,Chamblin:1999tk} with the angular velocities playing the role of charge chemical potentials.

For the sake of simplicity we will look at rotations along the diagonal $U(1)$ of the $U(1)^3$ Cartan of the five-sphere isometry group $SO(6)$. This abelian isometry corresponds to the fibre isometry when the ${\bf S}^5$ is taken to be the Hopf fibration ${\bf S}^1 \hookrightarrow {\bf S}^{5} \twoheadrightarrow \mathbb{CP}^2$.

%~~~~~~~~~~~~~~~~~~~~~~~~~~~~~~~~~~~~~~~~~~~~~~
\subsection{Consistent truncation} \label{subsec:consistent_truncation}
%~~~~~~~~~~~~~~~~~~~~~~~~~~~~~~~~~~~~~~~~~~~~~~

To describe the effective fluid degrees of freedom it turns out to be useful to perform a Kaluza-Klein reduction of the bosonic subsector of type IIB supergravity (respecting the fibration). Fortunately, this has already been worked out in detail in \cite{Maldacena:2008wh}. In addition to the usual metric, scalars and the gauge field $\mathcal{A}_\mu$ which arises from the fibre isometry, this truncation also incorporates a Proca vector field $\mathbf{A}_\mu$. It arises from the first massive level of the harmonic analysis of supergravity on ${\bf S}^5$ \cite{Kim:1985ez}, but may consistently be incorporated in the truncation.

We will apply this KK reduction and spectrum truncation to the metric and five-form profile of a stack of rotating D3-branes \cite{Cvetic:1999xp}, mildly adjusting the conventions of \cite{Maldacena:2008wh} by taking $\sim Q$ units of five-form flux $\int_{{\bf S}^5} F_5 = 2\, Q$ instead of $\int_{{\bf S}^5} F_5 = 4$. With this normalization, our choice of $Q$ corresponds to the one used in \cite{Emparan:2013ila}. In the following, we briefly review the effective action and equations of motion derived in \cite{Maldacena:2008wh}.

Using the ten-dimensional Einstein's equations and the closure of the five-form together with the reduction ansatz maintaining the fibration, one obtains the following five dimensional effective action for the resulting dynamics (see Appendix \ref{sec:rotating_D3_branes} for further details)\footnote{ In the following, we are going to take the liberty to multiply the Chern-Simons term with $4 \kappa$ to keep track of it in the analysis and leave more space for generality. Compared to the previous literature we have $\kappa = -\frac{\sqrt{3}}{2} \kappa^{(B)}$ where the latter $\kappa^{(B)}$ is the one used in the fluid/gravity analysis of \cite{Banerjee:2008th}.}
\begin{equation}
\label{MMT_action}
\begin{aligned}
   S={}&\frac{1}{2 \kappa_5^2}\; \int d^5 x \,\sqrt{-g}\, e^{4 U +V}
   \left[ R + 24 \, e^{-2 U}- 4 \, e^{-4 U + 2V}- 2 \,Q^2 \, e^{-8 U - 2V} \right. \\
   &\left. \qquad \qquad
   +\; 12\, \partial_\mu U \partial^\mu U + 8\, \partial_\mu U \partial^\mu V  -\frac{1}{4} \, e^{2 V} \mathcal{F}_{\mu\nu}\mathcal{F}^{\mu\nu} 
   -\frac{1}{8} \,Q^2\,e^{-4 U - 2 V} \mathbb{F}_{\mu\nu}\mathbb{F}^{\mu\nu}\right. \\ 
   & \left. \qquad \qquad
   - \; 2 \,Q^2 \,e^{-8 U} \mathbf A_\mu\mathbf A^\mu \right]
   +\; \frac{Q^2}{8\kappa_5^2} \left(4\kappa\right)\int \mathcal{A}\wedge\mathbb{F}\wedge\mathbb{F}\,.
\end{aligned}
\end{equation}
Apart from gravity, we have two scalar fields $U$ and $V$, the breathing modes of the $\mathbb{CP}^2$ base space and ${\bf S}^1$ fibre of the Hopf fibration. The vector fields with field strengths $d\mathcal{A}=\mathcal{F}$ and $d\mathbf{A} = \mathbf{F}$ are assembled into the combination
\begin{equation}
	\mathbb{F}=\mathcal{F}+\mathbf{F}\,.
\end{equation}

We claim that \eqref{MMT_action} incorporates the necessary ingredients to subsume the fluid/gravity discussions of {\cite{Banerjee:2008th,Erdmenger:2008rm} and the blackfold analysis of \cite{Emparan:2013ila}} and does so in a minimal fashion. 

First, our action reduces to the simpler action of \cite{Emparan:2013ila} once the gauge fields vanish and the dilatons are set equal, $U=V=\varphi$. This corresponds to the limit of vanishing charge density 
($q\rightarrow 0$).  Second, we can reduce \eqref{MMT_action} to the Einstein-Maxwell theory with negative cosmological constant discussed in \cite{Chamblin:1999tk}. For this, the Proca field $\mathbf{A}$ needs to be set to zero and the dilatons fixed to equal constant value. With the massive vector vanishing, we recover the gauge Chern-Simons term as desired. This reproduces then the effective action used in the analysis of 
\cite{Banerjee:2008th,Erdmenger:2008rm}. We will shortly also see how to compare the solutions used in these analyses in our language.

Let us record for completeness the equations of motion resulting from \eqref{MMT_action}. The Einstein equation is given by 
\begin{equation}
\label{MMT_Einstein_equation}
\begin{aligned}
   R_{\mu\nu}={}& 4 \left(\partial_\mu U \partial_\nu U+\nabla_\mu\partial_\nu U \right) +  \left(\partial_\mu V \partial_\nu V+\nabla_\mu\partial_\nu V \right)   - Q^2 \, e^{-8 U - 2 V}g_{\mu\nu}\\&
   +\frac{1}{2}\, e^{2 V}\mathcal{F}_{\mu\rho}\mathcal{F}\indices{_\nu^\rho} + \,Q^2 \, e^{-8 U} \left( 2 \mathbf{A}_\mu \mathbf{A}_\nu-g_{\mu\nu} \mathbf{A}_\rho\mathbf{A}^\rho\right)\\&+\frac{1}{16}  \,Q^2 \, e^{-4 U - 2V} \left( 4  \mathbb{F}_{\mu\rho}\mathbb{F}\indices{_\nu^\rho}-g_{\mu\nu} \mathbb{F}_{\rho\sigma}\mathbb{F}^{\rho\sigma}\right)\,.
\end{aligned}
\end{equation}
The (coupled) scalar equations of motion for the dilatons $U$ and $V$ are
\begin{align}
\begin{split}
  \square U+4\partial_\mu U \partial^\mu U +\partial_\mu U\partial^\mu V  ={}& 6\, e^{-2 U }- 2\, e^{-4 U + 2 V}\\&- Q^2\, e^{-8 U - 2 V}-Q^2\,e^{-8 U} \mathbf{A}_\mu\mathbf{A}^\mu\,,
\end{split}\\
\begin{split}
\square V+4\partial_\mu U \partial^\mu V +\partial_\mu V\partial^\mu V ={}& 4\, e^{-4 U + 2 V}  - Q^2\, e^{-8 U - 2 V}+\frac{1}{4}\,e^{2 V}\mathcal{F}_{\mu\nu}\mathcal{F}^{\mu\nu}\\
    &+Q^2\,e^{-8 U}\mathbf{A}_\mu\mathbf{A}^\mu-\frac{1}{16} \,Q^2\,e^{-4 U- 2 V}\mathbb{F}_{\mu\nu}\mathbb{F}^{\mu\nu}\,.
\end{split}
\end{align}
The last equation which arises from the ten-dimensional Einstein's equations is the equation of motion for the gauge field $\mathcal{A}$, which also involves a parity odd term
\begin{equation}
\label{MMT_Maxwell_equation}
  d\left(e^{4 U + 3 V}\star \mathcal{F} \right)= 4 \,Q^2 \, e^{-4 U + V} \star \mathbf{A}+Q^2\,\kappa \,\mathbb{F}\wedge\mathbb{F}\,.
\end{equation}
Additionally, we have some more equations from the closure of the self-dual five-form, $dF_5 = 0$, which are given by
\begin{align}
d\left(e^{-4 U + V} \star\mathbf{A}\right)&=0\\
\label{MMT_Maxwell_equation_massive}
d\left(e^{-V}\star\mathbb{F}\right)&=-8 e^{-4 U +V}\star \mathbf{A}+4\kappa\,\mathcal{F}\wedge\mathbb{F}\\
d\mathbb{F}&=0
\end{align}
The first is similar to a Lorentz gauge choice for $\mathbf{A}$. Furthermore, there is also a Maxwell equation for $\mathbb{F}$ and a simple Bianchi equation.

There is a remark in order. We see that both Maxwell's equations, \eqref{MMT_Maxwell_equation} and \eqref{MMT_Maxwell_equation_massive}, are sourced by the Proca field $\mathbf{A}$ (the contribution from the Chern-Simons terms is not relevant for the following discussion). To later obtain a conserved current from these fields we have to rotate to a basis, in which the kinetic terms for the Proca field $\mathbf{A}$ and a physical gauge field $A_\textrm{phys}$ are diagonal such that we end up with an equation of motion without source terms. It turns out that the linear combination
\begin{equation}
\label{physical_gauge_field}
\begin{split}
A_\textrm{phys} &=  \left(\frac{Q^2}{2}e^{-V} + e^{4U+3V}\right)^{-1}\left(\frac{Q^2}{2}e^{-V} \mathbb{A} + e^{4U+3V}\mathcal{A}\right)\\
&= \mathcal{A} + \left(\frac{Q^2}{2} e^{-V} + e^{4U+3V}\right)^{-1} \frac{Q^2}{2}\,e^{-V} \mathbf{A}
   \end{split}
\end{equation}
leads to such an equation of motion. Consequently the corresponding current $\mathcal{J}_\textrm{phys}$ will be conserved. The relevant part of the action \eqref{MMT_action} which we may rewrite in terms of the Proca field $\mathbf{A}$ and the physical gauge field $A_\textrm{phys}$ is given by
\begin{equation}
\label{MMT_action_rotated}
\begin{split}
&-\frac{1}{4} \, e^{2 V} \mathcal{F}_{\mu\nu}\mathcal{F}^{\mu\nu} 
   -\frac{1}{8} \,Q^2\,e^{-4 U - 2 V} \mathbb{F}_{\mu\nu}\mathbb{F}^{\mu\nu}- \; 2 \,Q^2 \,e^{-8 U} \mathbf A_\mu\mathbf A^\mu\\
 ={}& -\frac{1}{4}  \left(\frac{e^{-4U-V}}{\frac{Q^2}{2}e^{-V} + e^{4U+3V}}\right) \left(\frac{Q^2}{2}e^{-V}\mathbb{F}_{\mu\nu}+e^{4U+3V}\mathcal{F}_{\mu\nu}\right)^2\\ &- \frac{Q^2}{8}\left(\frac{ e^{V}}{\frac{Q^2}{2}e^{-V}+e^{4U+3V}}\right)\mathbf{F}_{\mu\nu}\mathbf{F}^{\mu\nu} - 2 Q^2 e^{-8U} \mathbf{A}_\mu\mathbf{A}^\mu\, .
   \end{split}
\end{equation}
However, we found it rather inconvenient to work in this rotated basis, due to the contributions from the dilatons. Thus in the following we will stick to the former choice~$\mathcal{A}$ and~$\mathbf{A}$ or depending on the context $\mathbb{A} = \mathcal{A} + \mathbf{A}$.

%~~~~~~~~~~~~~~~~~~~~~~~~~~~~~~~~~~~~~~~~~~~~~~
\subsection{Black brane background}
\label{sec:black_hole_background}
%~~~~~~~~~~~~~~~~~~~~~~~~~~~~~~~~~~~~~~~~~~~~~~

Starting from the stack of rotating D3-branes in \cite{Cvetic:1999xp} with all three angular momenta equal 
(cf., Appendix \ref{sec:rotating_D3_branes}), we may use the Kaluza-Klein reduction of \cite{Maldacena:2008wh} to arrive at a five-dimensional black hole space-time, which solves the equations of motion of \eqref{MMT_action}.

By construction, our lower-dimensional black hole solution is very much reminiscent of the Reissner-Nordstr\"om-\AdS{5} solution \cite{Chamblin:1999tk,Banerjee:2008th,Erdmenger:2008rm} and, in fact, is related to it via the near-horizon limit soon to be discussed. The metric reads
\begin{equation}
\label{background_metric}
   ds^2=H(r)^{-1/2}\left[-f(r) g(r) \,dt^2 + dx_1^2+dx_2^2+dx_3^2\right]+H(r)^{1/2}  f(r)^{-1}dr^2\,,
\end{equation}
where 
\begin{equation}
  \label{warp_factors}
   f(r)=1-\frac{r_0^4}{r^4}+\frac{q^2}{r^6}\,,\quad H(r)= 1+\frac{L^4}{r^4}\,,\quad g(r)=\frac{H(r)}{H(r)+\frac{q^2}{r^6}}\,.
\end{equation}
Clearly $f(r)$ is the emblackening factor incorporating the mass ($\sim r_0^4$) and charge ($\sim q$) 
of the Reissner-Nordstr\"om solution. $H(r)$ is the warp factor, which in the near-horizon limit behaves like $H(r)\rightarrow \frac{L^4}{r^4}$ changing the asymptotics of the near-horizon geometry to AdS.

The scalars $U$ and $V$ profiles are given by
\begin{equation}
\label{background_dilatons}
  e^{2 U}=   H(r)^{1/2} r^2  \,,\qquad e^{2 V} =  H(r)^{1/2} r^2 g(r)^{-1}\,,
\end{equation}
and we can see that $g(r)$ basically describes the relative squashing of base and fibre in the compact space we reduced on. The vector fields are given by
\begin{equation}
\label{background_gauge_fields}
    \mathcal{A}=-q \left(\frac{Q}{2 L^2}\right)\left(\frac{g(r)}{r^6 H(r)}\right)dt\,,\quad\mathbf{A}=-q \left(\frac{2  L^2}{Q}\right)\left( \frac{f(r)g(r)}{r^2 H(r)}\right)dt\,,
\end{equation}
where we have reintroduced $Q$ which relates to $L$ and $r_0$ via
\begin{equation}
\label{QtoL}
   \frac{Q}{2}= L^2 \sqrt{r_0^4+L^4} =4\pi g_s {\alpha^\prime}^2 N
\end{equation}
and thus represents the number of D3-branes $N$ in string units \cite{Kraus:1998hv}.

%~~~~~~~~~~~~~~~~~~~~~~~~~~~~~~~~~~~~~~~~~~~~~~
\subsection{The decoupling limit}
\label{decoupling_limit}
%~~~~~~~~~~~~~~~~~~~~~~~~~~~~~~~~~~~~~~~~~~~~~~

Of course, it is well known that for extremal non-rotating D3-branes the near-horizon geometry is \AdS{5} $\times\, {\bf S}^5$ \cite{Gibbons:1993sv}. For non-extremal spinning D3-branes a similar decoupling limit  relates the solution given in Eqs.~\eqref{background_metric}-\eqref{background_gauge_fields} to that analyzed 
in \cite{Banerjee:2008th,Erdmenger:2008rm}. It is simply given by \cite{Cvetic:1999xp}
\begin{equation}
\label{near_horizon_scaling}
   r_0\rightarrow \epsilon \,r_0\,,\quad L\rightarrow L\,,\quad r\rightarrow \epsilon\, r\,,\quad x^\mu \rightarrow \epsilon^{-1} x^\mu\,,\quad q\rightarrow \epsilon^3 \,q\,,
\end{equation}
taking $\epsilon\rightarrow 0$, which then implies that
\begin{equation}
\label{near_horizon_scaling_fhg}
 f(r)\rightarrow f(r)\,,\quad H(r)\rightarrow \frac{L^4}{\epsilon^4 r^4}\,,\quad g(r)\rightarrow 1  \,.
\end{equation}

In this limit, the metric \eqref{background_metric} does indeed reduce to the Reissner-Nordstr\"om-\AdS{5} black hole. Furthermore, the relative squashing between base and fibre is suppressed, as can be seen directly from \eqref{background_dilatons} using \eqref{near_horizon_scaling_fhg}. Also the dilatons are frozen to constant values $e^{2U}=e^{2V}=L^2$ accounting for the negative cosmological constant in \eqref{MMT_action}, where we retained the AdS radius $L$ explicitly.

The background profile of the Proca field $\mathbf{A}$ is suppressed by $\epsilon^4$ \eqref{background_gauge_fields}, but the gauge field $\mathcal{A}$ in the limit \eqref{near_horizon_scaling} precisely agrees with the one in \cite{Banerjee:2008th,Erdmenger:2008rm}  (modulo a trivial normalization factor). In particular, accounting for the differing normalization of the kinetic terms gives the relation $\frac{\sqrt{3}}{2}\mathcal{A}_\mu = A_\mu^{(B)}$, with $A_\mu^{(B)}$ being the gauge field of \cite{Banerjee:2008th}. 

Armed with these observations we are now prepared to make direct comparisons in the sequel with the fluid/gravity analysis of \cite{Banerjee:2008th,Erdmenger:2008rm}.
 
%~~~~~~~~~~~~~~~~~~~~~~~~~~~~~~~~~~~~~~~~~~~~~~
\subsection{Comments on the chargeless limit}
\label{chargeless_limit}
%~~~~~~~~~~~~~~~~~~~~~~~~~~~~~~~~~~~~~~~~~~~~~~

Taking the charge $q$ to zero of course, by construction, recovers the background solution of \cite{Emparan:2013ila}. However, \cite{Emparan:2013ila} uses a somewhat different choice of coordinates. Whilst it is simple to translate between the two at the level of the zeroth order solution \eqref{background_metric}, 
it notably introduces an important subtlety when we proceed to analyze fluctuations. 

Consider the coordinate change:
\begin{align}
\label{coordinate_change}
   \rho^4 = r^4 + L^4\,,\quad r_+^4 = r_0^4 + L^4\,,\quad r_- = L
\end{align}
under which the warp and emblackening factors $H(r)$ and $f(r)$ reduce to 
\begin{equation}
 \Delta_\pm (\rho)=1 -\frac{r_\pm^4}{\rho^4}  \,.
\end{equation}
This makes it clear that $r_0$ is related to the temperature or the deviation away from extremality of the D3-brane.
The analysis of \cite{Emparan:2013ila} employs the coordinate $\rho$ (where it is called $r$). 
To be clear below we will use $\rho = r_{\textrm{\scriptsize \cite{Emparan:2013ila}}}$ to avoid further loss in translation.
We however find it convenient for our purposes to stick to the coordinate chart displayed in~\eqref{background_metric}.

The main subtlety, which we want to emphasize, is that the coordinate change \eqref{coordinate_change} for the radial coordinates involves the parameter $L$. In the fluid/gravity and blackfold approaches, when we analyze fluctuations about equilibrium configurations, we have to take this parameter to depend on the world-volume coordinates, i.e., $L\rightarrow L(\sigma^a)$. We wish to do this however keeping the number  $N$ of D3-branes
or equivalently the parameter $Q$ as in \eqref{QtoL} fixed; this correlates the dependence of $L(\sigma^a)$ with that of $r_0(\sigma^a)$. 

However, the intrinsic dependence $L(\sigma^a)$ implies that constant $r$ and constant $\rho = r_{\textrm{\scriptsize \cite{Emparan:2013ila}}}$ surfaces are not isomorphic. The choice of cut-off surface in \cite{Emparan:2013ila}
was made to be an isodilatonic surface at constant $\rho= r_{\textrm{\scriptsize \cite{Emparan:2013ila}}} = P$, $e^{2U}=\rho^2$, which would correspond to a fluctuating surface in our radial coordinate $r$. 
We wish to impose boundary conditions at $r = R$ and owing to $e^{2U}=H^{1/2}r^2$, our iso(U)dilatonic surface
is not the same surface as used in the previous analysis. As a result comparison of results in the $q\to0$ limit requires some care which we will highlight when necessary.

% % Part 3 - Background
%!TEX root=main.tex

%~~~~~~~~~~~~~~~~~~~~~~~~~~~~~~~~~~~~~~~~~~~~~~~
\section{Black brane background and currents}
\label{sec:black_brane_background_and_currents}
%~~~~~~~~~~~~~~~~~~~~~~~~~~~~~~~~~~~~~~~~~~~~~~~

We now have the necessary ingredients to study the long-wavelength fluctuations on the world-volume of the rotating D3-branes. Before we analyze the perturbative solutions we first ensure that we convert our seed solution into a regular coordinate chart and extract the equilibrium thermodynamic data below. 

%~~~~~~~~~~~~~~~~~~~~~~~~~~~~~~~~~~~~~~~~~~~~~~~
\subsection{The seed metric}
%~~~~~~~~~~~~~~~~~~~~~~~~~~~~~~~~~~~~~~~~~~~~~~~

To set up the Dirichlet problem at finite $R$ in the spirit of the fluid/gravity correspondence and blackfold paradigm, we first transform our background \eqref{background_metric} and \eqref{background_gauge_fields} in ingoing Eddington-Finkelstein coordinates to make the non-singular nature of the outer horizon apparent.
To do so we introduce the ingoing coordinate $v=t+r_\star(r)$, with $r_\star$ being the tortoise coordinate for 
\eqref{background_metric}. In addition we make the choice of inertial frame at the zeroth order explicit, by introducing a boost velocity $u^a$ (normalized to $u^2=-1$). Using the projection tensor $P_{ab}=\eta_{ab}+u_a u_b$, which projects perpendicular to the velocity $u^a$, the Kaluza-Klein-reduced metric~\eqref{background_metric} derived in the previous section, reduces to\footnote{ Henceforth lowercase Latin indices refer to the world-volume; the early part of the alphabet $\{a,b,\cdots\}$ are world-volume covariant, while $\{i,j,\cdots\}$ refer to the spatial directions. Greek indices always refer to the bulk spacetime.}

\begin{equation}\label{eq:metric_rotD3_wo_rescaling}
   ds^2= - \frac{f(r)\, g(r)}{\sqrt{H(r)}}\,u_a u_b\, d\sigma^a d\sigma^b -2 \,\sqrt{g(r)} \, u_a\, d\sigma^a dr +\frac{1}{\sqrt{H(r)}}\,P_{ab}\, d\sigma^a d\sigma^b\,.
\end{equation}
In addition we have to transform the gauge fields \eqref{background_gauge_fields} accordingly. This leads to
\begin{align}
    \mathcal{A}&=q \left(\frac{Q}{2 L^2}\right)\left(\frac{g(r)}{r^6 H(r)}\right)u_a d\sigma^a\,,\\
    \mathbf{A}&=q \left(\frac{2  L^2}{Q}\right)\left( \frac{f(r)g(r)}{r^2 H(r)}\right)\left(u_a d\sigma^a - \frac{H(r)^{1/2}}{f(r) g(r)^{1/2}}dr\right)\,.
\end{align}

Note that we used the gauge freedom of the gauge field $\mathcal{A}$ to set the component $\mathcal{A}_r$ to zero. This is not possible for $\mathbf{A}$ owing to the absence of gauge invariance due to the mass term. If we view the Proca field as a gauge field together with a Stueckelberg scalar $\theta$, then we have already gauge fixed $\theta =0$, so no further choice is possible.

On the fluctuations of the various background fields we impose Dirichlet boundary conditions at a cut-off surface at a finite radial slice $r=R$ \cite{Brattan:2011my}. The resulting quasi-local stress-energy tensor and charge currents at $r=R$ for long-wavelength fluctuations along $\sigma^a$ are expected to be hydrodynamic. However, the induced metric on the $r=R$ surface whilst flat to leading order in these fluctuations is not manifest in the Minkowski form. We will therefore perform a further coordinate rescaling to achieve this.\footnote{ Note that as explained in \cite{Emparan:2012be}, the rescaling is not a necessity, but is
practically convenient. We could directly compute in the background \eqref{eq:metric_rotD3_wo_rescaling} at fixed $r=R$ without rescaling the coordinates, and subsequently translate to a manifestly flat space analysis (cf., appendices of~\cite{Brattan:2011my,Emparan:2012be} for further elaboration on this issue).} Therefore, as in~\cite{Emparan:2012be,Emparan:2013ila}, we redefine our space-time coordinates to make this manifest,\footnote{ We also supplement our various fields with an index $0$ since these are now the 
fields which are the seeds for our fluctuation  analysis.}
\begin{equation}
\label{seed_metric}
\begin{split}
   ds^2_0={}& - \frac{f(r)\, g(r) }{f_R \,g_R } \sqrt{\frac{H_R}{H(r)}}\,u_a u_b\, d\sigma^a d\sigma^b
    -2 \,\sqrt{\frac{g(r) H_R^{1/2}}{f_R\, g_R}} \, u_a\, d\sigma^a dr +\sqrt{\frac{H_R}{H(r)}}\,P_{ab}\, d\sigma^a d\sigma^b\,,
\end{split}
\end{equation}
where we have defined $f_R=f(R),\,g_R=g(R)$ and $H_R=H(R)$.
This automatically ensures that the theory on the cut-off surface $r=R$ is defined on a manifold with the  Minkowski metric.

The scalars being (bulk) diffeomorphism invariant are unchanged, viz., 
\begin{equation}
  e^{2 U_0}=   H(r)^{1/2} r^2  \,,\qquad e^{2 V_0} =  H(r)^{1/2} r^2 g(r)^{-1}\,,
\end{equation}
but the vector fields also undergo rescaling. To wit,
\begin{align}
	\label{seed_gauge_field}
    \mathcal{A}_0&=q \left(\frac{Q}{2 L^2}\right)\left(\frac{g(r)}{r^6 H(r)}\right)\sqrt{\frac{H_R^{1/2}}{f_R\, g_R}}\,u_a d\sigma^a\,,\\
	\label{seed_gauge_field_massive}
    \mathbf{A}_0&=q \left(\frac{2  L^2}{Q}\right)\left( \frac{f(r)g(r)}{r^2 H(r)}\right)\sqrt{\frac{H_R^{1/2}}{f_R\, g_R}}\left(u_a d\sigma^a - \frac{H(r)^{1/2}}{f(r) g(r)^{1/2}}dr\right)\,.
\end{align}
Note that the $dr$ terms of the metric and the gauge fields have not been rescaled. In~\eqref{seed_gauge_field_massive} the rescaling of $u_a d\sigma^a$ factors out, because the definition of Eddington-Finkelstein coordinates has to be modified accordingly.

%~~~~~~~~~~~~~~~~~~~~~~~~~~~~~~~~~~~~~~~~~~~~~~~
\subsection{Equilibrium thermodynamics}
%~~~~~~~~~~~~~~~~~~~~~~~~~~~~~~~~~~~~~~~~~~~~~~~

Given the seed metric \eqref{seed_metric}, we can compute entropy density and temperature of our stationary setup. Accounting for the non-vanishing scalar fields in \eqref{background_dilatons} or equivalently transforming to an Einstein frame Lagrangian, we get 
\begin{equation}
	S=\frac{A}{4G}= \left.\frac{2\pi}{\kappa_5^2} \int d^3 x \sqrt{-h}\, e^{4U+V}\,\right|_{r_+}
\end{equation}
which translates into an entropy density
\begin{equation}
\label{entropy_density}
	s= \frac{2\pi}{\kappa_5^2} H_R^{3/4} r_+^5 \sqrt{\frac{H(r_+)}{g(r_+)}}\,.
\end{equation}

\begin{figure}[ht]
 \centering
\includegraphics[width=0.9\linewidth]{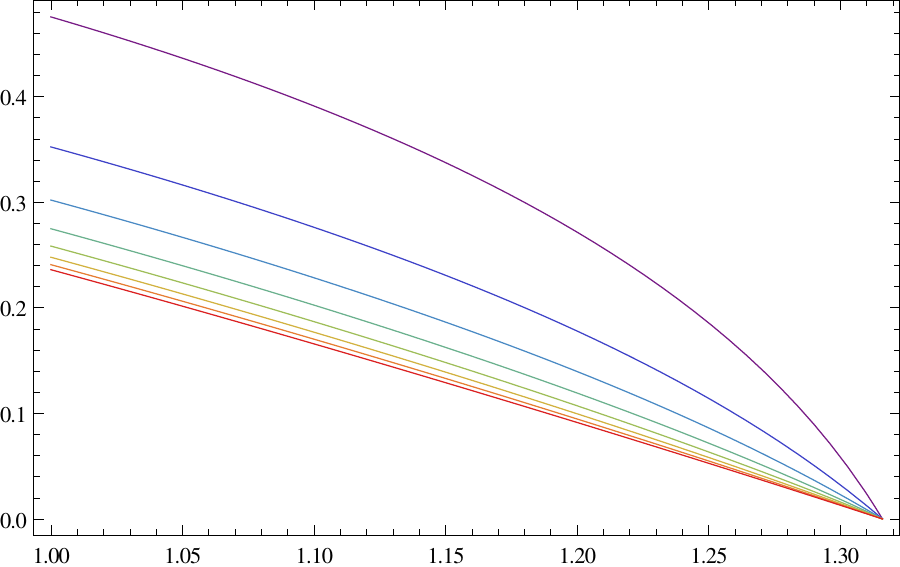}
\put(-393,230){$T$}\put(-12,-8){$r_0$}\put(-100,150){$R\,\downarrow$}
 \caption{We show the temperature versus the parameter $r_0$. To generate this plots we set $r_+=1$ and $L=1$. The different colours correspond to different values of $R$. The purple line corresponds to the smallest plotted value of $R$, while the red one to the biggest value.}
 \label{fig:temperature}
 \end{figure}

The black brane's temperature is easily calculated by computing the period of the Euclidean thermal circle in \eqref{seed_metric}. It is given by
\begin{equation}
\label{temperature}
 4 \pi T= \frac{H_R^{1/4}}{\left(f_R g_R\right)^{1/2}}\, f'(r_+) \frac{g(r_+)^{1/2}}{H(r_+)^{1/2}} \,.  
\end{equation}
If we use units, in which we measure length dimensions relative to $r_+$, i.e., if we effectively set $r_+ = 1$, we have $q^2 = r_0^4 - 1$. The physical condition $T \geq 0$ then translates into an interval for $r_0$ relative to $r_+$
\begin{equation}
   r_0 \in [0, 3^{1/4}]\,,
\end{equation}
which we will use in various plots later on. The behaviour of $T(r_0)$ is illustrated in Fig.~\ref{fig:temperature}.

From the gauge fields, we may read off the chemical potentials $\mu_\mathcal{J}$ and $\mu_\mathbb{J}$ which are conjugate to the respective charge densities.\footnote{ The notation with index $\mathcal{J}$ and $\mathbb{J}$ respectively will become clear in \S\ref{bf:charge_currents}, where we define these charge currents and their corresponding charge densities.} They are given as electric potential differences between the boundaries of our setup, i.e., the cutoff surface at $r=R$ and the black brane's outer horizon at $r=r_+$. The chemical potential $\mu_\mathcal{J}$ is given by
\begin{equation}
\label{bf:chemical_potential_massless}
\begin{split}
  \mu_\mathcal{J} &= \mathcal{A}_t(R)-\mathcal{A}_t(r_+)\\
		&= q\, \sqrt{L^4+r_0^4} \, \sqrt{\frac{H_R^{1/2}}{  f_R g_R  }}\left(\frac{g\left(r_+\right)}{r_+^6 H\left(r_+\right)} - \frac{g_R}{R^6 H_R} \right)
\end{split}
\end{equation}

The second chemical potential $\mu_\mathbb{J}$ may be computed from the gauge field $\mathbb{A} = \mathcal{A}+\mathbf{A}$, for which it is useful to note that $\mathbf{A}_t(r_+)=0$:
\begin{equation}
\label{bf:chemical_potential_mixed}
\begin{split}
  \mu_\mathbb{J} &= \mathbb{A}_t(R)-\mathbb{A}_t(r_+)\\
		& =  \mu_\mathcal{J}- \frac{q}{\sqrt{L^4+r_0^4}} \,\sqrt{\frac{ f_R g_R}{R^4 H_R^{3/2} }}
\end{split}
\end{equation}

%~~~~~~~~~~~~~~~~~~~~~~~~~~~~~~~~~~~~~~~~~~~~~~~
\subsection{The world-volume energy-momentum tensor}
\label{bf:energy_momentum_tensor}
%~~~~~~~~~~~~~~~~~~~~~~~~~~~~~~~~~~~~~~~~~~~~~~~

The general complication in determining the world-volume energy-momentum tensor lies in the fact 
there is no clean prescription to obtain the counter-terms on a finite cut-off surface  (see however \cite{Kraus:1999di}). While the boundary terms are fixed by a variational principle, the counter-terms are necessitated for finiteness which is not an issue with a rigid UV-cutoff. In what follows we will
basically restrict to the Brown-York procedure \cite{Brown:1992br}, which was also implemented in \cite{Emparan:2013ila} as our basic guiding principle.

We claim that the quasi-local stress-energy tensor takes the form:\footnote{ We will frequently identify this tensor with its pullback $T_{ab} = \frac{\partial x^\mu}{\partial \sigma^a}\frac{\partial x^\nu}{\partial \sigma^b} T_{\mu\nu}$ onto the world-volume with coordinates $\sigma^a$, given that we take $x^\mu = (\sigma^a,r) = (v,x^i,r)$ and a constant $r=R$ surface. Analogous statements apply for $h_{ab}$ and $K_{ab}$ and the connection $\hat{\nabla}_a$. }
\begin{equation}
\label{quasi_local_stress_energy_tensor}
\kappa_5^2\,T_{\mu\nu}= e^{4 U + V}\left(K_{\mu\nu}-K\, h_{\mu\nu}\right)+ \left(n_\rho \partial^\rho e^{4 U+V}-4\, e^{3U+V}-\, e^{4U}+Q\right)h_{\mu\nu}\,.
\end{equation}
Here $K_{\mu\nu}=-h\indices{_\mu^\rho} h\indices{_\nu^\sigma} \nabla_\rho n_\sigma$ is the extrinsic curvature of the surface in the direction of the outward pointing space-like normal $n_\sigma$.
$h_{\mu\nu}=g_{\mu\nu}-n_\mu n_\nu$ is the projection tensor which projects parallel to the surface and $K= g^{\mu\nu} K_{\mu\nu}$ is the trace of the extrinsic curvature.

The first terms which involve the extrinsic curvature tensor $K_{\mu\nu}$ and its trace $K$ are easy to understand: They constitute the usual terms which arise from the variation of the Gibbons-Hawking boundary term (present to ensure a consistent variational principle),
\begin{equation}
	\frac{1}{\kappa_5^2} \int d^4x\, \sqrt{-h}\, e^{4 U+V} K\,,
\end{equation}
with respect to the induced metric~$h^{\mu\nu}$. The origin of the third term involving the derivatives of the scalars $U,V$ owes to our choice of conformal frame for the action \eqref{MMT_action} (this was referred to as the naked frame in \cite{Emparan:2013ila}). Its origins can also be traced back to the Gibbons-Hawking term in ten dimensions, or by simply realizing that 
$4 U + V = 5\, \varphi_{\textrm{\scriptsize \cite{Emparan:2013ila}}}$. The last three terms are local counter-terms introduced to subtract off the curvature contributions from the internal space (the squashed ${\bf S}^5$) and the D3-brane tension.

We quickly review the derivation of terms. Differentiating the Einstein frame quantities by a bar in the following, consider evaluating the trace of the extrinsic curvature \mbox{$\bar{K}=-\bar{\nabla}_\mu n^\mu$}. One finds
\begin{equation}
\begin{aligned}
	\bar{K} &=-\left(e^{4U+V}\sqrt{-h}\right)^{-1} n^\mu \partial_\mu \left(e^{4U+V}\sqrt{-h}\right)\\
	&= K-\left(e^{4U+V}\right)^{-1}n^\mu \partial_\mu e^{4 U+V}\,.
\end{aligned}
\end{equation}
Including the rescaling of the world-volume volume element and using $\bar{K}_{\mu\nu}=K_{\mu\nu}$ we see that the frame translation precisely gives us the term of interest. The operative point is simply that in computing the trace, we have to account for the bending of the brane in the transverse directions, hence this is a genuine contribution from the curved internal space. 

The variational principle having been dealt with, we now have to worry about the counter-terms. The procedure of \cite{Brown:1992br} asks to embed the surface of interest into a reference space-time, which we take to be flat Minkowski space, compute the Brown-York tensor in that reference space-time and subtract it from the Brown-York tensor of interest. Thus, we would like to put the fibre bundle into flat space, including its breathing and squashing modes, and compute its extrinsic curvature tensor and the corresponding trace. A similar reasoning as in \cite{Emparan:2013ila} who accounted for the extrinsic curvature of a round ${\bf S}^5$, now generalized to 
allow for the squashing, leads to an contribution $- e^{4U+V}\left(4\,e^{-U} + e^{-V}\right)$, with the numerical coefficients being set by the dimensions of the base spaces and fibre respectively. It would be interesting to derive this from first principles, but we have not done so, being stymied in finding a proper embedding. One consistency check we can offer is that the term does  reduce, in the near-horizon decoupling limit, to the standard holographic renormalization counter-terms encountered in asymptotically AdS spacetimes.\footnote{ A more direct argument would also reveal whether we get higher order gradient contributions from the scalars and massive vectors for the stress tensor. Analogy with the forced fluid analysis of \cite{Bhattacharyya:2008ji}  suggests potential second order contributions of the form $\hat{\nabla}_a \phi\hat{\nabla}_b \phi$ or
$h_{ab}\,\hat{\nabla}^2 \phi$, with $\phi \in \{U,V\}$.}

The last term in \eqref{quasi_local_stress_energy_tensor} proportional to $ Q$ subtracts the energy-density of a stack of extremal D3-branes as in \cite{Emparan:2013ila}. Since we work with a fixed brane charge we remove this Lorentz invariant ground state contribution.

%~~~~~~~~~~~~~~~~~~~~~~~~~~~~~~~~~~~~~~~~~~~~~~~
\subsubsection{Energy density and pressure}
%~~~~~~~~~~~~~~~~~~~~~~~~~~~~~~~~~~~~~~~~~~~~~~~

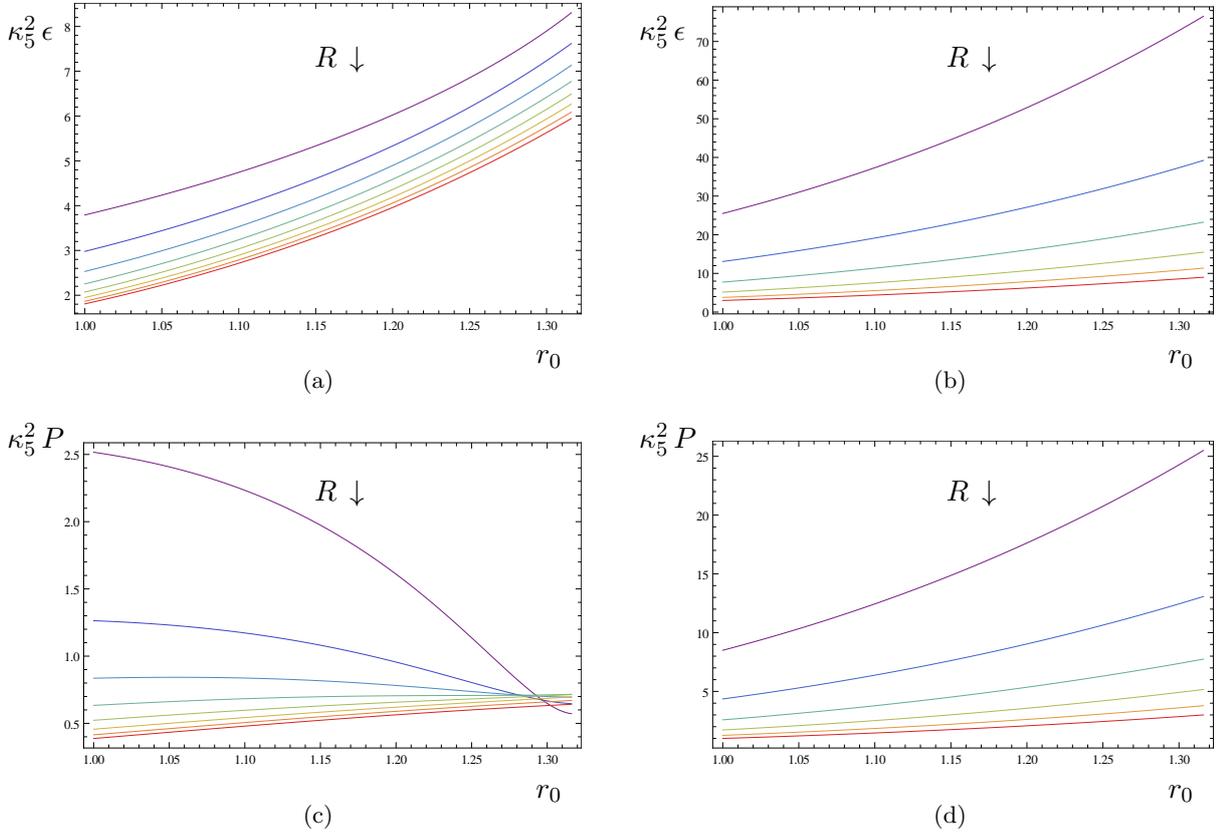
\begin{figure}[ht]
  \centering 
  \subfigure[]{\label{fig:energybl}\includegraphics[width=0.45\textwidth]{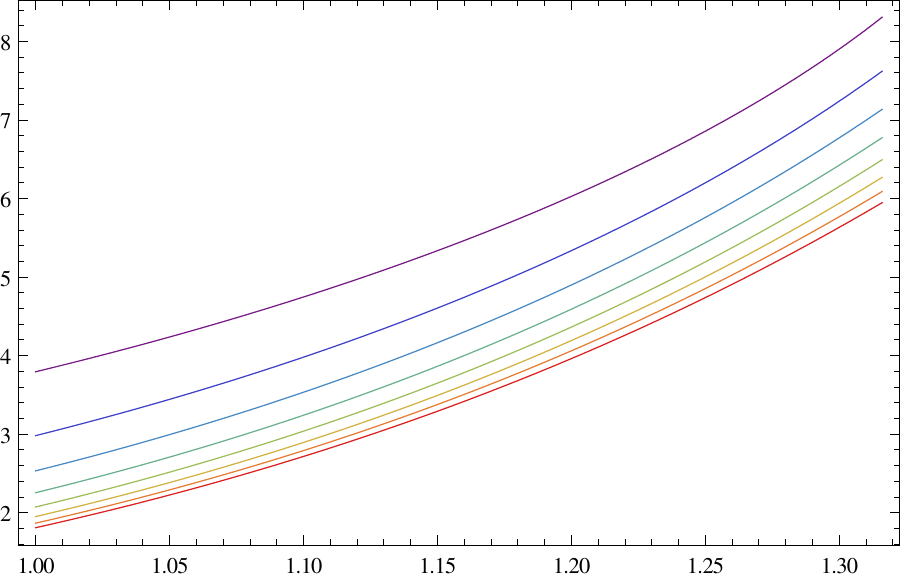}\put(-215,110){$\kappa _5^2\, \epsilon$}\put(-17,-12){$r_0$}}\put(-100,100){$R\,\downarrow$}
  \hfill 
  \subfigure[]{\label{fig:energyfg}\includegraphics[width=0.45\textwidth]{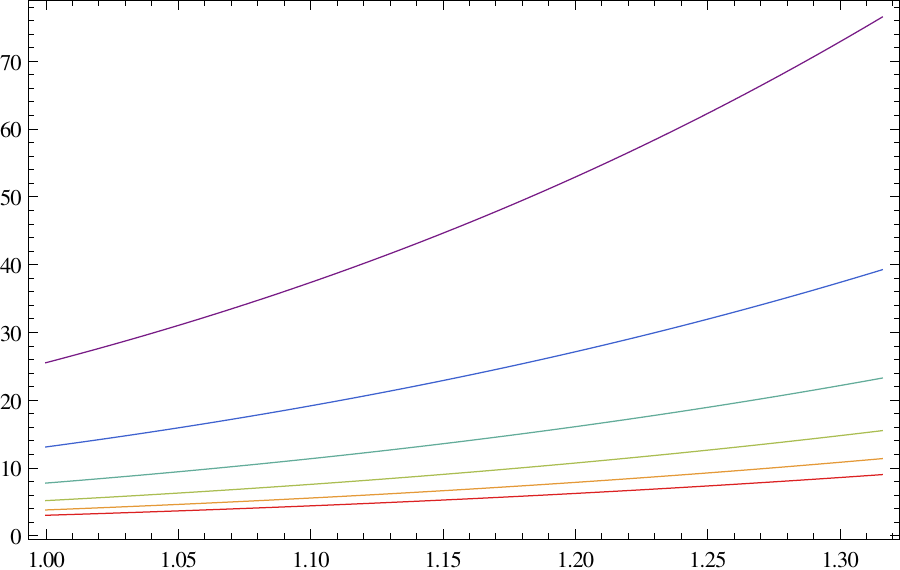}\put(-215,110){$\kappa _5^2\, \epsilon$}\put(-17,-12){$r_0$}}\put(-100,100){$R\,\downarrow$}              
  \hfill 
  \subfigure[]{\label{fig:pressurebl}\includegraphics[width=0.45\textwidth]{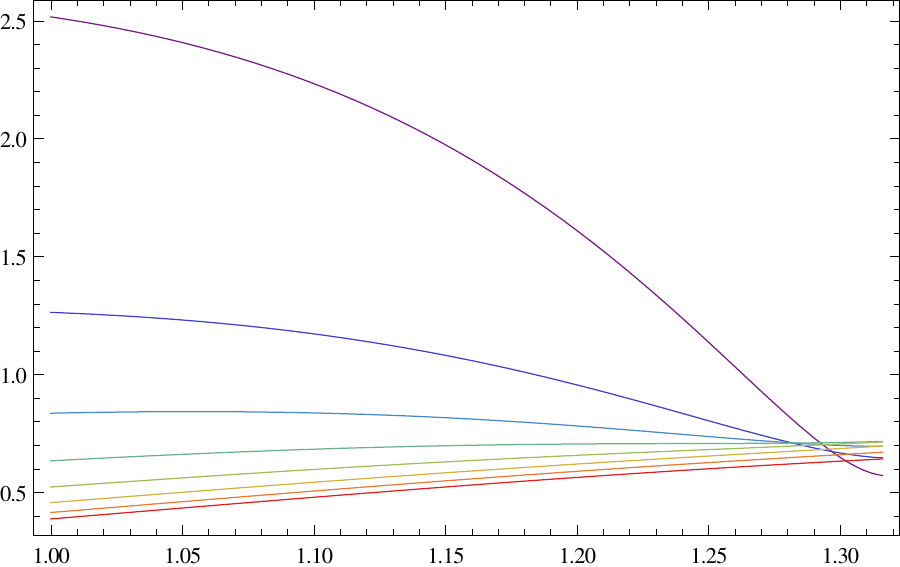}\put(-215,120){$\kappa _5^2\, P$}\put(-17,-12){$r_0$}}\put(-100,100){$R\,\downarrow$}
  \hfill 
  \subfigure[]{\label{fig:pressurefg}\includegraphics[width=0.45\textwidth]{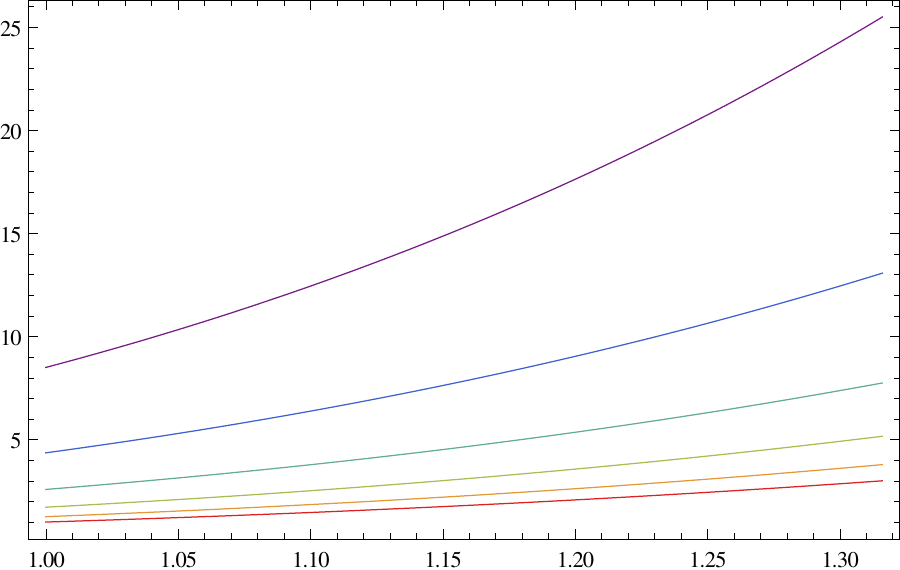}\put(-215,120){$\kappa _5^2\, P$}\put(-17,-12){$r_0$}}\put(-100,100){$R\,\downarrow$}
  \caption{We show the energy density \eqref{bf:energy_density} and pressure \eqref{bf:pressure} versus the parameter $r_0$. To generate these plots we set $r_+=1$ and choose $L=1$ in (a) and (c) and $L=1000$ in (b) and (d). In addition, the different colours correspond to different values of $R$. The purple line corresponds to the smallest plotted value of $R$, while the red one to the biggest value. Note that $L=1000$ approximates the decoupling limit and we obtain the fluid/gravity results for $L\approx R \gg r_+,\, r_0$, i.e., the red line on the figures (b) and (d).} 
  \label{fig:energypressure} 
\end{figure}

From the expression of the quasi-local energy-momentum tensor \eqref{quasi_local_stress_energy_tensor}, we may easily extract energy density $\epsilon$ and pressure $P$. As clear by construction, the zeroth order energy-momentum tensor is of ideal fluid form
\begin{equation}
T^{(0)}_{ab}=\epsilon\, u_a u_b + P P_{ab}\,.
\label{idealT}
\end{equation}
with energy density given by
\begin{equation}
\label{bf:energy_density}
\begin{aligned}
\kappa _5^2\, \epsilon ={}& R^4 H_R\left(1+\frac{4}{\sqrt{g_R}}\right)-Q\\
&\qquad +\sqrt{\frac{f_R}{g_R}}\left(-3 L^4-5 R^4 H_R-\frac{5}{4} R^5  H'(R)+ \frac{R^5 H_R}{2g_R}  g'(R)\right)\,.
\end{aligned}   
\end{equation}
The pressure is
\begin{equation}
\label{bf:pressure}
\begin{aligned}
P ={}& - \epsilon + \frac{\sqrt{g_R}}{\kappa_5^2 R^{12}\sqrt{f_R} H_R} \left(L^8 \left(2 r_0^4 R^4-3 q^2 R^2\right)+r_0^4 R^6 \left(2
   R^6-q^2\right)\right.\\
&\qquad\qquad\left.+L^4 \left(-2 q^4+q^2 \left(r_0^4 R^2-5 R^6\right)+4 r_0^4 R^8\right)\right)\,.
\end{aligned}   
\end{equation}
Both expressions as expected agree with \cite{Emparan:2013ila} if we take $q\rightarrow 0$ and perform the appropriate coordinate change (cf., \S\ref{chargeless_limit}). In the near-horizon limit, we obtain (setting $L =1$)
\begin{equation}
	\epsilon = \frac{3 r_0^4}{2\kappa_5^2}\,,\qquad P = \frac{r_0^4}{2\kappa_5^2}\,,
\end{equation}
These are the expressions for energy density and pressure of the charged fluid plasma in the fluid/gravity decoupling limit \cite{Banerjee:2008th}. The behaviour of these in different regimes is displayed in 
Fig.~\ref{fig:energypressure}.

Note that the trace of the world volume energy-momentum tensor \eqref{quasi_local_stress_energy_tensor} is non-vanishing:
\begin{equation}
\label{bf:trace_em_tensor}
   T\indices{^a_a} = -\epsilon + 3 P \neq 0\,.
\end{equation}
This is not surprising since by taking our cut-off surface at finite $R$ we introduce a length-scale into the theory such that it should not be conformal. However we record this explicitly since at first order in the perturbative analysis of the charge currents we will soon observe structures which are apparently constrained by Weyl covariance nevertheless.

%~~~~~~~~~~~~~~~~~~~~~~~~~~~~~~~~~~~~~~~~~~~~~~~
\subsection{The world-volume charge currents}
\label{bf:charge_currents}
%~~~~~~~~~~~~~~~~~~~~~~~~~~~~~~~~~~~~~~~~~~~~~~~

We now turn to the computation of the Dirichlet surface currents following \cite{Brown:1992br}. They may be read off from the boundary terms after an integration by parts of the action \eqref{MMT_action}, possibly using \eqref{MMT_action_rotated}. Due to the interaction of the gauge field and Proca field different bases for the two currents may be used (see also the discussion at the end of \S\ref{subsec:consistent_truncation}). We may work in the basis in which one current,
\begin{equation}
\mathcal{J}_\textrm{phys}^\mu = \left(\frac{\sqrt{-h}}{2\kappa_5^2} \right) \,  \left(e^{4U+3V}  {\mathcal F}^{\mu\nu}  + \frac{Q^2}{2} e^{-V}  {\mathbb F}^{\mu\nu}\right)n_\nu-\frac{Q^2}{2 \kappa_5^2}\,\kappa\, \epsilon\indices{^\mu^\nu^\rho^\sigma^\lambda}n_\nu\,\mathcal{A}_\rho\,\mathbb{F}_{\sigma\lambda}\,,
\end{equation}
is dual to the massless gauge field $A^\mu_\textrm{phys}$ defined in \eqref{physical_gauge_field} and up to an anomaly term conserved by means of the equations of motion. The other (non-conserved) current ought to be understood as the expectation value of a vector operator $\mathbf{J}^\mu$ dual to the Proca field $\mathbf{A}^\mu$. It is defined via
\begin{equation}
   \mathbf{J}^\mu =\left(\frac{e^{4(U+V)}}{\frac{Q^2}{2} + e^{4(U+V)}}\right)\left[\left(\frac{\sqrt{-h}}{2\kappa_5^2} \right)  \,  \frac{Q^2}{2}  e^{-V} {\mathbf F}^{\mu\nu}n_\nu-\frac{Q^2}{2 \kappa_5^2}\,\kappa\, \epsilon\indices{^\mu^\nu^\rho^\sigma^\lambda}n_\nu\,\mathcal{A}_\rho\,\mathbb{F}_{\sigma\lambda}\right]\,.
\end{equation}
Clearly this different interpretation originates from the non-existing gauge redundancy for the Proca field $\mathbf{A}^\mu$ (we fixed the gauge by choosing the Stueckelberg scalar $\theta = 0$), which will be reflected in the non-conservation of~$\mathbf{J}^\mu$.

We however find it convenient to treat both vectors simultaneously and work with the following basis instead:
\begin{align}
\label{bf:conserved_currents_massless}
   \mathcal{J}^\mu &= \left(\frac{\sqrt{-h}}{2\kappa_5^2} \right) \,e^{4 U + 3 V}  \mathcal{F}^{\mu\nu }n_\nu\,,\\
\label{bf:conserved_currents_mixed}
   \mathbb{J}^\mu &= \left(\frac{\sqrt{-h}}{2\kappa_5^2} \right) \frac{Q^2}{2} \,e^{- V}  \mathbb{F}^{\mu\nu }n_\nu-\frac{Q^2}{2 \kappa_5^2}\,\kappa\, \epsilon\indices{^\mu^\nu^\rho^\sigma^\lambda}n_\nu\,\mathcal{A}_\rho\,\mathbb{F}_{\sigma\lambda}\,.
\end{align}
This basis is in fact closer to the fields used in the original Lagrangian \eqref{MMT_action} which results in slightly simplified expressions, e.g., only one of the currents receives contributions from the Chern-Simons term. Neither of the currents $\mathcal{J}^\mu$ and $\mathbb{J}^\mu$ is actually conserved as we will show later, but we may simply get the physical i.e., (anomalous) conserved current from these via $\mathcal{J}^\mu+\mathbb{J}^\mu = \mathcal{J}_\textrm{phys}^\mu$.

Evaluating the expressions for the currents on the solution of interest we find:\footnote{ Again, we use $\mathcal{J}_\mu$ and the expression $\mathcal{J}_a = \frac{\partial x^\mu}{\partial \sigma^a} \mathcal{J}_\mu$, i.e., the pullback to the world-volume, interchangeably -- likewise for $\mathbb{J}_a$.}
\begin{align}
\label{bf:background_current_massless}
 2 \kappa_5^2 \, \mathcal{J}_a &= 2q\, \sqrt{L^4+r_0^4}\left(\frac{ L^4+3 R^4}{H_R^{1/4}  R^4}\right)u_a
 \\
\label{bf:background_current_mixed}
 2 \kappa_5^2 \, \mathbb{J}_a &=  \frac{4 q\, L^4 \sqrt{L^4+r_0^4}}{H_R^{1/4}  R^{4}}\, u_a
\end{align}
which defines the charge densities via $\mathcal{J}_a = n_\mathcal{J}\, u_a$ and $\mathbb{J}_a = n_\mathbb{J}\, u_a$. In the near-horizon limit with the cut-off surface located on the AdS boundary (\mbox{$R\to \infty$}) we recover the  result of \cite{Banerjee:2008th} (after taking into account the different normalizations).

Furthermore, we may use the obtained values for the entropy density~\eqref{entropy_density}, temperature~\eqref{temperature}, energy density~\eqref{bf:energy_density}, pressure~\eqref{bf:pressure}, chemical potentials~\eqref{bf:chemical_potential_massless}, \eqref{bf:chemical_potential_mixed} and charge densities~\eqref{bf:background_current_massless}, \eqref{bf:background_current_mixed} to verify the following Smarr relation
\begin{align}
   \epsilon + P = T s + \mu_\mathcal{J} \, n_\mathcal{J} + \mu_\mathbb{J} \, n_\mathbb{J}\,.
\end{align}
It is noteworthy that the Smarr relation does also include a non-vanishing contribution from the massive gauge field; we may also write it as
\begin{align}
   \epsilon + P = T s + \mu_{\mathcal{J}_\textrm{phys}} \,n_{\mathcal{J}_\textrm{phys}} + \mu_\mathbf{J} \,n_\mathbf{J}\,,
\end{align}
with charge densities and chemical potentials defined in the analogous way as before.

It is worth remarking that the spinning black D3-brane satisfies a simple Smarr relation in ten dimensions with angular chemical potentials. In the process of dimensional reduction coupled with the imposition of the Dirichlet boundary condition at $r=R$, we have made a certain choice for the sources and expectation values of the various operators. While our choice for the metric  and physical gauge field degrees of freedom was motivated by the physical necessity of agreement with the familiar results on the AdS boundary in the decoupling limit, the choice for the Proca field and the scalars was based on pragmatism.  As a result we find a non-trivial source and expectation value for the operator dual to the Proca field ${\mathbf J}^\mu$. This can and does enter the Ward identity determining the equation of state on the Dirichlet surface, and hence we see it explicitly contributing to the Smarr relation. It would be interesting to derive the corresponding Ward identity directly by examining the counter-term 
Lagrangian, a task we leave for future analysis.
% % Part 4 - Background pertrubations
%!TEX root=main.tex

%~~~~~~~~~~~~~~~~~~~~~~~~~~~~~~~~~~~~~~~~~~~~~~~
\section{Long wavelength perturbations}
\label{sec:blackfold_setup}
%~~~~~~~~~~~~~~~~~~~~~~~~~~~~~~~~~~~~~~~~~~~~~~~

Now, the usual logic of the fluid/gravity correspondence and blackfold paradigm asks to promote the a priori constant velocity $u^a$, parameters $L$, $r_0$ and charge $q$ to \emph{fields} depending on the world-volume coordinates $\sigma^a$.
\begin{equation}
   u^a,\, L,\, r_0,\, q\quad\Rightarrow \quad u^a(\sigma^b),\, L(\sigma^a),\, r_0(\sigma^a),\, q(\sigma^a)\,.
\end{equation}

In general, the equations of motion for the gravitational setup are not satisfied anymore. However, if we expand our parameters $u^a,\, L,\, r_0,\, q$ in a hydrodynamic-like derivative expansion with respect to its world-volume dependence, we may order by order in this derivative expansion \emph{impose} the gravitational equations of motion onto the setup. This then constrains the perturbations of $u^a,\, L,\, r_0,\, q$ in a particular way which can be interpreted as the constitutive relations and current conservation equations of the familiar hydrodynamics of specific fluids. Our setup was designed to describe the single $U(1)$ charge generalization of the setup described in \cite{Emparan:2013ila} and in a particular limit reduce to \cite{Banerjee:2008th,Erdmenger:2008rm}.

We compute the first order variations from \eqref{seed_metric} by working in the vicinity of $\sigma^a = 0$ with the choice of local rest frame set by $u^a =\{1,0,0,0\}$. Thus, the variation of the velocity to first order in the derivative expansion is given by
\begin{equation}
   u_a d\sigma^a =-dv+ \sigma^a\, \partial_a \beta_i \, dx^i\,, \quad \textrm{or}\quad \delta u^0=0\quad \textrm{and}\quad \delta u^i=\sigma^a \,\partial_a \beta^i\,.
\end{equation}
The parameters $r_0,\, L,\, q$ are varied as
\begin{equation}
   \delta r_0 = \sigma^a\, \partial_a r_0\,,\quad
   \delta L = \sigma^a \,\partial_a L\,,\quad
   \delta q = \sigma^a \, \partial_a q\,.
\end{equation}
Note that we would like to describe the intrinsic dynamics of a \emph{fixed number of D3-branes}, i.e.,$\delta Q = 0$. Thus, from \eqref{QtoL} we learn of the following relation between $\delta L$ and $\delta r_0$:
\begin{equation}
\label{constraint_delta_L}
   \delta L =-\frac{L r_0^3}{2 L^4+r_0^4} \,\delta r_0\,.
\end{equation}
In the following we will always trade $\delta L$ by $\delta r_0$ using this relation.

The total metric which is inserted into the equations of motion is computed from the background seed metric \eqref{seed_metric} as
\begin{equation}
  \label{fluid_gravity_perturbations_metric}
   ds^2=ds^2_0+\left(\frac{\delta}{\delta u^a} ds^2_0\right) \delta u^a+\left(\frac{\delta}{\delta r_0} ds^2_0\right) \delta r_0+\left(\frac{\delta}{\delta L} ds^2_0\right) \delta L+\left(\frac{\delta}{\delta q} ds^2_0\right) \delta q\,.
\end{equation}
We proceed likewise with the gauge fields
\begin{equation}\label{fluid_gravity_perturbations_gaugefields}
\begin{split}
   \mathcal{A}&=\mathcal{A}_0+\left(\frac{\delta}{\delta r_0} \mathcal{A}_0\right) \delta r_0+\left(\frac{\delta}{\delta L} \mathcal{A}_0\right) \delta L+\left(\frac{\delta}{\delta q} \mathcal{A}_0\right) \delta q\,,\\
   \mathbf{A}&=\mathbf{A}_0+\left(\frac{\delta}{\delta r_0} \mathbf{A}_0\right) \delta r_0+\left(\frac{\delta}{\delta L} \mathbf{A}_0\right) \delta L+\left(\frac{\delta}{\delta q} \mathbf{A}_0\right) \delta q\,,
\end{split}
\end{equation}
and dilatons
\begin{equation}
   U=U_0+\left(\frac{\delta}{\delta L} U_0\right) \delta L\,,\qquad
   V=V_0+\left(\frac{\delta}{\delta L} V_0\right) \delta L+\left(\frac{\delta}{\delta q} V_0\right) \delta q\,,
\end{equation}
where we have made apparent that the background profiles of the dilatons depend only on certain parameters.

The fluctuations of the seed metric have to be compensated by explicit correction terms to ensure that we have a solution to the equations of motion at first order in the gradient expansion. These fluctuations are nicely categorized by their transformation under the spatial $SO(3)$ symmetry which leaves fixed our inertial frame choice: $u^a =\{1,0,0,0\}$. By symmetry, the different irreducible representations decouple.  We will display explicit results for the tensor and vector fluctuations in what follows. The scalar sector has proven quite intransigent for explicit analysis, thus despite the presence of a non-trivial transport coefficient in that sector, we will ignore it in the sequel.

% Part 5 - Perturbations
%!TEX root=main.tex

%~~~~~~~~~~~~~~~~~~~~~~~~~~~~~~~~~~~~~~~~~~~~~~~
\section{Perturbations in the tensor sector of 
\texorpdfstring{$SO(3)$}{SO(3)}}
\label{sec:tensor_sector}
%~~~~~~~~~~~~~~~~~~~~~~~~~~~~~~~~~~~~~~~~~~~~~~~

The tensor sector is as usual the easiest one to deal with, since there is only a contribution from the metric. Moreover, this tensor perturbation obeys a minimally coupled scalar equation of motion (in non-Einstein frame) and is therefore relatively simple to integrate.

We start off with a tensor fluctuation in the background \eqref{fluid_gravity_perturbations_metric} with the following normalization
\begin{equation}
   ds^2_T= \sqrt{\frac{H_R}{H(r)}}\, \alpha_{ij}(r)\, dx^i dx^j\,,
\end{equation}
in which $\alpha_{ij}$ is a symmetric traceless tensor of the spatial $SO(3)$ symmetry group. The equation of motion for this fluctuation reads
\begin{equation}
   \frac{d}{dr} \left(r^5 f(r)  \frac{d}{dr} \alpha_{ij}\right)=-2 \sqrt{\frac{f_R g_R}{H_R^{1/2}}} \,\frac{d}{dr}\left(C_{ij}+r^5\sqrt{\frac{H(r)}{g(r)}} \sigma_{ij}\right)\,,
\end{equation}
with $\sigma_{ij}=\partial_{(i} \beta_{j)} - \frac{1}{3} \delta_{ij} \partial^k \beta_k$ being the tensorial part of the fluid/gravity perturbations described in \eqref{fluid_gravity_perturbations_metric}. Evidently, it only comes from the spatial dependence of the boost parameters $u^a$. The equation is a minimally coupled scalar equation of motion with shear source; the left hand side may be written as $\partial_\mu \left(g^{\mu\nu} e^{4U + V} \sqrt{-g} \,\partial_\nu \,\alpha_{ij}\right)$. Note that the appearance of the dilatons is due to the non-Einstein frame. It is easy to see that this directly generalizes the equivalent equation of motion of the fluid/gravity correspondence~\mbox{\cite{Banerjee:2008th,Erdmenger:2008rm}}. In the near-horizon limit \eqref{near_horizon_scaling}, \eqref{near_horizon_scaling_fhg}, we get perfect agreement with the known results.

We may easily integrate this expression and fix the integration constant by imposing regularity at the future horizon: 
\begin{equation}
   \partial_r \alpha_{ij} =-2 \sqrt{\frac{f_R g_R}{H_R^{1/2}}} \, \left(\frac{r^5\sqrt{\frac{H(r)}{g(r)}}-r_+^5 \sqrt{\frac{H(r_+)}{g(r_+)}}}{r^5 f(r)}\right) \; \sigma_{ij}\,.
\end{equation}
This analysis is greatly facilitated by working in manifestly regular coordinates in the vicinity of the horizon.

The second integration is also straightforward and can be expressed in terms of the tortoise coordinate $r_\star$. The corresponding integration constant is fixed by requiring that at the cut-off $r=R$ the tensor perturbation  vanish. This ensures that the induced metic on the cut-off surface at $r=R$ remains the Minkowski metric to first order in the perturbations. Implementing this we arrive at the solution for the metric fluctuation
\begin{equation}
  \label{solution_tensor_sector}
   \alpha_{ij}(r)= -2 \,\sqrt{\frac{f_R g_R}{H_R^{1/2}}}\, \left(r_\star - R_\star  +r_+^5 \sqrt{\frac{H(r_+)}{g(r_+)}}\int_r^R \frac{dr^\prime}{{r^\prime}^5 f(r^\prime)}\right)\sigma_{ij}\,.
\end{equation}

Compared to \cite{Emparan:2013ila} not much has changed. Our solution just accounts for the appropriate generalization of the metric functions \eqref{warp_factors}, to account for the charge density due to the spin  of the branes. This is responsible for the presence of $g(r)\neq 1$ in the result above, since there is a relative squashing of the $U(1)$ fibre. We may thus also expect the shear viscosity $\eta$ of our setup to trivially generalize the chargeless result of \cite{Emparan:2013ila}.

Also the comparison with \cite{Banerjee:2008th,Erdmenger:2008rm} (or the AdS cut-off setup \cite{Bai:2012ci}) matches perfectly. The combination of integrals and boundary conditions in \eqref{solution_tensor_sector} exactly reproduces their results, when we take the near-horizon limit of \S\ref{decoupling_limit} and put the Dirichlet surface from finite $R$ to $R\rightarrow \infty$, i.e., the AdS boundary. Altogether we expect then the shear viscosity to be proportional to the entropy density and satisfy the KSS bound \cite{Kovtun:2004de}.

%~~~~~~~~~~~~~~~~~~~~~~~~~~~~~~~~~~~~~~~~~~~~~~~
\section{Perturbations in the vector sector of 
\texorpdfstring{$SO(3)$}{SO(3)}}
\label{sec:vector_sector}
%~~~~~~~~~~~~~~~~~~~~~~~~~~~~~~~~~~~~~~~~~~~~~~~

The relative simplicity of the tensor sector belies the complications inherent in the problem. These become immediately manifest when we turn to the vector sector. Now we are to deal with the contributions from the vector fields in addition to the metric. Decoupling the resulting equations requires some significant amount of work as we shall now demonstrate.

%~~~~~~~~~~~~~~~~~~~~~~~~~~~~~~~~~~~~~~~~~~~~~~~
\subsection{Perturbation ansatz}
%~~~~~~~~~~~~~~~~~~~~~~~~~~~~~~~~~~~~~~~~~~~~~~~

As already outlined, there are several perturbations of importance which transform as vectors under the spatial $SO(3)$ symmetry. First of all, there are the two kinds of vector perturbations of the metric:
\begin{equation}
\label{Vector_Perturbation_ansatz}
ds^2_V=2\sqrt{\frac{H_R}{H(r)}}\left(1-\frac{f(r)\, g(r) }{f_R \,g_R } \right)w_i(r)\, dx^i dv+z_i(r)\, dx^i dr\,,
\end{equation}
where the way we have written the first perturbations $w_i(r)$ is analogous to the way $u_a$ appears in \eqref{seed_metric}. This will be of importance later on, when we identify the physical information encoded in the integration constants which we are going to get. A shift in $w_i(r)$ can then be absorbed into a redefinition of the fluid velocity, whose overall constant value is of course a free parameter, cf., Appendix~\ref{sec:fix_int_const}.

The perturbation  $z_i(r)$ on the other hand does not contribute to the equations of motion. Therefore it is pure gauge and may be set to zero which we henceforth do.

Apart from the metric perturbations both gauge fields contribute further vector perturbations. In the same way we constructed the perturbation $w_i(r)$ above to appear as $u_a$ in~\eqref{seed_metric}, equations \eqref{seed_gauge_field}, \eqref{seed_gauge_field_massive} tell us how we have to incorporate $w_i(r)$ in the gauge fields. In addition we parametrize the independent perturbations $v_i$ and $\mathbf{v}_i$ as follows
\begin{align}
\mathcal{A}_V&=-\frac{q Q}{2L^2}\,\frac{g(r)}{r^6 H(r)}\,\sqrt{\frac{ H_R^{1/2}}{f_R\, g_R}}\,w_i(r)\, dx^i+v_i(r) \,dx^i\,,\label{bf:first_order_gauge_fields}\\
\mathbf{A}_V&=-\frac{2  L^2 q}{Q}\,\frac{f(r) g(r)}{r^2 H(r)}\,\sqrt{\frac{ H_R^{1/2}}{f_R\, g_R}}\,w_i(r)\, dx^i+\mathbf{v}_i(r) \,dx^i\,.\label{bf:first_order_gauge_fields1}
\end{align}

Next, we plug all perturbations just introduced into the Einstein \eqref{MMT_Einstein_equation} and Maxwell equations \eqref{MMT_Maxwell_equation}, \eqref{MMT_Maxwell_equation_massive} of motion. The set of three coupled second order non-homogeneous ODE's is rather complicated but may be solved exactly in a step by step procedure. Let us denote the Einstein equations by $E_{\mu\nu}=0$, the first Maxwell equation (from \eqref{MMT_Maxwell_equation}) by $\mathcal{M}_{\mu}=0$ and the equation of motion derived by variation of the action with respect to $\delta\mathbb{A}^\mu$ (from \eqref{MMT_Maxwell_equation_massive}) by $\mathbb{M}_{\mu}=0$.

%~~~~~~~~~~~~~~~~~~~~~~~~~~~~~~~~~~~~~~~~~~~~~~~
\subsection{Constraint equation}
%~~~~~~~~~~~~~~~~~~~~~~~~~~~~~~~~~~~~~~~~~~~~~~~

The constraint equation in the vector sector may be computed directly from the Einstein equations by the linear combination
\begin{equation}
g^{rr}E_{ri}+g^{rv}E_{vi}=0\,.
\end{equation}
Explicitly we obtain
\begin{equation}
\label{vector_constraint}
\begin{split}
\beta_{i,v}+\frac{q}{R^6}\left(\frac{g_R}{H_R}-\frac{1}{f_R}\right)\,q_{,i}+ \left[\frac{2 L^4 r_0^4}{R^4 f_R} -\frac{2}{f_R} \left(2 L^4+r_0^4\right)+\frac{L^4 q^2 }{R^6}\left(\frac{g_R}{H_R}-\frac{3}{f_R}\right)\right.\\ \left.+R^4H_R-\frac{2 L^4 R^4}{r_0^4}\right]\frac{r_0^3}{R^4(2 L^4+r_0^4)}\,r_{0,i}=0\,.
\end{split}
\end{equation}
which in the near-horizon limit \eqref{near_horizon_scaling} reduces to $r_{0,i} + r_0\beta_{i,v} = 0$ as in \cite{Banerjee:2008th}.

The equation \eqref{vector_constraint} is the vectorial component of the energy-momentum conservation equation $\hat{\nabla}_\mu T^{\mu\nu} = {\cal S}^\nu$, where the ${\cal S}^\nu$ encodes the external work done on the system by external sources for other operators. For instance, if we had a spatially varying chemical potential then we would expect to see a Joule heating term $F_{\mu\nu} J^\nu$; typically ${\cal S}^\mu = \left(\text{source} \times \text{vev}\right)^\mu$.

Ideally, our holographic renormalization procedure would have indicated how to treat the sources and vevs for various fields. Since we have not done a thorough analysis we are mostly going to aim for a consistency check with the conservation Ward identity (which should be derivable from the counter-term action). Taking the divergence of the stress tensor \eqref{quasi_local_stress_energy_tensor} with respect to the world-volume metric $h_{\mu\nu}$ we find an expression for ${\cal S}^\mu$ in terms of the other fields
\begin{align}
\kappa_5^2 \hat{\nabla}^\mu T_{\mu i}&=\left( \hat{\nabla}^\mu e^{4 U +V}\right)\left(K_{\mu i}-h_{\mu i} K\right)+\hat{\nabla}_ i\left(n^\mu \nabla_\mu e^{4 U+V}\right)-\hat{\nabla}_ i\left(4 e^{3U+V}+e^{4U}\right)\nonumber\\
&-4 e^{4 U+V} (\hat{\nabla}_ i U)n^\rho \nabla_\rho U -4 e^{4 U+V}  n^\mu h\indices{_ i^\rho} \nabla_\mu \nabla_\rho U \label{conservation_equation}\\
&- e^{4 U+V} (\hat{\nabla}_ i V)n^\rho \nabla_\rho V - e^{4 U+V}  n^\mu h\indices{_ i^\rho} \nabla_\mu \nabla_\rho V\nonumber\\
&- e^{4 U+V}\left(\frac{1}{2} e^{2 V} n^\mu \mathcal{F}_{\mu\rho}\mathcal{F}\indices{_ i^\rho}+\frac{Q^2}{4} e^{-4 U -2 V} n^\mu \mathbb{F}_{\mu\rho}\mathbb{F}\indices{_ i^\rho}+ 2 Q^2 e^{-8 U} n^\mu \mathbf{A}_\mu \mathbf{A}_ i\right) \,.\nonumber
\end{align}
In this expression, $\hat{\nabla}_\mu$ is the connection on that surface compatible with $h_{\mu\nu}$, i.e., for any tensor $T^{\nu\cdots}_{\rho\cdots}$, we have $\hat{\nabla}_\mu T^{\nu\cdots}_{\rho\cdots}=h\indices{_\mu^\sigma} h\indices{^\nu_\lambda} h\indices{_\rho^\tau} \cdots \nabla_\sigma T^{\lambda \cdots}_{\tau \cdots}$ (cf., Lemma 10.2.1 of \cite{Wald:1984rg}). 
Apart from the definition of the quasi-local energy-momentum tensor we used the Gauss-Codazzi equation 
\begin{equation}
 \hat{\nabla}^\mu\left(K_{\mu\nu}-h_{\mu\nu} K\right)=-h\indices{_\nu^\rho} \,R_{\rho\sigma}\, n^\sigma 
\end{equation}
along with the equation of motion \eqref{MMT_Einstein_equation} to arrive at \eqref{conservation_equation}.

It is clear that on the cut-off surface the operators dual to the scalars $\{U,V\}$ and the Proca field ${\mathbf A}$  acquire non-vanishing vevs. Should we in addition know the source terms, we would be able to isolate the contributions and infer the external work done.

In the absence of this knowledge we mainly focus on checking that the content of~\eqref{vector_constraint} can be assembled into the form required. This is relatively simple; for the quasi-local stress-energy tensor $T_{ab}$ of the fluid living on the cut-off surface at $r=R$ obtained, one can compute the divergence at leading order in the gradient expansion, i.e., for the ideal fluid stress tensor \eqref{idealT}. This gives a contribution (up to a  factor of $\kappa_5^2$)
\begin{equation}
	(\epsilon+P)\beta_{i,v}+ \partial_i P\,,
\end{equation}
where we have to account for the world-volume dependence arising from $r_0(\sigma^a)$, $L(\sigma^a)$ and~$q(\sigma^a)$ respectively. This part is to be identified from \eqref{vector_constraint} as the energy-momentum gradient and the reminder associated with the work done on the system.

It is important when comparing \eqref{vector_constraint} to the corresponding equation in \cite{Emparan:2013ila}
that we account for the coordinate changes described in \S\ref{chargeless_limit}. As the two choices of cut-off surfaces are unequal we have again the problem of dealing with the dilaton source terms. In \cite{Emparan:2013ila} this obstacle was circumvented by fixing the cut-off to be an isodilatonic surface, to ensure absence of scalar fluctuations, but we do not have the luxury of ensuring this owing to the increased complexity of the system. 

%~~~~~~~~~~~~~~~~~~~~~~~~~~~~~~~~~~~~~~~~~~~~~~~
\subsection{Dynamical equations}
%~~~~~~~~~~~~~~~~~~~~~~~~~~~~~~~~~~~~~~~~~~~~~~~

We are now in the position to state the dynamical equations and then solve them step by step. There are three independent dynamical equations coming from the Einstein equations (either $E_{ri}=0$ or $E_{vi}=0$) and the two Maxwell equations $\mathcal{M}_i=0$ and $\mathbb{M}_i=0$. The overall normalization of these dynamical equations was chosen to make the check that in the near-horizon limit of the geometry \eqref{seed_metric} they reduce to the ones stated in \cite{Banerjee:2008th} as simple as possible.

The one coming from the Einstein equations is
\begin{align}
\label{Einstein_equation}
\begin{split}
0=&\,\frac{r^{11} H(r)}{g(r)}\Big(f_R g_R-f(r) g(r)\Big) w_i''(r)\\
&+2q\,r^2\,\sqrt{L^4+r_0^4}\,\sqrt{\frac{f_R g_R}{{H_R^{1/2}}}}\left(3\,r^4 H(r)\,v_i'(r)+2 L^4\,\mathbf{v}_i'(r)-\frac{8L^4}{r H(r)}\,\mathbf{v}_i(r)\right)\\
&+\frac{f_R g_R}{H(r)} \left(5 L^8 r^2+L^4 \left(3 q^2+10 r^6\right)+r^4 \left(5 r^6-q^2\right)\right)w_i'(r)\\
&+r^4H(r)\left(q^2-5 r^6-3 r_0^4 r^2\right)w_i'(r)-2\left(\frac{f_R g_R }{\sqrt{H_R}}\right)\left(\frac{r^{11} H(r)^2 }{f(r) g(r)} \right)S_{i,1}(r)\,,
\end{split}
\end{align}
where $S_{i,1}(r)$ is a source term completely fixed by the background.\footnote{ The explicit expressions for the source terns $S_{i,1}(r)$, $S_{i,2}(r)$ and $S_{i,3}(r)$ are relegated to Appendix \ref{ap:source_terms}.} One can show that at the horizon, this source term vanishes: $S_{i,1}(r_+)=0$; the differential equation is regular at $r=r_+$. In addition we observe that the perturbation $v_i$ only appears through its first derivative.

Next, we look at the Maxwell equation $\mathcal{M}_i=0$:
\begin{align}
0=&\,-\sqrt{3} q\,\sqrt{L^4+r_0^4}\, r^7 f(r)\, w_i''(r)+\sqrt{3}\,\sqrt{\frac{f_R g_R}{ H_R^{1/2}}}\left(\frac{ r^{13} f(r) H(r)}{g(r)}\right)v_i''(r)\nonumber\\
&+\sqrt{3} q \,\sqrt{L^4+r_0^4}\left(2\,f_R g_R\,\frac{ r^2\left(L^4+3
   r^4\right)}{H(r)}-r^6-3 r_0^4 r^2+5 q^2\right)\,w_i'(r)\nonumber\\
&+\sqrt{3}\,\sqrt{\frac{f_R g_R}{ H_R^{1/2}}}\,\Big(r^4H(r)(3 r^8+ r_0^4 r^4-3 q^2 r^2 )\Big.\label{Maxwell_equation_1}\\
&\Big.\qquad+4 r^{12}f(r)+q^2(r^6+3 r_0^4 r^2-5q^2)\Big) v_i'(r)\nonumber\\
&+16 \sqrt{3}\,L^4\left(L^4+r_0^4\right)\,\sqrt{\frac{f_R g_R}{ H_R^{1/2}}}\,\frac{r^3}{H(r)}\,\mathbf{v}_i(r)-\sqrt{3}\left(\frac{f_R g_R}{H_R^{3/4}}\right)r^6 S_{i,2}(r)\nonumber\,,
\end{align}
where $S_{i,2}(r)$ is the source. In this equation it is noteworthy that the perturbation $\mathbf{v}_i$ appears with no derivative. Thus, in principle we could use this equation to eliminate $\mathbf{v}_i$ in the other two.

The last dynamical equation is the Maxwell equation $\mathbb{M}_i=0$:
\begin{align}
0=&\,-\frac{\sqrt{3}\,q  }{\sqrt{L^4+r_0^4}}\,r^7 f(r)\,w_i''(r)+\sqrt{3}\,\sqrt{\frac{f_R g_R}{ H_R^{1/2}}}\,
   r^9 f(r) \Big(v_i''(r)+\mathbf{v}_i''(r)\Big)\nonumber\\
&+\frac{\sqrt{3} \,q}{\sqrt{L^4+r_0^4}} \left(\frac{2 r^{6} f_R g_R }{g(r)^2}-r^6
   -3 r_0^4 r^2+5 q^2\right.\nonumber\\
&\left.\qquad+\frac{4 r^{12} f(r)-q^2 \left(3 r^6+r_0^4 r^2-3 q^2\right)}{r^6 H(r) }\right)\,g(r)\,w_i'(r)\label{Maxwell_equation_2} \\
&+\sqrt{3}\,\sqrt{\frac{f_R g_R}{ H_R^{1/2}}}\,\left(3 r^8+r_0^4 r^4-3 q^2 r^2\right.\nonumber\\
&\left.\qquad-\frac{  4 r^{12} f(r)+q^2 \left(-5 r^6+r_0^4 r^2+q^2\right) }{r^4 H(r) }\right)\,g(r)\,\Big(v_i'(r)+\mathbf{v}_i'(r)\Big)\nonumber\\
&-8 \sqrt{3}\,\sqrt{\frac{f_R g_R}{ H_R^{1/2}}}\,\frac{ r^7 }{g(r)
   }\,\mathbf{v}_i-\sqrt{3}\,\left(\frac{  f_R g_R  }{ H_R^{3/4}}\right)\left(\frac{r^{10}H(r)}{g(r)}\right)\,S_{i,3}(r)\nonumber\,,
\end{align}
with $S_{i,3}(r)$ being the source. Here, the appearance of $v_i(r)$ and $\mathbf{v}_i(r)$ reflects the fact that the combination $\mathbb{F}=\mathcal{F}+\mathbf{F}$ is the natural object to consider. However, we also see a contribution proportional to $\mathbf{v}_i(r)$, i.e., without derivatives, which comes from the mass term of the Proca field.

In the following we combine \eqref{Einstein_equation}, \eqref{Maxwell_equation_1} and \eqref{Maxwell_equation_2} in a way that the three perturbations decouple and we integrate the resulting equations.

%~~~~~~~~~~~~~~~~~~~~~~~~~~~~~~~~~~~~~~~~~~~~~~~
\subsection{Solution of the dynamical equations}
%~~~~~~~~~~~~~~~~~~~~~~~~~~~~~~~~~~~~~~~~~~~~~~~

Our strategy for integrating the dynamical equations just derived is as follows. Firstly, we integrate \eqref{Einstein_equation} -- \eqref{Maxwell_equation_2} or combinations thereof as many times as possible, without eliminating any  of the three functions $w_i(r), v_i(r), {\mathbf v}_i(r)$. Subsequently, we derive a second order differential equation for $w_i(r)$ from the integrals obtained which we integrate explicitly. This procedure leads to a total of six integration constants which we fix by imposing several conditions, which  include regularity at the horizon and at the Dirichlet cut-off surface, as well as a redefinition of the fluid velocity and choosing the Landau frame for the energy-momentum tensor.

The homogeneous part of the Einstein equation \eqref{Einstein_equation} may be integrated once,
\begin{align}
\label{Integrated_Einstein_equation}
\begin{split}
0=&\,\frac{r^{5} }{g(r)}\Big(f_R g_R-f(r) g(r)\Big) w_i'(r)-4\, r_0^4\, w_i(r) \\
&+6\,q\,\sqrt{L^4+r_0^4}\,\sqrt{\frac{ f_R g_R }{H_R^{1/2}}}\, \left(\, v_i(r)+\widetilde{C}_{i,1}+\frac{2\, L^4}{3 r^4 H(r)} \,\mathbf{v}_i(r)\right)\\
&-2\,\left( \frac{f_R g_R}{\sqrt{H_R}}\right)\int\frac{r^5 H(r)}{f(r)g(r)}\, S_{i,1}(r) \,dr\,,
\end{split}
\end{align}
where we introduced the integration constant $\widetilde{C}_{i,1}$.

The structure of the two Maxwell equations, \eqref{MMT_Maxwell_equation_massive} and \eqref{MMT_Maxwell_equation}, already suggests that a particular sum of them may be directly integrated, given that the Chern-Simons contributions are total derivatives. We thus add \eqref{Maxwell_equation_1} and \eqref{Maxwell_equation_2} in such a way that the contributions proportional to $\mathbf{A}_i$, the mass term, cancel. Then we integrate to obtain
\begin{align}
0=&\,\frac{q}{2 L^4  \sqrt{L^4+r_0^4}}\left(\frac{r^6+3 L^4 r^2+q^2}{r^4}\right)w_i'(r)-\sqrt{\frac{f_R g_R} {H_R^{1/2}}}\,\mathbf{v}_i'(r)\nonumber\\
&-\frac{1}{2 L^4 \left(L^4+r_0^4\right)}\,\sqrt{\frac{f_R g_R} {H_R^{1/2}}}\, \left(2 q^2 r^2 H(r)+3 L^8+2 L^4 \left(r^4+r_0^4\right)+r^8+\frac{q^4}{r^4}\right) v_i'(r)\nonumber\\
&-\left(\frac{3\, q\,  f_R g_R }{L^4  \sqrt{L^4+r_0^4}}\right)\left(\frac{r H(r)}{f(r) g(r)}\right)\Big(w_i(r)+\widetilde{C}_{i,2}\Big)\label{Integrated_Maxwell_Sum}\\
&+\left(\frac{f_R g_R}{H_R^{3/4}}\right)\left(\frac{r  H(r)}{f(r) g(r) }\right)\int \left(\frac{S_{i,2}(r)}{2\, L^4\left(L^4+r_0^4\right)}+S_{i,3}(r)\right)\,dr\nonumber\,.
\end{align}
The new integration constant appearing above is $\widetilde{C}_{i,2}$. In Appendix \ref{ap:source_terms}, we also show that $S_{i,2}(r)$ and $S_{i,3}(r)$ are total derivatives which allows us to integrate the relevant term in~\eqref{Integrated_Maxwell_Sum} in a closed form.

To derive the third independent equation we combine the Einstein equation \eqref{Einstein_equation} and the first Maxwell equation \eqref{Maxwell_equation_1} in such a way that the terms proportional to $\mathbf{v}_i(r)$ cancel. In addition we use \eqref{Integrated_Maxwell_Sum} to eliminate $\mathbf{v}_i'(r)$. Note that the homogeneous part of this equation can be integrated twice. Integrated once this combination leads to

\begin{align}
\label{Relation_Between_V_and_VM_integrated_once}
\begin{split}
0=&\,\left(\frac{\sqrt{L^4+r_0^4}}{q}\,\sqrt{\frac{H_R^{1/2}  }{f_R g_R}}\right)\,\frac{d}{dr}\left[\left(\frac{f_R g_R-f(r)}{f(r)}\right)\,w_i(r)\right]+v^\prime_i(r)\\
&+6 q \left(H_R^{1/4} \sqrt{f_R g_R} \,\sqrt{L^4+r_0^4}\right) \,\widetilde{C}_{i,2} \left(\frac{1}{r^7 f(r)^2}\right)\\
&+\frac{S_{i,4}(r)}{r^5 f(r)^2} - \frac{4 r_0^4}{q}\left( H_R^{1/4} \sqrt{f_R g_R }\sqrt{L^4+r_0^4}\right) \, C_{i,3}\left(\frac{1}{r^5 f(r)^2} \right)\,,
\end{split}
\end{align}
in which the new integration constant $C_{i,3}$ appears.

Up to this point we have obtained three independent integration constants $\widetilde{C}_{i,1}$, $\widetilde{C}_{i,2}$ and $C_{i,3}$. We would like one of them, or rather a particular linear combination, to describe the freedom to shift $w_i(r)$ by a constant. This shift would correspond to a redefinition of the fluid velocity $u_i$ and can thus be absorbed (see discussion in Appendix \ref{sec:fix_int_const}).
Defining
\begin{equation}
\begin{split}
   \widetilde{C}_{i,2} &= C_{i,2} + C_{i,3}\,,\\
   \widetilde{C}_{i,1} &= C_{i,1} - 4 r_0^4 \left(6\,q\,\sqrt{L^4+r_0^4}\,\sqrt{\frac{ f_R g_R }{H_R^{1/2}}}\right)^{-1} C_{i,3}
\end{split}
\end{equation}
we explicitly find that $C_{i,3}$ corresponds to this shift freedom: Subject to these redefinitions, the integration constant always appears in the combination $ w_i(r)+C_{i,3}$ in \eqref{Integrated_Einstein_equation} -- \eqref{Relation_Between_V_and_VM_integrated_once}. Integrating \eqref{Relation_Between_V_and_VM_integrated_once} again, we obtain
\begin{align}
\label{Relation_Between_V_and_VM}
\begin{split}
0=&\,\frac{\sqrt{L^4+r_0^4}}{q}\,\sqrt{\frac{H_R^{1/2}  }{f_R g_R}}\,\left(\frac{f_R g_R-f(r)}{f(r)}\right)\Big(w_i(r)+C_{i,3}\Big)+v_i(r)-C_{i,4}\\
&+6 q H_R^{1/4} \sqrt{f_R g_R} \,\sqrt{L^4+r_0^4} \; C_{i,2} \int \frac{1}{r^7 f(r)^2}+\int \frac{S_{i,4}(r)}{r^5 f(r)^2} \, dr\,.
\end{split}
\end{align}
The source term $S_{i,4}(r)$ is defined in Appendix \ref{ap:source_terms}.

Note that in \eqref{Integrated_Einstein_equation} -- \eqref{Relation_Between_V_and_VM}, $v_i(r)$ appears as $v_i(r)+C_{i,1}$ or $v_i(r)-C_{i,4}$, i.e., also with shifts by a constant. Since $v_i(r)$ describes the vector fluctuation of the gauge field $\mathcal{A}_i(r)$, we recognize that a combination of $C_{i,1}$ and $C_{i,4}$ will not have any effect on the physical observables given that only the gauge-invariant quantity $\mathcal{F}_{\mu\nu}$ is relevant. The other, linearly independent combination of $C_{i,1}$ and $C_{i,4}$ will however be important influencing the solution $w_i(r)$, as we will see shortly.\par\vspace{\baselineskip}

This concludes the first step described at the beginning of this section. Next we solve the equation \eqref{Integrated_Einstein_equation} for $\mathbf{v}_i(r)$ and use it to eliminate it from \eqref{Integrated_Maxwell_Sum}. In the resulting expression we use \eqref{Relation_Between_V_and_VM} to replace $v_i(r)$ arriving at a second order ODE fo $w_i(r)$. Solving this equations allows for a full determination of $v_i(r)$ and $\mathbf{v}_i(r)$ via \eqref{Relation_Between_V_and_VM} and \eqref{Integrated_Einstein_equation} as well.

The homogeneous part of the final ODE may be written as the following differential operator
\begin{equation}
\label{Second_order_ODE_for_w_hom}
   0=\frac{d }{dr}\left[-\widetilde{C}_{i,5}+r^3 \left(3 r^4-r_0^4\right)^2 f(r) \frac{d}{dr}\left(
  -\widetilde{C}_{i,6}+ \frac{r^4 H(r)\left[f_R g_R-f(r) g(r)\right] }{\left(3 r^4-r_0^4\right) f(r) g(r)}\,w_{i}^{(\textrm{hom})}(r)\right)\right]\,.
\end{equation}
It is clear that we have to worry about regularity of 
\begin{equation}
   w_i(r)\sim \frac{1}{f_R g_R-f(r) g(r)}
\end{equation}
at $r=R$ and the regularity of the entire inner derivative term $\frac{d}{dr}\big(\ldots w_i(r)\big)\sim f(r)^{-1}$ at the horizon, where $f(r_+)=0$. Imposing regularity in both cases fixes $\widetilde{C}_{i,5}$ and $\widetilde{C}_{i,6}$, as we show below.

The solution to this ODE is
\begin{align}
	w_i(r) &=-C_{i,3}-\frac{4}{3} L^4 q  \sqrt{L^4+r_0^4}\left(\frac{ \sqrt{f_R g_R}}{  H_R^{1/4} }\right)\left(\frac{\left(3 r^4-r_0^4\right) f(r) g(r)}{r^4 H(r)\left[f_R g_R-f(r) g(r)\right]}\right)\nonumber\\
	&\quad\times
	\left[-C_{i,6}+\int \frac{S_{i,5}(r)-C_{i,5}}{r^3 \left(3 r^4-r_0^4\right)^2 f(r)} \, dr\right.\nonumber\\
	&\quad\left.\qquad-\frac{3\sqrt{f_R g_R}}{L^4 q H_R^{1/4} \sqrt{L^4+r_0^4}}\int \left(\frac{r^3}{(3 r^4 -r_0^4)^2}\int \frac{r^5 H(r) }{f(r) g(r)}\,S_{i,1}(r) \, dr\right) dr\right.\label{full_solution_w}\\
	&\quad\left.\qquad+\frac{3}{4 L^4
   \left(3 r^4-r_0^4\right)} \left(C_{i,1}-C_{i,4}+\int \frac{S_{i,4}(r)}{r^5 f(r)^2} \, dr\right.\right.\nonumber\\
   &\quad\qquad\qquad\qquad\left.\left.+6  q \,\sqrt{f_R g_R}\,H_R^{1/4} \,\sqrt{L^4+r_0^4}\;  C_{i,2}\int \frac{1}{r^7 f(r)^2} \, dr\right)\right]\nonumber\,,
\end{align}
with the two solutions to the homogeneous part of the ODE parametrized by $C_{i,5}\sim\widetilde{C}_{i,5}$ and $C_{i,6}\sim\widetilde{C}_{i,6}$. In this expression, we recognize some of the structures of~\eqref{Integrated_Einstein_equation} and~\eqref{Relation_Between_V_and_VM}, which leads to a specific combination of source terms, integration constants and integrals. By performing an integration by parts we could easily transform the double integral into two single integrals. However, the following analysis is not simplified by this procedure, therefore we refrain from doing so. We can use this expression to compute explicit expressions for~$v_i(r)$ via \eqref{Relation_Between_V_and_VM} and~$\mathbf{v}_i(r)$ via \eqref{Integrated_Einstein_equation}.

The next step we have to deal with is to restrict our most general solution to one which allows for the description of a sensible hydrodynamic system. This we obtain by imposing physical conditions on the perturbations which fix the integration constants in the following way:

Of the integration constants we obtained in \eqref{full_solution_w}, $C_{i,3}$ may directly be set to zero since it corresponds to a shift in the fluid velocity as already remarked earlier. We may just absorb it into a redefinition of the fluid velocity $u_i-C_{i,3}\rightarrow u_i$. The integration constants $C_{i,2}$ and $C_{i,5}$ are fixed by imposing regularity at the horizon $r=r_+$ on particular combinations of the vector fluctuations and their derivatives. $C_{i,6}$ is fixed by demanding regularity for $w_i(r)$ at $r=R$. Using this, we preserve a Minkowski metric at $r=R$ given that the off-diagonal metric component $g_{vi}$ behaves like $g_{vi}\propto (r-R)\, w_i(r)$ for $r\approx R$, cf., equation~\eqref{Vector_Perturbation_ansatz}. Since we are dealing with a charged fluid, we also have the fluid frame ambiguity choosing Landau or Eckart frame. It is convenient to choose the Landau frame which will effectively determine the combination $C_{i,1}-C_{i,4}$ in \eqref{full_solution_w}. The combination $C_
{
i,1}+C_{i,4}$ need not be fixed. It corresponds to a residual gauge freedom of the gauge field~$\mathcal{A}$. For the explicit computations and values of the corresponding integration constant see Appendix~\ref{sec:fix_int_const}.

To summarize, we integrated the equations of motion in the vector sector of our setup. The metric perturbation $w_i(r)$ is read off from \eqref{full_solution_w}, from which we may deduce $v_i(r)$ and $\mathbf{v}_i(r)$ using \eqref{Relation_Between_V_and_VM} and \eqref{Integrated_Einstein_equation}. In total, we had six integration constants $C_{i,1}, \ldots, C_{i,6}$. Out of these only five linearly independent combinations appeared in \eqref{full_solution_w}. One integration constant is irrelevant since it corresponds to a shift of the gauge field $\mathcal{A}$ by a constant or in other words to the Dirichlet condition $v_i(R)$ given the perturbation ansatz \eqref{bf:first_order_gauge_fields}. The corresponding Dirichlet condition $\mathbf{v}_i(R)$ for the other gauge field $\mathbf{A}$ however is not arbitrary because of the appearance of explicit mass terms $\mathbf{A}_\mu \mathbf{A}^\mu$ in \eqref{MMT_action}. Effectively it is fixed by the Landau frame choice. The other constants are fixed by imposing 
regularity for the fluctuations at the horizon and Dirichlet cut-off surface at $r=R$ and a redefinition of the fluid velocity.

%~~~~~~~~~~~~~~~~~~~~~~~~~~~~~~~~~~~~~~~~~~~~~~~
\section{Comments on the perturbations in the scalar sector of \texorpdfstring{$SO(3)$}{SO(3)}}
\label{sec:scalar_sector}
%~~~~~~~~~~~~~~~~~~~~~~~~~~~~~~~~~~~~~~~~~~~~~~

The scalar sector is by far the most challenging sector, even surpassing the vector sector's complexity. Since we have so far not been able to integrate its dynamical equations, which we would need for determining the system's bulk viscosity, we restrict to a short outline of the relevant perturbations and an overview of the (current) constraint equations.

%~~~~~~~~~~~~~~~~~~~~~~~~~~~~~~~~~~~~~~~~~~~~~~~
\subsection{Perturbation ansatz}
%~~~~~~~~~~~~~~~~~~~~~~~~~~~~~~~~~~~~~~~~~~~~~~

A priori, the sector consists of eight coupled scalar perturbations. Three perturbations, $k(r)$, $j(r)$ and $h(r)$, stem from the metric. We parametrize them, similarly to \cite{Emparan:2013ila}, as follows
\begin{equation}
ds^2_S = k(r)\, dv^2+2 \,j(r)\, dv dr + h(r)\, dx^i dx^i\,.
\end{equation}
Of these, only two perturbations will be truely dynamical and it will be possible to pick a gauge, in which one particular linear combination is gauge fixed to zero (in addition to $g_{rr}=0$, which has already been used above).

Additionally, there are the gauge field perturbations. For presenting these, we recall that for $\mathcal{A}$, we have chosen an axial gauge, in which the radial component vanishes $\mathcal{A}_r =0$. Thus, we only get one further perturbation from $\mathcal{A}$
\begin{equation}
\mathcal{A}_S = a(r)\, dv\,.
\end{equation}

The second gauge field $\mathbf{A}$, the Proca field, does not have any gauge freedom left anymore, which we could use to gauge fix $\mathbf{A}_r=0$. We rather have to account for this degree of freedom also, which can basically be thought of as the St\"{u}ckelberg scalar. Therefore, we have
\begin{equation}
\mathbf{A}_S = \mathbf{a}(r)\, dv+\mathbf{s}(r)\, dr
\end{equation}
Furthermore, there are two scalar perturbations originating from the dilatons
\begin{equation}
U_S = u(r)\,,\qquad V_S = v(r)\,.
\end{equation}
In \cite{Emparan:2013ila}, it was possible to define the cut-off surface as being isodilatonic such that no dilaton perturbation would arise in the scalar sector. As alluded to already several times, this is not possible for our system.

So, as we have seen, we have to deal with effectively eight coupled scalar perturbations, which surpasses the complexity of \cite{Emparan:2013ila,Banerjee:2008th,Erdmenger:2008rm} significantly.

\subsection{Constraint equations}

Of the eight equations, which originate from the Einstein equations
\begin{equation}
   E_{vv} = E_{vr} = E_{rr} = E_{ii} = 0\,,
\end{equation}
the two Maxwell equations 
\begin{equation}
   \mathcal{M}_v =  \mathbb{M}_v = \mathcal{M}_r =  \mathbb{M}_r = 0
\end{equation}
and the two dilaton equations of motion, we expect four constraint equations.

The first constraint equation may be computed by taking a particular linear combination of $\mathcal{M}_r = 0$ and $\mathbb{M}_r = 0$. From this we obtain
\begin{equation}
\label{bf:scalar_sector_first_constraint}
  0= q_{,v}+q \, \beta _{i,i}-\left(\frac{q\, r_0^3}{2 L^4+r_0^4}\right)\left(\frac{ 3 L^4-2 R^4 H_R}{R^4 H_R }\right) r_{0,v}\,.
\end{equation}

The second linearly independent constraint equation which originates from $\mathcal{M}_r = 0$ and $\mathbb{M}_r = 0$ may be computed by using \eqref{bf:scalar_sector_first_constraint} in $\mathcal{M}_r = 0$. This implies
\begin{equation}
\label{bf:scalar_sector_second_constraint}
\begin{split}
  0 = \mathbf{a}(r)+\frac{f_R g_R}{\sqrt{H_R}}\left(\frac{  \sqrt{H(r)} }{f(r) g(r)
   }\right) k(r)+j(r)-\mathbf{s}(r)\\
   -\left(\frac{r_0^3}{2 L^4+r_0^4}\right)\sqrt{\frac{f_R g_R}{H_R^{1/2}}} \left(\frac{r   \sqrt{H(r)} }{f(r) \sqrt{g(r)} }\right)  r_{0,v}\,,
\end{split}
\end{equation}
which basically shows that we had not fixed the gauge completely. 

If we furthermore use $E_{vv} = 0$ along with $E_{vr} = 0$, we may use our two previous constraint equations \eqref{bf:scalar_sector_first_constraint} and \eqref{bf:scalar_sector_second_constraint} to arrive at
\begin{equation}
\label{bf:scalar_sector_third_constraint}
  0 = r_{0,v} + r_0 \left(2 L^4+r_0^4\right) \left(\frac{ R^4 H_R  }{R^4 H_R \left(6 L^4+5 r_0^4\right)-3 L^4 r_0^4}\right) \beta _{i,i}\,.
\end{equation}

Moreover, we may combine $E_{rr}=0$, the dilaton equations of motion and $E_{ii} = 0$ in a particular way. This leads to a constraint equation, which only involves up to first derivatives of various perturbations along with terms proportional to $q_{,v}$, $r_{0,v}$ and $\beta_{i,i}$. We will however not state this rather complicated constraint equation here for space reasons since we do not use it anyway.

These four constraint equations reduce to the appropriate ones in \cite{Banerjee:2008th} in the scaling limit \eqref{near_horizon_scaling}.

We may confront these constraint equations with our expressions for the quasi-local stress-energy tensor \eqref{quasi_local_stress_energy_tensor}, charge currents \eqref{bf:conserved_currents_massless} and \eqref{bf:conserved_currents_mixed} and their conservation equations in the scalar sector. Similiarly to \eqref{conservation_equation} the corresponding (non-)conservation equation of the quasi-local stress-energy tensor in the scalar sector reads
\begin{align}
\kappa_5^2 \hat{\nabla}^\mu T_{\mu v}&=\left( \hat{\nabla}^\mu e^{4 U +V}\right)\left(K_{\mu v}-h_{\mu v} K\right)+\hat{\nabla}_ v\left(n^\mu \nabla_\mu e^{4 U+V}\right)-\hat{\nabla}_ v\left(4 e^{3U+V}+e^{4U}\right)\nonumber\\
&-4 e^{4 U+V} (\hat{\nabla}_ v U)n^\rho \nabla_\rho U -4 e^{4 U+V}  n^\mu h\indices{_ v^\rho} \nabla_\mu \nabla_\rho U \label{em_tensor_conservation_equation_scalar_sector}\\
&- e^{4 U+V} (\hat{\nabla}_ v V)n^\rho \nabla_\rho V - e^{4 U+V}  n^\mu h\indices{_ v^\rho} \nabla_\mu \nabla_\rho V\nonumber\\
&- e^{4 U+V}\left(\frac{1}{2} e^{2 V} n^\mu \mathcal{F}_{\mu\rho}\mathcal{F}\indices{_ v^\rho}+\frac{Q^2}{4} e^{-4 U -2 V} n^\mu \mathbb{F}_{\mu\rho}\mathbb{F}\indices{_ v^\rho}+ 2 Q^2 e^{-8 U} n^\mu \mathbf{A}_\mu \mathbf{A}_ v\right) \,.\nonumber
\end{align}

The conservation equations for the charge currents defined in \eqref{bf:conserved_currents_massless} and \eqref{bf:conserved_currents_mixed} may be simplified using the Maxwell equations \eqref{MMT_Maxwell_equation} and \eqref{MMT_Maxwell_equation_massive}. They are given by the following relations
\begin{align}
   2\kappa_5^2 \hat{\nabla}^\mu \mathcal{J}_\mu &= - 4 Q^2 e^{-4 U + V} n_\mu \mathbf{A}^\mu - \frac{Q^2}{4}\, \kappa\, n_\mu \epsilon^{\mu\nu\rho\sigma\lambda}\mathbb{F}_{\nu\rho}\mathbb{F}_{\sigma\lambda}\,, \label{current_massless_conservation_equation_scalar_sector}\\
   2\kappa_5^2 \hat{\nabla}^\mu \mathbb{J}_\mu &= 4 Q^2 e^{-4 U + V} n_\mu \mathbf{A}^\mu\,. \label{current_mixed_conservation_equation_scalar_sector}
\end{align}

We observe that $\mathbb{J}^\mu$ is not conserved due to the additional degree of freedom of the Proca field $\mathbf{A}$. Also the current $\mathcal{J}^\mu$ is not conserved, however it contains the whole contribution from the Chern-Simons term. Since these terms vanish for our background configuration, we find that the conserved current is given by $\mathcal{J}^\mu + \mathbb{J}^\mu = J^\mu_\textrm{phys}$.

Indeed, it is assuring to observe that the constraint equations \eqref{bf:scalar_sector_first_constraint}, \eqref{bf:scalar_sector_second_constraint} and \eqref{bf:scalar_sector_third_constraint} are identical to the current conservation equations \eqref{em_tensor_conservation_equation_scalar_sector}, \eqref{current_massless_conservation_equation_scalar_sector} and \eqref{current_mixed_conservation_equation_scalar_sector} to first order in the derivative expansion where the left-hand side of the former involves the derivatives of the first order results \eqref{idealT}, \eqref{bf:background_current_massless} and \eqref{bf:background_current_mixed}. These are given by
\begin{equation}
   -\epsilon_{,v}-\left(\epsilon + P\right) \beta_{i,i}
\end{equation}
for the stress-energy tensor and
\begin{equation}
\begin{split}
   &n_{\mathcal{J},v} + n_\mathcal{J} \beta_{i,i}\\
  & n_{\mathbb{J},v}+ n_\mathbb{J} \beta_{i,i}
\end{split}
\end{equation}
for the charge currents given the world-volume dependence of $r_0(\sigma^a)$, $L(\sigma^a)$ and $q(\sigma^a)$.

\subsection{Dynamical equations}

The dynamical equations in the scalar sector will be very tedious to solve. In contrast to the earlier analysis of \cite{Emparan:2013ila}, we have three more perturbations which couple. But, in total, it might still be possible to reach results here. As was convincingly shown in \cite{Torabian:2009qk}, it might suffice to perform an educated guess of the solution in the scalar sector, motivated by previous analysis, and argue in favour of its uniqueness up to coordinate reparametrizations. This does not sound unreasonable also for our case. But the simplification in the setup of~\cite{Torabian:2009qk} allowed for (correctly) guessing that most of the perturbations vanish as in \cite{Banerjee:2008th,Erdmenger:2008rm}, which, given the analysis of \cite{Emparan:2013ila}, seems unlikely in our case.

In solving the scalar equations of motion, previous works have repeatedly used quite nice simplification arguments. The appearing integration constants may often be reabsorbed into a redefinition of the charge $q$ and the parameter $r_0$ \cite{Banerjee:2008th}. This also seems possible in our case. Furthermore, integration constants might be fixed by requiring regularity at the horizon and at the Dirichlet cut-off $r=R$ additional to a further Landau frame constraint.

% Part 6 - EM Tensor and Charged Currents
%!TEX root=main.tex
%~~~~~~~~~~~~~~~~~~~~~~~~~~~~~~~~~~~~~~~~~~~~~~~
\section{Physical results}
\label{sec:physical_results}
%~~~~~~~~~~~~~~~~~~~~~~~~~~~~~~~~~~~~~~~~~~~~~~~

In this section we derive the hydrodynamical transport coefficients and diffusion constants of the theory on the cut-off surface, i.e., we derive the energy-momentum tensor and the charge currents up to first order in the derivative expansion of $q,\,r_0$ and $u_a$. 

To set the stage for our discussion let us recall that for a simple charged fluid in an external electric field, for which 
\begin{equation}
    \hat{\nabla}^a T_{ab} = F_{ba} {\mathcal J}^a_\text{phys} \,, \qquad \textrm{and}\qquad 
    \hat{\nabla}_a {\mathcal J}_\text{phys}^a = -\frac{C}{8} \epsilon^{abcd}F_{ab} F_{cd}\,,
\end{equation}
at first order in the gradient expansion, we have the energy-momentum and charge currents given by
\begin{align}
T_{ab} &= \epsilon\, u_a\, u_b + P\, P_{ab} - \eta\, \sigma_{ab} - \zeta\, \Theta\, P_{ab} 
\nonumber \\
{\mathcal J}_\text{phys}^a &= n_\mathcal{J_\text{phys}}\, u^a +\sigma\left(F^{ab}u_b - 
T\, P^{ab}\,{\hat \nabla}_b\left(\frac{\mu_{\mathcal{J}_\text{phys}}}{T}\right)\right) + \xi_{{\cal J}_\text{phys}} \, \epsilon^{abcd}
u_b\, {\hat \nabla}_c u_d 
\label{hydrocur}
\end{align}
where $\eta$, $\zeta$ are the shear and bulk viscosities, $\sigma$ the electric conductivity, while $\xi_{{\cal J}_\text{phys}}$ is the chiral vortical coefficient. 

However, our setup is more complicated. The conservation equations of the quasi-local stress-energy tensor \eqref{conservation_equation} and \eqref{em_tensor_conservation_equation_scalar_sector} involve various forcing terms from the dilatons and the Proca field $\mathbf{A}$ which are expected to modify \eqref{hydrocur}. Given our limited understanding of how the expression for ${\mathcal J}_\text{phys}^a$ in \eqref{hydrocur} is generalized in our case -- we would require a proper understanding of sources and vacuum expectation values of our cutoff surface operators along with their proper renormalization -- we only derive $T_{ab}$ and ${\mathcal J}_\text{phys}^a$ from our gravitational setup and leave a proper interpretation in terms of various first order transport coefficients within ${\mathcal J}_\text{phys}^a$ for future work.

Nevertheless, we may deduce the shear viscosity which arises from the tensorial perturbations of the metric. From the vector sector we find the contributions which involve the sum of charge diffusion constants and forcing terms and in the parity breaking sector we compute the chiral anomaly terms at the cut-off surface. The scalar sector which contains the bulk viscosity has been ignored in our analysis owing to its complexity. In addition we also have an expectation value for the vector operator that couples to ${\mathbf A}$ or equivalently ${\mathbb A}$, which we called ${\mathbb J}^a$ defined in \eqref{bf:conserved_currents_mixed}. All the results are consistent with~\cite{Banerjee:2008th,Erdmenger:2008rm} in the near-horizon limit (when we additionally put the Dirichlet surface at the AdS boundary).

%~~~~~~~~~~~~~~~~~~~~~~~~~~~~~~~~~~~~~~~~~~~~~~~
\subsection{Shear viscosity}
%~~~~~~~~~~~~~~~~~~~~~~~~~~~~~~~~~~~~~~~~~~~~~~~

In \S\ref{sec:tensor_sector}, we have computed the tensor perturbation by solving the corresponding equations of motion, imposing regularity at the future horizon for $\partial_r \alpha_{ij}$ and the Dirichlet condition $\alpha_{ij}(R)=0$ to retain Minkowski space of the perturbed metric at $r=R$. This is enough information to compute the shear viscosity of the system. Although we know on general grounds that it will correspond to the usual universal value, it is an important non-trivial check if the chosen boundary conditions, the energy-momentum tensor with its regularization and the units are consistent.

From the expression \eqref{quasi_local_stress_energy_tensor} with the solution \eqref{solution_tensor_sector} inserted we may compute the symmetric traceless part of the energy-momentum tensor at first order:
\begin{equation}
	T_{ij}^{(1)}=-\frac{1}{\kappa _5^2}\, H_R^{3/4} \,r_+^5\, \sqrt{\frac{H\left(r_+\right)}{g\left(r_+\right)}}\, \sigma_{ij}\,.
\end{equation}
This allows us to read off the shear viscosity
\begin{equation}
\label{shear_viscosity}
	\eta=\frac{1}{2\kappa _5^2}\, H_R^{3/4} \,r_+^5\, \sqrt{\frac{H\left(r_+\right)}{g\left(r_+\right)}}\,.
\end{equation}
It perfectly matches the results of \cite{Banerjee:2008th} in the near-horizon limit, in which we put the cut-off surface to the AdS boundary via $R\rightarrow\infty$.\footnote{ Note that we use unconventional units:  $\kappa_5^{-2}=\kappa_{10}^{-2}\,\textrm{Vol}(S^5)$ with a sphere of radius $1$. The length scale of the internal space is absorbed into the scalars.} From \eqref{shear_viscosity} and the earlier result for the entropy density \eqref{entropy_density} we get the usual universal ratio
\begin{equation}
	\frac{\eta}{s}=\frac{1}{4\pi}\,.
\end{equation}

%~~~~~~~~~~~~~~~~~~~~~~~~~~~~~~~~~~~~~~~~~~~~~~~
\subsection{Diffusion constant and chiral vortical conductivity}
%~~~~~~~~~~~~~~~~~~~~~~~~~~~~~~~~~~~~~~~~~~~~~~~

We now analyze the contributions from the vector sector to the physical transport. Since we choose the Landau frame, this information is encoded in the charge currents $\mathcal{J}_i$ and $\mathbb{J}_i$ only, for which we may use \eqref{bf:conserved_currents_massless} and \eqref{bf:conserved_currents_mixed}. We insert our ansatz \eqref{bf:first_order_gauge_fields} and \eqref{bf:first_order_gauge_fields1} into these expressions at an arbitrary radial position~$r$ and obtain
\begin{equation}
\label{bf:conserved_currents_pert}
\begin{aligned}
	2 \kappa_5^2~ \mathcal{J}_i ={}&2 \kappa_5^2~ \tilde{\mathcal{J}}_i-\frac{H_R}{\sqrt{f_R g_R}}\left(\frac{r^7 f(r) H(r)^{1/2}}{g(r)}\right)v_i'(r)\\
	&+q\,\sqrt{L^4+r_0^4}\left(\frac{ H_R^{5/4} }{f_R g_R }\right)\left(\frac{r f(r)}{H(r)^{1/2}}\right) w_i'(r) +q\,\sqrt{L^4+r_0^4}\left(\frac{ H_R^{5/4} }{f_R g_R}\right)\\
	&\qquad \times \left(\frac{f(r)}{g(r) H(r)^{3/2}}\right) \left(r H(r) g'(r)-g(r) \left(r H'(r)+6 H(r)\right)\right) w_i(r)\,,\\
	2 \kappa_5^2~ \mathbb{J}_i ={}&2 \kappa_5^2~ \tilde{\mathbb{J}}_i+2 L^4 \left(L^4+r_0^4\right) \left(\frac{  H_R }{ \sqrt{f_R g_R} }\right) \left(\frac{f(r) g(r)}{r H(r)^{3/2}}\right) \left(v_i'(r) + \mathbf{v}_i'(r)\right)\\
	&-2  q \,L^4\sqrt{L^4+r_0^4}\left(\frac{ H_R^{5/4} }{f_R g_R}\right)\left(\frac{ f(r) g(r)  }{r^4  H(r)^{3/2}}\right)\left(-2 w_i(r)+r w_i'(r)\right)\,.
\end{aligned}
\end{equation}
The terms $\tilde{\mathcal{J}}_i$ and $\tilde{\mathbb{J}}_i$ contain the terms which explicitly depend on $r_{0,i}$, $q_{,i}$ and $\epsilon_{ijk}\beta_{j,k}$. These arise from the long wavelength perturbations of the background gauge fields, eq.~\eqref{fluid_gravity_perturbations_gaugefields}.

We now first focus on the computation of $\mathcal{J}_i$. We may use the relation \eqref{Relation_Between_V_and_VM} to eliminate $v'_i(r)$ and evaluate the resulting expression at $r = R$. The term proportional to $w'_i(R)$ drops out, where we implicitly use the fact that $w'_i(r)$ is regular at $r=R$ after we fix the integration constants as detailed in Appendix \ref{sec:fix_int_const}. The remaining terms are then
\begin{align}
   2 \kappa_5^2 \, \mathcal{J}_i &= 2 \kappa_5^2~ \tilde{\mathcal{J}}_i+6 q\, \sqrt{L^4+r_0^4} \left(\frac{ H_R^{7/4}  }{f_R g_R} \right) C_{i,2}+R^2 \left(\frac{H_R}{f_R g_R}\right)^{3/2} S_{i,4}(R)\\
   &\quad - \frac{\sqrt{L^4+r_0^4}}{q }\left(\frac{H_R^{3/4}}{g_R}\right) \left(
   R^7 H_R \frac{f'(R)}{f_R }+q^2  \left(R\frac{  H'(R)}{H_R}-R \frac{g'(R)}{g_R}+6  \right)\right) w_i(R)\,.\nonumber
\end{align}
In order to evaluate this expression in terms of $r_{0,i}$, $q_{,i}$ and $\epsilon_{ijk}\beta_{j,k}$ we use the Landau frame choice, eq.~\eqref{bf:landau_frame}, for $w_i(R)$ and the values for $C_{i,2}$ and $S_{i,4}(R)$ from Appendices \ref{sec:fix_int_const} and \ref{ap:source_terms}, respectively. In addition, the explicit terms in $\tilde{\mathcal{J}}_i$ are used. The final result is
\begin{equation}\label{bf:charge_current_end_result}
\begin{split}
	\mathcal{J}_i = & -\left(\frac{ H_R  }{f_R g_R}\right)^{3/2}\frac{\left(3 L^4+r_0^4\right)\left(r_+^4+r_0^4\right)}{4 \kappa _5^2\, r_+ r_0^4}\left( q_{,i}-3 \frac{q}{r_0}\, r_{0,i}\right)\\&-\frac{H_R^{3/2}}{R^{4} \left(f_R g_R\right)^2}\left(3 R^4-r_0^4\right)\frac{2\kappa  L^4 q^2}{\kappa _5^2 r_0^4}\,\epsilon_{{ijk}} \beta_{j,k}\\
	 = & - \mathcal{D}\, \left( q_{,i}-3 \frac{q}{r_0}\, r_{0,i}\right) + \xi_\mathcal{J}\,\epsilon_{{ijk}} \beta_{j,k} \,.
\end{split}
\end{equation}

It would be desirable to translate this expression into electrical conductivity and external forcing contributions (see discussion around eq.~\eqref{hydrocur}) using e.g., our earlier expressions for chemical potential \eqref{bf:chemical_potential_massless} and temperature \eqref{temperature}. We will however leave it in the present form since currently it is unclear how the various external forcing terms modify the general structure of eq.~\eqref{hydrocur}. This has the obvious advantage of retaining the manifest simplicity visible in eq. \eqref{bf:charge_current_end_result}. Likewise this form makes it clear how to take the decoupling limit and recover the familiar AdS values.  We thus use the expression~${\mathcal D}$, which is rather similar to the charge diffusion constant, as the physical measure of charge transport in the system.

In the near-horizon limit \eqref{near_horizon_scaling} together with the cut-off surface taken to the AdS boundary $R\rightarrow\infty$, we recover the results of \cite{Banerjee:2008th,Erdmenger:2008rm} (modulo the issue about differing conventions mentioned earlier). For a quick check note in the scaling regime \eqref{near_horizon_scaling} the constraint equation \eqref{vector_constraint} reduces to $r_{0,i}+ r_0 \beta_{i,v}=0$ as required.

A similar approach allows a computation of the one-point function of the vector operator dual to the Proca field $\mathbb{J}_i$. In this case, additionally to the equations mentioned above, we need to use eq.~\eqref{Integrated_Maxwell_Sum} for eliminating $\mathbf{v}'_i(r)$. We obtain
\begin{equation}
\begin{split}
  2 \kappa_5^2~ \mathbb{J}_i = &2 \kappa_5^2~ \tilde{\mathbb{J}}_i + 6  q\, \left(L^4+r_0^4\right)^{3/2}\left(\frac{ H_R^{3/4} }{R^4 f_R}\right) C_{i,2} + R^2  \left(\frac{H_R}{f_R g_R}\right)^{3/2} S_{i,4}(R)\\
  +& \left. \int \left(\frac{S_{i,2}(r)}{2 L^4 \left(L^4+r_0^4\right)}+S_{i,3}(r)\right) \, dr \right|_R -\frac{2 \sqrt{L^4+r_0^4}}{q\, R^{10}
   f_R H_R^{1/4}}\Big(L^8 \left(2 r_0^4 R^4-3 q^2 R^2\right) \Big.\\
 &   \Big.+ L^4 \left(-2 q^4+q^2 \left(r_0^4 R^2-5 R^6\right)+4 r_0^4 R^8\right)+r_0^4 R^6 \left(2 R^6-q^2\right)\Big)  w_i(R)\,.
\end{split}
\end{equation}
Again we insert all the constants and integrals into this equation and use the Landau frame choice. We then arrive at the final result
\begin{equation}\label{bf:charge_current_mixed_end_result}
\begin{split}
	\mathbb{J}_i = &-\left(\frac{H_R^{1/2}}{R^4 f_R^{3/2} g_R^{1/2}}\right)\frac{\left(L^4+r_0^4\right) \left(3 L^4+r_0^4\right) \left(r_0^4+r_+^4\right)}{4 \kappa_5^2\,r_0^4\, r_+} \left( q_{,i}-3 \frac{q}{r_0}\, r_{0,i}\right)\\&-\frac{H_R^{1/2}}{R^8 f_R^2 g_R} \left(3 R^4-r_0^4\right)\left(L^4+r_0^4\right)\frac{2 \kappa L^4 q^2}{\kappa_5^2\,r_0^4}\,\epsilon_{{ijk}} \beta_{j,k}\\
	 = & - \mathbb{D}\, \left( q_{,i}-3 \frac{q}{r_0}\, r_{0,i}\right) + \xi_\mathbb{J}\,\epsilon_{{ijk}} \beta_{j,k}\,,
\end{split}
\end{equation}
which curiously takes the form of a conserved current similar to $\mathcal{J}_i$ (despite the absence of any boundary global symmetry).
In the near-horizon limit with the Dirichlet surface at the AdS boundary $R\rightarrow \infty$ the current $\mathbb{J}_i$ vanishes.

\begin{figure}[ht]
  \centering 
  \subfigure[]{\label{fig:diffusionAbl}\includegraphics[width=0.45\textwidth]{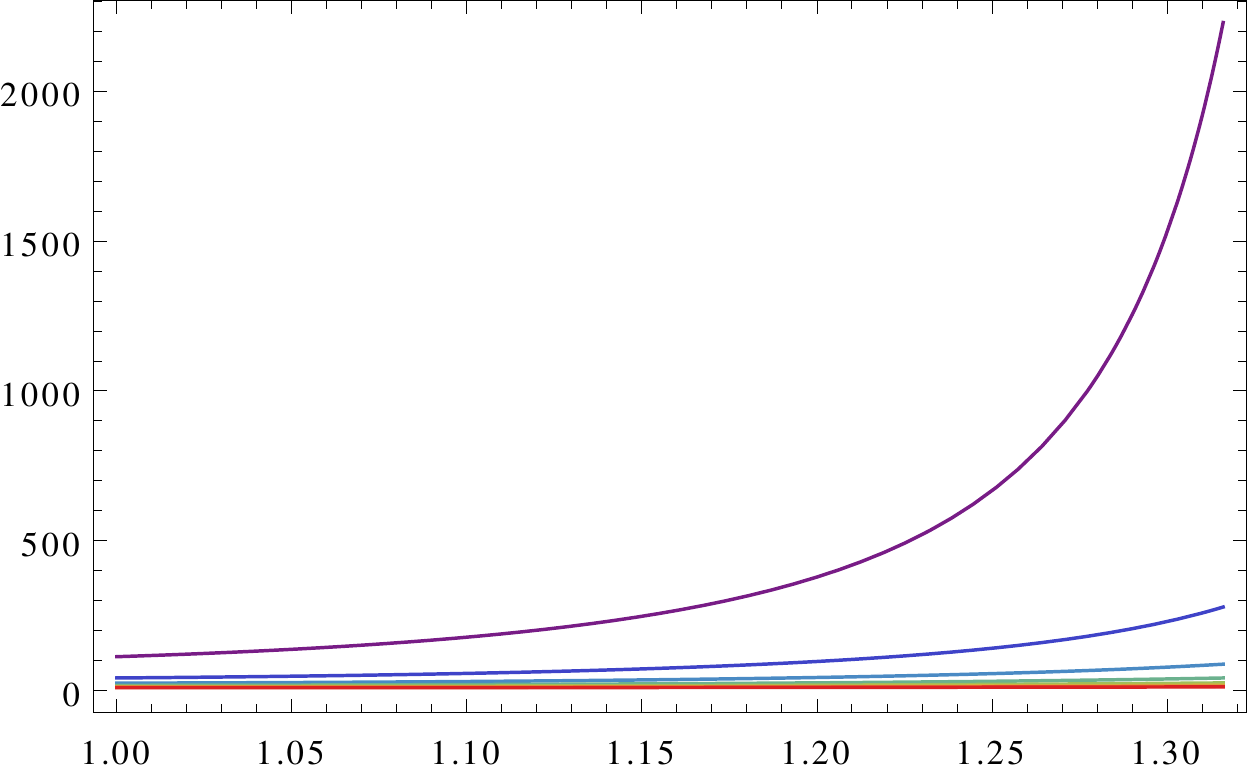}\put(-218,110){$4\kappa _5^2\, \mathcal{D}$}\put(-17,-8){$r_0$}}\put(-100,100){$R\,\downarrow$}
  \hfill 
  \subfigure[]{\label{fig:diffusionAfg}\includegraphics[width=0.45\textwidth]{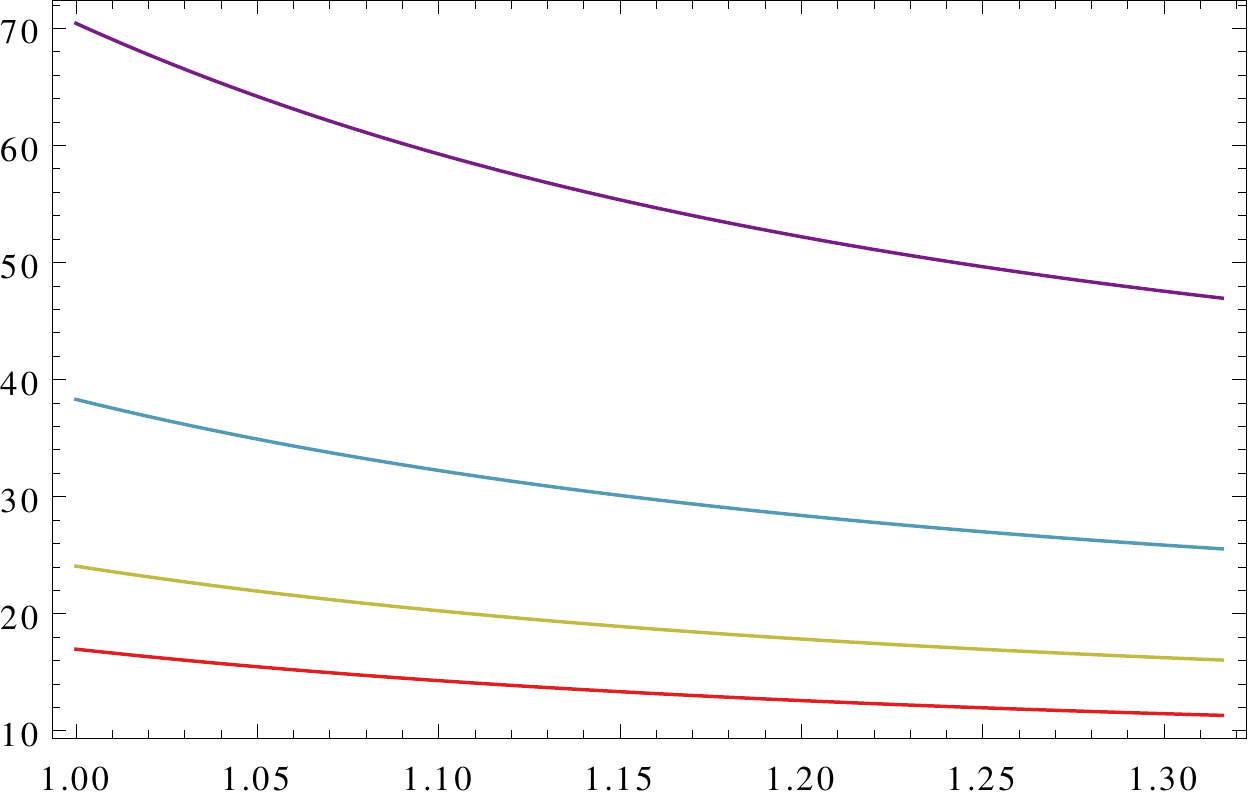}\put(-222,110){$4\kappa _5^2\, \mathcal{D}$}\put(-17,-8){$r_0$}}\put(-100,100){$R\,\downarrow$}              
  \hfill 
  \subfigure[]{\label{fig:diffusionAAbl}\includegraphics[width=0.45\textwidth]{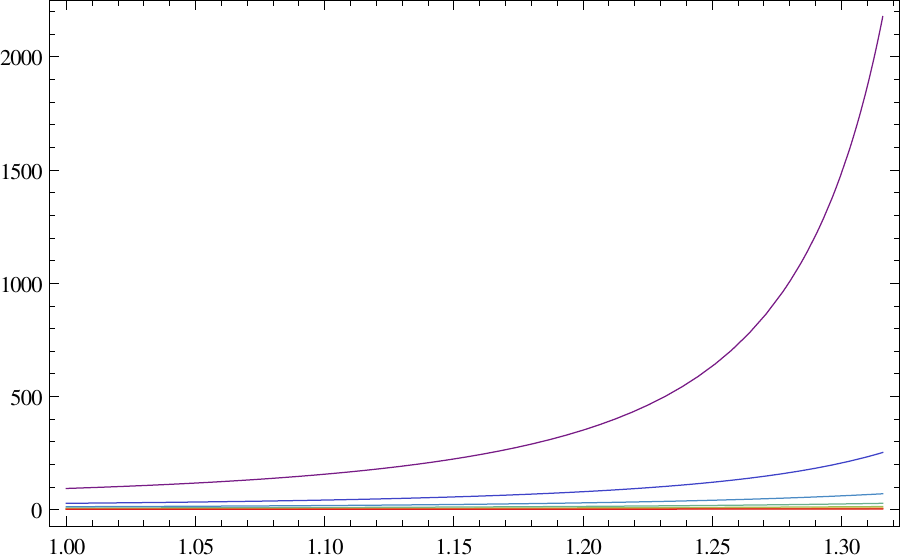}\put(-220,110){$4\kappa _5^2\, \mathbb{D}$}\put(-17,-8){$r_0$}}\put(-100,100){$R\,\downarrow$}
  \caption{We show the diffusion constants $\mathcal{D}$ (cf., \eqref{bf:charge_current_end_result}) and $\mathbb{D}$ (cf., \eqref{bf:charge_current_mixed_end_result}) versus the parameter $r_0$. To generate these plots we set $r_+=1$ and choose $L=1$ in (a) and (c) and $L=1000$ in (b). In addition, the different colours correspond to different values of $R$. The purple line corresponds to the smallest plotted value of $R$, while the red one to the biggest value. Note that $L=1000$ approximates the decoupling limit and we obtain the fluid/gravity results for $L\approx R \gg r_+,\, r_0$, i.e., the red line on the figures (b). $\mathbb{D}$ is suppressed in this limit therefore we show no plot for this component.} 
  \label{fig:diffusionA} 
\end{figure}

\begin{figure}[ht]
  \centering 
  \subfigure[]{\label{fig:diffusionBbl}\includegraphics[width=0.45\textwidth]{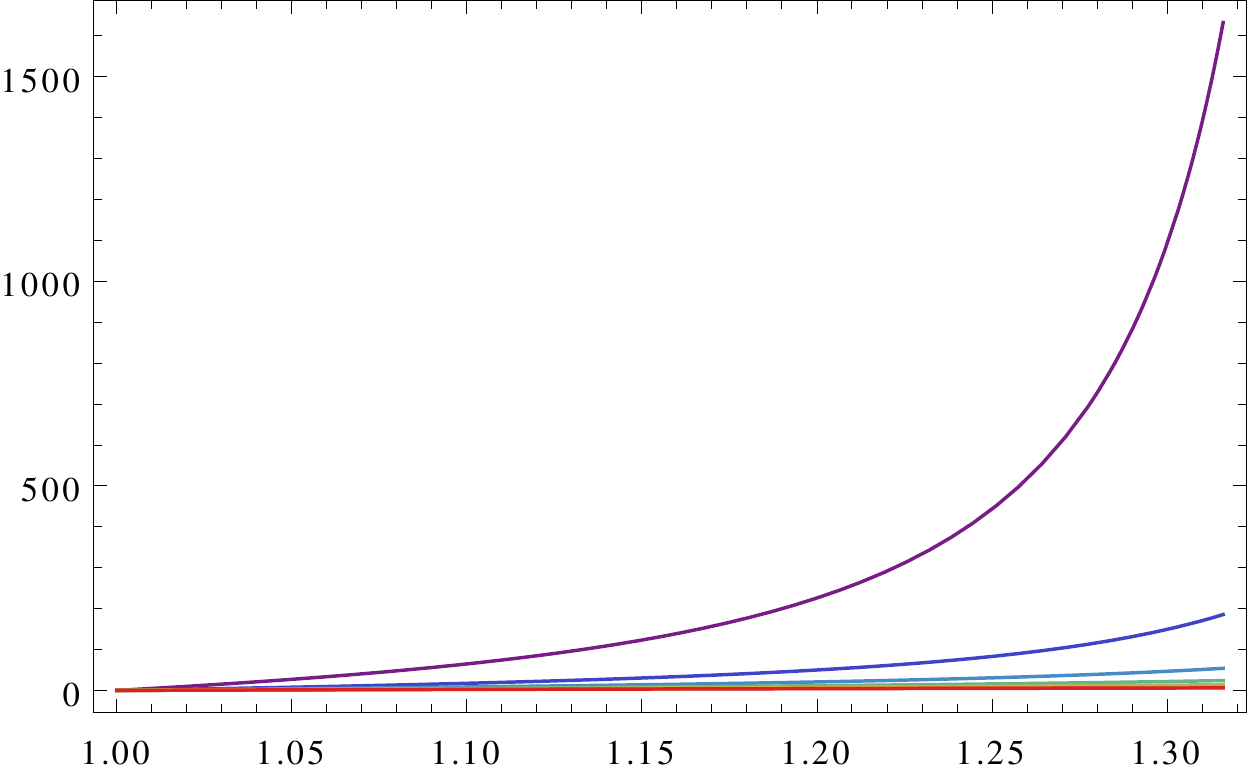}\put(-218,113){$4\kappa _5^2\, \xi_\mathcal{J}$}\put(-17,-8){$r_0$}}\put(-100,100){$R\,\downarrow$}
  \hfill 
  \subfigure[]{\label{fig:diffusionBfg}\includegraphics[width=0.45\textwidth]{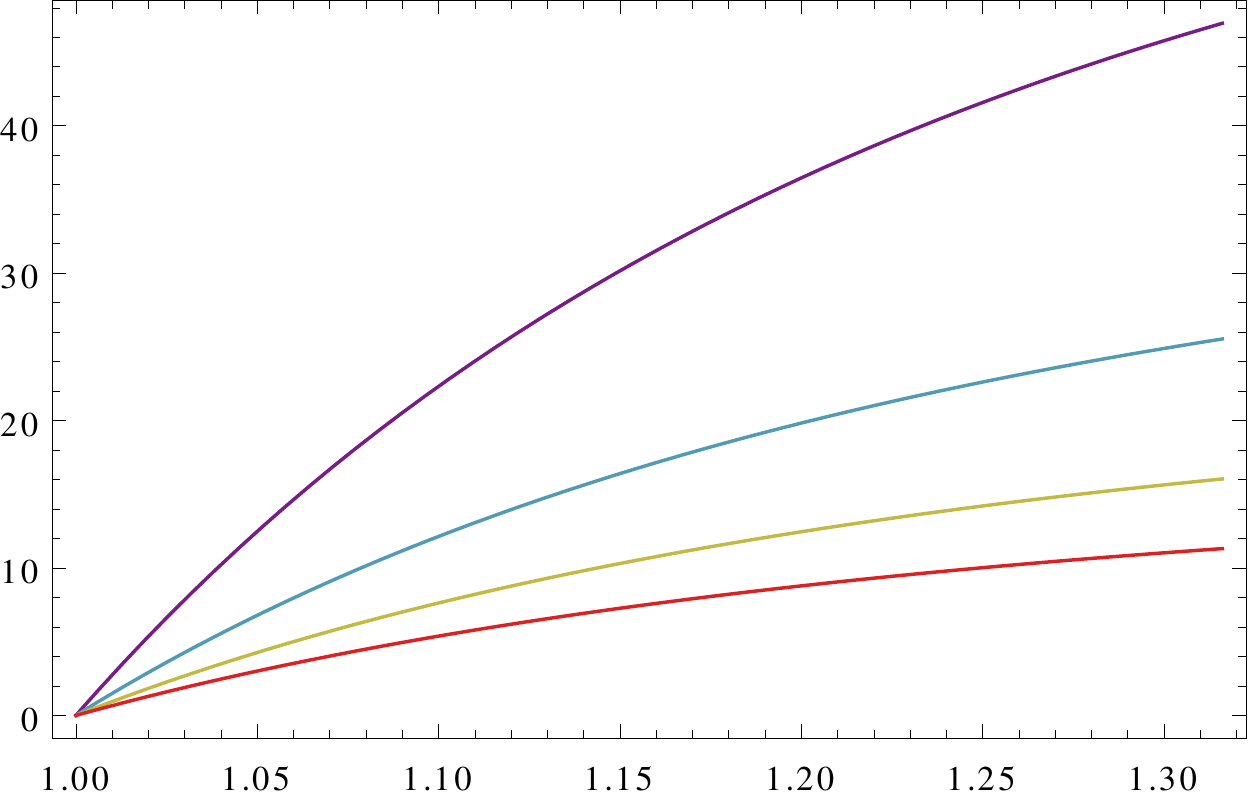}\put(-222,113){$4\kappa _5^2\, \xi_\mathcal{J}$}\put(-17,-8){$r_0$}}\put(-100,100){$R\,\downarrow$}              
  \hfill 
  \subfigure[]{\label{fig:difussionBBbl}\includegraphics[width=0.45\textwidth]{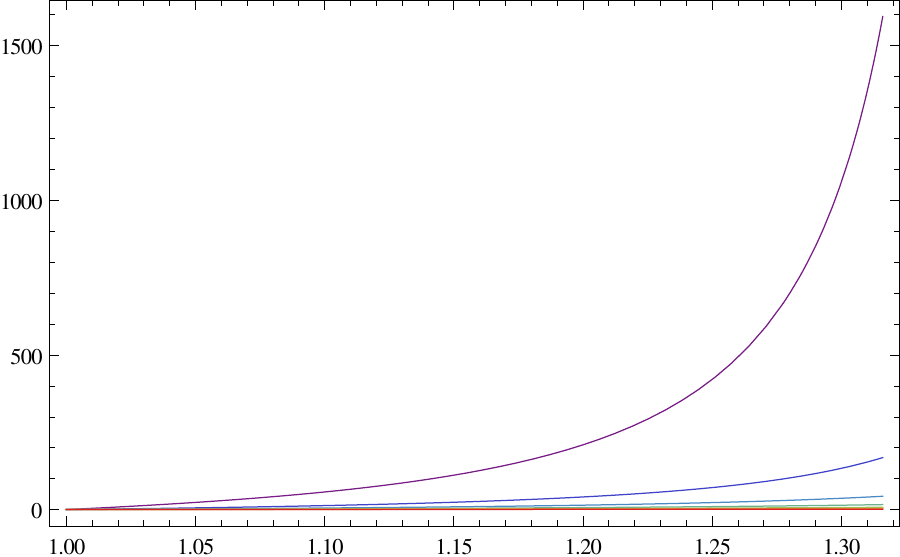}\put(-220,113){$4\kappa _5^2\, \xi_\mathbb{J}$}\put(-17,-8){$r_0$}}\put(-100,100){$R\,\downarrow$}
  \caption{We show the diffusion constants $\xi_\mathcal{J}$ (cf., \eqref{bf:charge_current_end_result}) and $\xi_\mathbb{J}$ (cf., \eqref{bf:charge_current_mixed_end_result}) versus the parameter $r_0$. To generate these plots we set $r_+=1$ and choose $L=1$ in (a) and (c) and $L=1000$ in (b). In addition, the different colours correspond to different values of $R$. The purple line corresponds to the smallest plotted value of $R$, while the red one to the biggest value. Note that $L=1000$ approximates the decoupling limit and we obtain the fluid/gravity results for $L\approx R \gg r_+,\, r_0$, i.e., the red line on the figures (b). $\xi_\mathbb{J}$ is suppressed in this limit therefore we show no plot for this component.} 
  \label{fig:diffusionB} 
\end{figure}

We show the behaviour of the diffusion coefficients and chiral conductivities in Fig.~\ref{fig:diffusionA} and 
Fig.~\ref{fig:diffusionB} for the conserved current ${\mathcal J}^a$ and the vector operator ${\mathbb J}^a$ respectively.

The astute reader will be curious that the combination $q_{,i}-3 \frac{q}{r_0} r_{0,i}$ is reminiscent of scale invariance. Indeed as can be noted from \cite{Banerjee:2008th} this combination is the Weyl covariant derivative acting on the charge. Since  scale invariance in our system is broken by the explicit cut-off surface, e.g., we verified  in~\eqref{bf:trace_em_tensor} that $T^\mu_\mu \neq 0$, this is curious. This particular structure is also encountered in the charged fluid computation of \cite{Bai:2012ci} in a cut-off AdS spacetime. So a-priori, this combination is not the result of an approximate scale symmetry emerging.\footnote{ While we have not computed the bulk viscosity the similarities with \cite{Emparan:2013ila} suggest that this will be non-vanishing for generic cut-off (it most likely will vanish in the AdS throat as in \cite{Bai:2012ci} and to leading order away from extremality).}  
Nor does it seem plausible to demand that the combination is enforced by the form of the charge current in \eqref{hydrocur} 
owing to our limited understanding of the source terms. It would be useful to study the dynamics of a charged fluid forced by 
(non-conserved) vector and scalar operators to understand the origins of this scale invariant form.

% Part 6 - Conclusions
%!TEX root=main.tex

%~~~~~~~~~~~~~~~~~~~~~~~~~~~~~~~~~~~~~~~~~~~~~~~
\section{Discussion}
\label{sec:discuss}
%~~~~~~~~~~~~~~~~~~~~~~~~~~~~~~~~~~~~~~~~~~~~~~
   
In this paper, we have developed the intrinsic sector of the long-wavelength effective theory of rotating D3-branes in flat space. To technically aid the computation, we used a particular Kaluza-Klein reduction of ten-dimensional type IIB supergravity to an effective five-dimensional gravitational theory. We then determined how the dynamics of the gravitational system imprints itself onto an effective theory on a radial cut-off surface, which we describe via its stress-energy tensor, a conserved (anomalously) charge current and vector operator that couples to the Proca field up to first order in a derivative expansion. For the fluctuations of the stationary black brane background we imposed Dirichlet boundary conditions on that cut-off surface and investigated the equations of motion in the tensor and vector sector of the setup. 

The physical situation we considered, which is ana\-lyzed within the blackfold paradigm, allows for an interpolation between different areas of gravitational physics which may be interpreted as effective hydrodynamical theories -- the membrane paradigm, the fluid/gravity correspondence and more simple AdS cut-off systems. The main result is the determination of the transport coefficient which is related to chiral anomalies in~\eqref{bf:charge_current_end_result} and~\eqref{bf:charge_current_mixed_end_result}, which perfectly interpolates between all the previously known scenarios. Furthermore, in these equations we derived a term which is the sum of electrical conductivity term and force contributions. In this term an interesting derivative structure was found which is related to the Weyl-covariant derivative of the charge in \cite{Banerjee:2008th}, although our setup is in fact not conformal; a similar phenomenon occurs in the related setup of \cite{Bai:2012ci}. Observations of this kind were already 
found in \cite{Brattan:2011my} though in a much simpler setup, which does not include the asymptotically flat region.

The stress-energy tensor which we obtained by a Brown-York procedure is not conserved due to the space-time dependence of the scalar fields and a (non-conserved) vector operator which do work on the cut-off surface fluid. 
As a consequence  we deal with a forced fluid analogous to the fluid/gravity discussion of \cite{Bhattacharyya:2008ji}. It is rather curious that extending away from the AdS throat results in a more complex dynamics involving two scalars and a vector operator. 

We argued that the desired Ward identities can be derived from the counter-term effective action we considered, though, we presented only a minimal such construction consistent with various limiting cases. It would be an interesting exercise to derive this explicitly and set-up the dictionary for the sources for the different fields, perhaps employing the general construction of
the works of \cite{Kraus:1999di}. We note in passing that the analysis of~\cite{Caldarelli:2012hy,Caldarelli:2013aaa}, which was able to embed the construction of blackfold conserved currents into an effective AdS spacetime, doesn't seem to a-priori provide any clues; in fact even the non-spinning D3-brane analysis of \cite{Emparan:2013ila} as far as we are aware does not admit an embedding into this framework.

A major lacuna in our analysis is the absence of an explicit solution in the scalar sector. To see the full dynamics we would need to obtain these as well.  It fortunately is reasonably simple to extract the constraint equations in this sector; we have the temporal equation for the stress tensor (non-)conservation and charge conservation which serve to check the consistency of our set-up. However, the intricate dynamical equations of motion in the scalar sector have to be disentangled and integrated with appropriate physical boundary conditions that determine the corresponding integration constants. While we have attempted this exercise, the equations have not availed themselves to simplifications; we therefore have refrained from reporting them.

\section*{Acknowledgements}
We are grateful to Niels Obers, Kostas Skenderis and Marika Taylor for discussions. HZ is grateful to the University of Southampton for hospitality during the final stages of this work. This work was supported in part by  {\it The Cluster of Excellence for Fundamental Physics - Origin and Structure of the Universe}.
MR was supported by the STFC Consolidated Grants ST/J000426/1 and ST/L000407/1.

\begin{appendix}
%!TEX root=main.tex

%~~~~~~~~~~~~~~~~~~~~~~~~~~~~~~~~~~~~~~~~~~~~~~~
\section{KK reduction of rotating D3-branes}
\label{sec:rotating_D3_branes}
%~~~~~~~~~~~~~~~~~~~~~~~~~~~~~~~~~~~~~~~~~~~~~~~

In this Appendix we would like to present some more details of the consistent truncation~\cite{Maldacena:2008wh} as applied to spinning D3-branes.

Since D3-branes in flat space are codimension six objects, we may rotate them in three different planes, giving them up to three different angular momenta $l_1, l_2, l_3$.

The metric for the $l_2=l_3=0$ case was obtained in \cite{Russo:1998mm} and extended to all angular momenta non-vanishing in \cite{Kraus:1998hv} (see also \cite{Russo:1998by}) using previous results of \cite{Cvetic:1996dt}. We are going to take the results of \cite{Cvetic:1999xp} as a starting point here since it corrected some typos in the aforementioned literature. Additionally, in that work, the decoupling limit of such configurations was stated, which will be useful for us. In particular, it was shown in \cite{Cvetic:1999xp} that this near-horizon geometry can be reduced to the STU black holes \cite{Behrndt:1998jd} under a Kaluza-Klein ansatz, which in general describes the consistent truncation of ten dimensional type IIB supergravity \cite{Schwarz:1983qr} on the five-sphere \cite{Gunaydin:1984fk,Kim:1985ez} to $\mathcal{N}=8$, $SO(6)$ gauged supergravity \cite{Pernici:1984xx,Gunaydin:1984qu}, further truncated to five dimensional~$\mathcal{N}=2$,~$U(1)^3$ gauged supergravity.

The metric for the full rotating D3-brane (with all $l_i\neq 0$) is \cite{Cvetic:1999xp}
\begin{equation}
\label{rotating_D3_brane}
\begin{split}
   ds^2_{10}={}& \tilde{H}^{-1/2}\left[-\left(1-\frac{2 m}{\tilde{r}^4 \Delta}\right)dt^2+dx_1^2+dx_2^2+dx_3^2\right]\\
   &+\tilde{H}^{1/2}\left[\frac{\Delta d\tilde{r}^2}{H_1 H_2 H_3-\frac{2m}{\tilde{r}^4}}+\tilde{r}^2\sum_{i=1}^3 H_i\left(d\mu_i^2+\mu_i^2 d\phi_i^2\right)\right.\\
   &\left.-\frac{4 m \cosh \alpha }{\tilde{r}^4 \tilde{H} \Delta} dt \left(\sum_{i=1}^3 l_i \mu_i^2 d\phi_i\right) +\frac{2m}{\tilde{r}^4 \tilde{H} \Delta}\left(\sum_{i=1}^3 l_i \mu_i^2 d\phi_i\right)^2\right]\,,
\end{split}
\end{equation}
where we have used
\begin{equation}
   \Delta(\tilde{r})=H_1 H_2 H_3 \sum_{i=1}^3 \frac{\mu_i^2}{H_i}\,,\quad \tilde{H}(\tilde{r}) = 1+\frac{2m \sinh^2 \alpha}{\tilde{r}^4 \Delta}\,,\quad H_i(\tilde{r})=1+\frac{l_i^2}{\tilde{r}^2}\,.
\end{equation}
In the notation of \cite{Myers:1986un}, the five-sphere metric is given by $d\Omega_5^2=\sum_{i=1}^3 \left(d\mu_i^2+\mu_i^2 d\phi_i^2\right)$, in which the $\mu_i$ parametrize a two-sphere
\begin{equation}
	\mu_1=\sin \theta,\,\quad\mu_2=\cos\theta\sin\psi,\,\quad\mu_3=\cos\theta\cos\psi\,.
\end{equation}

 The self-dual five-form $F_5=\star_{10} F_5$ stems from $G_5=d B_4$ via $F_5 = \left(1+\star_{10}\right)G_5$ and the four-form
\begin{equation}
   B_4=\frac{1-\tilde{H}^{-1}}{\sinh \alpha}\left(-\cosh\alpha \,dt + \sum_{i=1}^3 l_i \mu_i^2 d\phi_i\right)\wedge d^3 x\,.
\end{equation}
We now slightly adjust the conventions of \cite{Cvetic:1999xp} via
\begin{equation}
   2m=r_0^4\,,\quad 2m \sinh^2\alpha = L^4\quad\Rightarrow \quad2m \cosh\alpha = r_0^2 \sqrt{r_0^4+L^4}
\end{equation}
and take all angular momenta equal $l_i=l$. Then metric and five-form exactly fit into the Kaluza-Klein ansatz of \cite{Maldacena:2008wh}, which is given by\footnote{ $\star$ denotes the five-dimensional Hodge star operator.}
\begin{align}
\label{MMT_ansatz_metric}
ds^2_{10}=&\;ds^2(M)+e^{2U} ds^2(B_{KE})+e^{2V} (\eta+\mathcal{A})^2\,,\\
\label{MMT_ansatz_form}
\begin{split}
F_5=&\;\frac{Q}{2}\left(4 e^{-4U-V}\mathrm{vol}(M)+4 e^{-4U-V}(\eta+\mathcal{A})\wedge \star \mathbf{A}+e^{-V} \omega \wedge \star \mathbb{F}\right.\\&\left.+2 \omega^2\wedge(\eta+\mathcal{A})+2\omega^2\wedge \mathbf{A}-\omega\wedge(\eta+\mathcal{A})\wedge \mathbb{F}\right)\,.
\end{split}
\end{align}

To give some more details, note that we may write the five-sphere as a Hopf fibration~\mbox{$\mathbf{S}^1 \hookrightarrow \mathbf{S}^{2n +1} \twoheadrightarrow \mathbb{CP}^n$} with $n=2$. It is straightforward to show ($H_i=H_1$ for all $i$ since~$l_i=l$)
\begin{equation}
\label{five_sphere_metric_as_Hopf_fibration}
\tilde{r}^2\sum_{i=1}^3 H_i\left(d\mu_i^2+\mu_i^2 d\phi_i^2\right)=\tilde{r}^2 H_1 \left(\eta^2+2 g_{i\bar{j}} dz^i d\bar{z}^{\bar{j}}\right)\,,
\end{equation}
where $\eta=d\phi+P$ and $g_{i\bar{j}}=\frac{1}{2} \partial_i \bar{\partial}_{\bar{j}}\mathcal{K}$ is the Fubini-Study metric on $\mathbb{CP}^2$ with K\"{a}hler potential $\mathcal{K}$ and K\"{a}hler form $\omega= d\eta/2$ given by
\begin{equation}
\mathcal{K}=\log\left(1+|z_1|^2+|z_2|^2\right)\,,\quad \omega = 2 i g_{i\bar{j}} dz^i \wedge d\bar{z}^{\bar{j}}\,,\quad P=\frac{i}{2}\frac{z^i d\bar{z}^i-\bar{z}^i dz^i}{1+|\vec{z}|^2}\,.
\end{equation}
As before, the $\mu_i$ parametrize a two-sphere (with $\sum_i \mu_i^2=1$). All angular coordinates of the five-sphere are then related to the complex coordinates $z_1$, $z_2$ of $\mathbb{CP}^2$ and the fibre angle~$\phi$ by
\begin{equation}
   |z_1|^2=\frac{\mu_1^2}{\mu_3^2}\,,\quad|z_2|^2=\frac{\mu_2^2}{\mu_3^2}\,,\quad\phi_1=\phi +\arg z_1\,,\quad \phi_2=\phi +\arg z_2\,,\quad\phi_3=\phi\,.
\end{equation}
Another easy to derive relation from this coordinate change is $\sum_i\mu_i^2 d\phi_i=\eta$.

From these relations it is clear, that the last term in \eqref{rotating_D3_brane} together with the ones rewritten as in \eqref{five_sphere_metric_as_Hopf_fibration} yield a relative squashing of fibre and base metrics whose respective volumes were parametrized in \eqref{MMT_ansatz_metric} by the scalars $U$ and $V$. They thus receive profiles\footnote{ Note that these acquire somewhat unusual length dimensions as in \cite{Emparan:2013ila}. We could introduce a length scale for the radius of the Sasaki-Einstein space, but do not necessarily need to.}
\begin{equation}
  e^{2 U}=   \tilde{H}^{1/2} \tilde{r}^2 H_1 \,,\qquad e^{2 V} = \tilde{H}^{1/2} \left(\tilde{r}^2 H_1 + \frac{r_0^4 l^2}{\tilde{r}^4 \tilde{H} \Delta}\right)=\frac{ \tilde{H}^{1/2} \tilde{r}^2}{\Delta \tilde{g}} \,.
\end{equation}
where we have introduced $\tilde{g}(\tilde{r})=\tilde{H} \left(H_1^3 \tilde{H}+\frac{r_0^4 l^2}{\tilde{r}^6}\right)^{-1}$. The (electric) gauge field profile can be read off from the off-diagonal term proportional to $dt \left(\sum_{i} l_i \mu_i^2 d\phi_i\right)$. It is given by
\begin{equation}
   \mathcal{A}=-l\,r_0^2 \sqrt{r_0^4+L^4} \left(\frac{\tilde{g}}{\tilde{r}^6 \tilde{H}}\right)\,dt\,.
\end{equation}
Now, the only term which needs to be read off from \eqref{MMT_ansatz_metric} is the background metric. It is given by
\begin{equation}
ds^2(M)=\tilde{H}^{-1/2}\left[-\tilde{f} \tilde{g} \,dt^2 + dx_1^2+dx_2^2+dx_3^2\right]+\tilde{H}^{1/2} \Delta \tilde{f}^{-1}d\tilde{r}^2\,
\end{equation}
where we have defined $\tilde{f}(\tilde{r})=H_1^3-\frac{r_0^4}{\tilde{r}^4}$. Now, we only need to check the various components of the five-form \eqref{MMT_ansatz_form} and read off the profile of the gauge field $
\mathbf{A}$. It is given by
\begin{equation}
   \mathbf{A}=-l\,r_0^2 \,L^2 \left(\frac{2}{Q}\right)\left( \frac{\tilde{f}\tilde{g}}{\tilde{r}^2 H_1 \tilde{H}}\right)dt\,.
\end{equation}
The coordinates, in which we have worked so far, are the ones which directly descend from the ones in \eqref{rotating_D3_brane}, but are not the most convenient ones e.g., for comparisons with \cite{Banerjee:2008th,Erdmenger:2008rm}. The coordinate change is however very easy to implement, namely $r^2=\tilde{r}^2+l^2$. The charge is then related to the angular momentum via $q \equiv l \, r_0^2$. Like this, we recover the results in \S\ref{sec:black_hole_background}.

\section{Source terms in the vector sector}
\label{ap:source_terms}
Here we collect some of the quite non-trivial source terms which appear in \S\ref{sec:vector_sector}.

The first source term $S_{i,1}(r)$ is the one which appears in the Einstein equation \eqref{Einstein_equation}. It is given by
\allowdisplaybreaks
\begin{align*}
 S_{i,1}(r)={}&
 -\left(\frac{H_R^{1/4}}{\sqrt{f_R g_R}}\right)\left(\frac{f(r) g(r)^{3/2}}{2 r^7H(r)^{3/2}}\right)\left(3 L^4 r^2+2 q^2+5 r^6\right) \beta_{i,v}\\
 & -\left(\frac{H_R^{1/4}}{\sqrt{f_R g_R}}\right)\left(\frac{f(r) g(r)^{5/2}}{2 r^{13}  H(r)^{5/2}}\right) q  \left(2 q^2-r^2 \left(3
   L^4+r^4\right)\right) q_{,i}\\
   &-\left(\frac{f_R g_R-H_R}{R^6H_R^{3/4} f_R \sqrt{f_R g_R}}\right)\left(\frac{f(r) g(r)^{5/2}}{2  r^{13} H(r)^{5/2}}\right)\\*
   &\qquad\times
 q   \left(q^2 \left(5 L^4 r^2+7
   r^6\right)+r^4 \left(3 L^8+8 L^4 r^4+5 r^8\right)+2 q^4\right) q_{,i}\\
   & +\left(\frac{H_R^{1/4}}{\sqrt{f_R g_R}}\right)\frac{ L^4 r_0^3 f(r) g(r)^{5/2} }{\left(2 L^4+r_0^4\right)r^{11} H(r)^{5/2} }\left(5
   L^4 r^2+10 q^2+7 r^6\right)r_{0,i}\\
   &-\left(\frac{f_R L^4 \left(2 g_R
   q^2-H_R R^6\right)+2 H_R^2 R^6 \left(2 L^4+r_0^4\right)}{R^{10} f_R^{3/2}
   g_R^{1/2} H_R^{7/4}\left(2 L^4+r_0^4\right)}\right)\left(\frac{ r_0^3 f(r) g(r)^{5/2}  }{2  r^{13} H(r)^{5/2}}\right)\\*
   &\qquad\times\left(q^2 \left(5 L^4 r^2+7
   r^6\right)+r^4 \left(3 L^8+8 L^4 r^4+5 r^8\right)+2 q^4\right)r_{0,i}
\end{align*}
In the near-horizon limit (with $R\rightarrow \infty$) the given expression hugely simplifies to
\begin{equation}
- \left(\frac{3   r }{2 L^2}\right)f(r)\beta_{i,v}\,.
\end{equation}
With the additional prefactors in \eqref{Einstein_equation}, this then exactly reproduces the source term in the Einstein equation of \cite{Banerjee:2008th}.

The first Maxwell equation $\mathcal{M}_i=0$ \eqref{Maxwell_equation_1} includes the source term $S_{i,2}(r)$
which may actually be written as a total derivative
\begin{align*}
 S_{i,2}(r)={}& \frac{d}{dr}\left[
 -\left(\frac{\sqrt{H_R}}{f_R g_R}\right)\frac{4 \kappa  L^4 q^2 }{r^4 }  \epsilon_{ijk} \beta_{j,k}-\sqrt{\frac{H_R}{f_R g_R}}\,\sqrt{\frac{ r^2 H(r)  }{g(r) }}\,\sqrt{L^4+r_0^4}\, q \beta_{i,v}\right.\\
 &+\left(\frac{\sqrt{H_R}}{R^6 f_R^{3/2}\sqrt{g_R}}\right)\sqrt{\frac{r^2 H(r)   }{g(r) }} \,\sqrt{L^4+r_0^4}\,q^2 q_{,i}\\
 &+\sqrt{\frac{g_R H_R}{f_R}}\,\sqrt{\frac{r^2 H(r)   }{g(r) }} \,\sqrt{L^4+r_0^4}\, q_{,i}\\
 &-2\,\sqrt{\frac{H_R}{f_R g_R}}\, \sqrt{r^2 g(r)H(r)}   \sqrt{L^4+r_0^4} q_{,i}\\
 &-2\left(\frac{\sqrt{g_R}}{R^{10} \sqrt{f_R}H_R^{3/2}}\right)\sqrt{\frac{r^2 H(r)   }{g(r) }}\frac{   \sqrt{L^4+r_0^4} }{   \left(2
   L^4+r_0^4\right)}\,L^4 q^3  r_0^3 \, r_{0,i}\\
 &-2\,\sqrt{\frac{H_R}{f_R g_R}}\,\sqrt{\frac{g(r)}{r^{10} 
   H(r)}} \left(3 L^4 r^2+q^2+r^6\right)\frac{  \sqrt{L^4+r_0^4} }{ \left(2 L^4+r_0^4\right)}\, q r_0^3 \,r_{0,i}\\
  &+\left(\frac{1}{R^4\sqrt{f_R g_R H_R}}\right)\sqrt{\frac{ r^2H(r) }{ g(r)}}\frac{  \sqrt{L^4+r_0^4} }{ \left(2 L^4+r_0^4\right)}\,L^4 q  r_0^3\,r_{0,i}\\
 &\left.-2\left(\frac{\sqrt{H_R}}{R^4 f_R^{3/2}\sqrt{g_R}}\right)\sqrt{\frac{ r^2H(r) }{ g(r)}}  \sqrt{L^4+r_0^4}\, q r_0^3 r_{0,i}
 \right]
\end{align*}
In this expression, one may notice the anomaly related term $\sim \epsilon_{ijk}\beta_{j,k}$. Its near-horizon, $R\rightarrow\infty$ limit is also quite simple:
\begin{equation}
\frac{16 \kappa  L^4  }{r^5}\,q^2\epsilon_{ijk}\beta_{j,k}+\frac{ L^4}{r^2}\left( q \beta_{i,v}+q_{,i}\right)
\end{equation}
and reproduces the terms in \cite{Banerjee:2008th}.

The second Maxwell equation \eqref{Maxwell_equation_2} includes a source term $S_{i,3}(r)$, which may also be written as a total derivative
\begin{align*}
 S_{i,3}(r)={}& \frac{d}{dr}\left[-4\left(\frac{\sqrt{H_R}}{f_R g_R}\right)\frac{ g(r) 
   }{r^8  H(r)} \,\kappa  q^2\epsilon_{ijk} \beta _{j,k}\right.
   -\sqrt{\frac{H_R}{f_R g_R}}\,\sqrt{\frac{ g(r)  }{r^6  H(r) }}\,\frac{q}{\sqrt{L^4+r_0^4}}\,\beta_{i,v}\\
   &
   +\left(\frac{H_R-f_R g_R}{R^6 f_R^{3/2} \sqrt{g_R H_R}}\right)\sqrt{\frac{ g(r)  }{r^6  H(r)}  }\frac{q^2}{\sqrt{L^4+r_0^4}}\,q_{,i}\\
   &-\sqrt{\frac{H_R}{f_R g_R}} \,\frac{g(r)^{3/2}
    }{r^9 H(r)^{3/2} }\left(r^2 g(r) \left(L^4+r_0^4\right)+q^2\right)\frac{1}{\sqrt{L^4+r_0^4}}\, q_{,i}\\
   &-2\left(\frac{\sqrt{g_R}}{R^{10} \sqrt{f_R}H_R^{3/2}}\right)\sqrt{\frac{  g(r)  }{r^6  H(r)}  }\,\frac{L^4 q^3 r_0^3}{\sqrt{L^4+r_0^4} \left(2
   L^4+r_0^4\right)}\,r_{0,i}\\
   &+\left(\frac{1}{R^4\sqrt{f_R g_R H_R}}\right) \sqrt{\frac{g(r) }{r^6  H(r)} }\frac{L^4 q r_0^3}{\sqrt{L^4+r_0^4} \left(2
   L^4+r_0^4\right)}\,r_{0,i}\\
   &-2\left(\frac{\sqrt{H_R}}{R^4 f_R^{3/2}\sqrt{ g_R }}\right)\sqrt{\frac{  g(r) }{r^6  H(r)} }\frac{q r_0^3}{\sqrt{L^4+r_0^4}}\, r_{0,i}\\
   &+2\,\sqrt{\frac{H_R}{f_R g_R}}\,\frac{  g(r)^{3/2}  }{r^7 H(r)^{3/2}}\frac{q r_0^3}{ \sqrt{L^4+r_0^4}}\, r_{0,i}\\
   &\left.-2\,\sqrt{\frac{H_R}{f_R g_R}}\,\frac{  g(r)^{5/2}
    }{r^{11}  H(r)^{5/2} }\,\frac{L^4 q r_0^3 \sqrt{L^4+r_0^4}}{\left(2 L^4+r_0^4\right)}\,r_{0,i}
 \right]\,.
\end{align*}
In the near-horizon limit, we reproduce the source term in near-horizon limit of the first Maxwell equation, however with an additional factor of $1/L^8$.

From these source terms, we define two more source terms, which capture specific combinations and integrals of the above ones. We have already made clear, that $S_{i,2}$ and $S_{i,3}$ may be integrated.  The expression one obtains like that also appears together with $S_{i,1}$ and $S_{i,2}$ in a way, which one may integrate even a further time,
\begin{align*}
   S_{i,4}(r)={}&\sqrt{\frac{f_R g_R}{H_R}}\int\left[ -2\left(\frac{ H_R^{1/4} \sqrt{L^4+r_0^4}  }{q }\right)r^5 H(r)\,S_{i,1}-\left(\frac{f(r) g(r)  }{r^2 H(r)}\right) \,S_{i,2}(r)\right.\\
   &\left.+\frac{4  L^4 \left(L^4+r_0^4\right)
   }{r^3}\int \left(\frac{S_{i,2}(r)}{2\, L^4\left(L^4+r_0^4\right)}+S_{i,3}(r)\right)\,dr\right]\,dr
   \intertext{or explicitly}
   ={}&\frac{4 \kappa  L^4 }{r^6 \sqrt{f_R g_R}}\,q^2\, \epsilon_{ijk} \beta _{j,k}+ \left(\frac{g(r)^{3/2}}{r^{13}H(r)^{3/2}}\right) \left(2 L^8 r^4+2 L^4 q^2 r^2+4 L^4 r^8\right.\\
   &\qquad\left.-q^4+q^2 r^6+q^2 r_0^4 r^2+2 r^{12}\right)\,\sqrt{L^4+r_0^4}\, q_{,i}\\
   &- \left(\frac{g(r)^{3/2}}{r^{13}H(r)^{3/2}}\right) \left(2 L^{12} q^2 r^4+2 L^{12} r^{10}+6 L^{12} r_0^4
   r^6+4 L^8 q^4 r^2+8 L^8 q^2 r^8\right.\\
   &\qquad\left.+L^8 q^2 r_0^4 r^4+4 L^8 r^{14}+9 L^8 r_0^4 r^{10}+7 L^8 r_0^8 r^6+2 L^4 q^6+6 L^4 q^4 r^6\right.\\
   &\qquad\left.-10 L^4 q^4 r_0^4 r^2+6 L^4 q^2 r^{12}-8 L^4 q^2 r_0^4 r^8+12 L^4 q^2
   r_0^8 r^4+2 L^4 r^{18}\right.\\
   &\qquad\left.+2 L^4 r_0^4 r^{14}+12 L^4 r_0^8 r^{10}-3 q^6 r_0^4-7 q^4 r_0^4 r^6+3 q^4 r_0^8 r^2\right.\\
   &\qquad\left.-5 q^2 r_0^4 r^{12}+8 q^2 r_0^8 r^8-r_0^4 r^{18}+5 r_0^8 r^{14}\right)\,\frac{\sqrt{L^4+r_0^4}}{q r_0
    \left(2 L^4+r_0^4\right)}\, r_{0,i} \,.
\end{align*}
For actually performing the integral, one should eliminate $\beta_{i,v}$ in the integrand first using~\eqref{vector_constraint}. 

Its near-horizon limit is given by
\begin{equation*}
   \frac{ 4 \kappa L^4  }{r^6}\, q^2\,\epsilon_{ijk} \beta_{j,k}+\frac{L^4 \left(q^2+r^6+3 r_0^4 r^2\right) }{q r^3 }\beta_{i,v}+\frac{2L^4 }{r^3}\,q_{,i}\,.
\end{equation*}

In a slightly different combination, it also comes up in the following integral, which one may perform analytically. For actually performing the integration it is advisable to use the constraint equation \eqref{vector_constraint} to eliminate $\beta_{i,v}$ from the integrand.
\begin{align*}
   S_{i,5}(r)&=\int\left[3\left(\frac{\sqrt{f_R g_R}}{H_R^{1/4}q L^4  \sqrt{L^4+r_0^4}}\right)\frac{ r^{11} H(r)  }{ g(r) }\,S_{i,1}\right.\\
   &\left.+3\,\sqrt{\frac{f_R g_R}{H_R}}\left(\frac{ 3 r^4-r_0^4}{4 L^4 \left(L^4+r_0^4\right)}\right) \,S_{i,2}(r)+\frac{3
   \left(3 r^4-r_0^4\right) }{4 L^4 \left(L^4+r_0^4\right)}\,\frac{d}{dr}\left(r^2 S_{i,4}(r)\right)\right]\,dr\,,
   \intertext{explicitly,}
   ={}&-\frac{3 g(r)^{3/2}}{4 L^4 r^9 H(r)^{3/2} \sqrt{L^4+r_0^4}}  \left(r^6 H(r) \left(L^4 \left(r^4+r_0^4\right)-r^8+3 r_0^4 r^4\right)\right.\\*
   &\qquad\left.+q^2 \left(2 L^4 r_0^4+2 q^2
   r^2+r^8+r_0^8\right)\right)\,q_{,i}\\
   &\frac{3 \left(\frac{g(r)}{H(r)}\right)^{3/2}}{4 L^4
   q r^9 r_0 \sqrt{L^4+r_0^4} \left(2 L^4+r_0^4\right)} \left(L^{12} \left(4 q^2 r^6-2 r^4 \left(r^8+2 r_0^4 r^4-3 r_0^8\right)\right)\right.\\
   &\left.+L^8 r^2 \left(8 q^4 r^2+q^2 \left(4 r^8-2 r_0^4 r^4+8
   r_0^8\right)-4 r^{14}-5 r_0^4 r^{10}+2 r_0^8 r^6+7 r_0^{12} r^2\right)\right.\\
   &\left.+2 L^4 r^2 \left(2 q^6+3 q^4 r^2 \left(r^4+r_0^4\right)+q^2 r_0^4 \left(3 r^8-5 r_0^4 r^4+6 r_0^8\right)\right)\right.\\
   &\left.- 2 L^4 r^8\left(r^{12}+5
   r_0^8 r^4-6 r_0^{12}\right)+r_0^4 \left(-2 q^6 r^2-3 q^4 \left(r^8-r_0^8\right)\right)\right.\\
   &\left.+r_0^4\left(q^2 \left(8 r^6 r_0^8-6 r^{10} r_0^4\right)+\left(r^8-6 r_0^4 r^4+5 r_0^8\right) r^{12}\right)\right) r_{0,i}
\end{align*}
This expression reduces to
\begin{equation*}
-\frac{3 r \left(2 q^2 r^2-r^8-2 r_0^4 r^4+3 r_0^8\right) }{4 L^4 q}\,\beta _{i,v}-\frac{3 \left(r^4+r_0^4\right) }{4 L^4 r}\,q_{,i}
\end{equation*}
in the near-horizon limit.

\section{Fixing the integration constants}\label{sec:fix_int_const}

\subsection{Fixing \texorpdfstring{${C_{i,2}}$}{C2} by regularity at the horizon}

We may derive a relation which determines $C_{i,2}$ in terms of source terms evaluated at the horizon. The physical requirement we get this from is \emph{regularity at the horizon} for a particular combination of the first derivatives of our perturbations $v_i^\prime(r)$, $\mathbf{v}_i^\prime(r)$, $w_i^\prime(r)$.

For doing so, we take equation \eqref{Integrated_Maxwell_Sum} and add the first derivative of \eqref{Relation_Between_V_and_VM} with a prefactor such that in the resulting equation the coefficient of $w_i(r)$ vanishes. From this we get an expression which only contains the first derivative terms $v_i^\prime(r)$, $\mathbf{v}_i^\prime(r)$, $w_i^\prime(r)$ and further source and integration constant terms. We now require regularity at the horizon for this particular combination\footnote{ We require it for all these functions individually, therefore it must also hold for the combination.} of $v_i^\prime(r)$, $\mathbf{v}_i^\prime(r)$, $w_i^\prime(r)$; since the other terms do contain a pole $\sim 1/f(r)$ at the horizon, we require its residue to vanish. This gives us the following relation
\begin{equation}
\begin{aligned}
0={}& 2 L^4 \left(L^4+r_0^4\right) \left(2 r^2 r_0^4-3 q^2\right) \sqrt{f_R g_R} \int \left(\frac{S_{i,2}(r)}{2\, L^4\left(L^4+r_0^4\right)}+S_{i,3}(r)\right) \, dr\\
&-3 q
   r^2 \sqrt{H_R} \left(q\, S_{i,4}(r)+4  r_0^4 H_R^{1/4} \sqrt{f_R g_R}\,\sqrt{L^4+r_0^4}\,C_{i,2}\right)\,,
\end{aligned}
\end{equation}
which has to be evaluated at the horizon. To get a more compact expression, we invoke the definition of ${S_{i,4}}^\prime(r)$ given in Appendix \ref{ap:source_terms} to eliminate the integral term. In this definition the prefactor of $S_{i,2}$ vanishes at the horizon while $S_{i,2}$ itself is regular; additionally, we have $S_{i,1}(r_+)=0$. In this way, we arrive at the relation
\begin{equation}
\label{bf:fixed_CMi}
C_{i,2}=\frac{\left(2 r_+^3 r_0^4-3 q^2 r_+\right) {S_{i,4}}^\prime\left(r_+\right)-6 q^2 \,S_{i,4}(r_+)}{24 q r_0^4 \sqrt{f_Rg_R} H_R^{1/4} \sqrt{L^4+r_0^4}}\,,
\end{equation}
in which
\begin{multline}
	\left(2 r_+^3 r_0^4-3 q^2 r_+\right) {S_{i,4}}^\prime\left(r_+\right)-6 q^2 S_{i,4}(r_+)\\
	=-\frac{48 \kappa  L^4 q^2 }{\sqrt{f_R g_R}}\, \epsilon_{ijk} \beta_{j,k}-\frac{2 \left(r_+^4+r_0^4\right) \left(3 L^4+r_0^4\right) }{r_+ r_0}\,\left(r_0 \,q_{,i}-3 q\,
   r_{0,i}\right)\,.
\end{multline}
We see that these are terms of the structure expected from the near-horizon limit \cite{Banerjee:2008th}. We have a term which stems from the Chern-Simons term; additionally the combination $r_+^4+r_0^4\propto\left(1+M\right)$, in the notation of \cite{Banerjee:2008th}, appears as a prefactor of $\left(r_0 \,q_{,i}-3 q\, r_{0,i}\right)$ which in the near-horizon limit reduces to the unique first order Weyl-covariant derivative \mbox{$\mathcal{D}_i q = q_{,i}+3 q\, \beta_{i,v}$} used in \cite{Banerjee:2008th}, given that in this limit the vector constraint \eqref{vector_constraint} simplifies to $r_{0,i}+r_0 \beta_{i,v}=0$. However in our full setup we get more naturally the structure in terms of $q_{,i}$ and $r_{0,i}$ with only the prefactor depending on $R$.

\subsection{Fixing \texorpdfstring{${C_{i,5}}$}{C5} by regularity at the horizon}

We may also fix the integration constant $C_{i,5}$ by imposing regularity at the horizon on a particular combination of $w_i(r)$, $w_i^\prime(r)$ and $v_i^\prime(r)$: Note that in \eqref{full_solution_w} one may eliminate the particular combination of expressions involving
\begin{equation}
	\sim \int \frac{S_{i,4}(r)}{r^5 f(r)^2} \, dr\qquad\textrm{and}\qquad \sim\, C_{i,2}\int \frac{1}{r^7 f(r)^2} \, dr\,,
\end{equation}
which could potentially complicate considerations at the horizon, where $f(r_+)=0$, in favour of $v_i(r)$ using \eqref{Relation_Between_V_and_VM}.
Going back to the second order ODE we obtained for $w_i(r)$ alone, i.e., \eqref{Second_order_ODE_for_w_hom} along with its inhomogeneous pieces, we may use the same relation \eqref{Relation_Between_V_and_VM} to express part of the inhomogeneous terms in that ODE in favour of $v_i(r)$. Thus the ODE may be written as a differential equation for a combination of $w_i(r)$ and $v_i(r)$. The only potentially diverging term at $r=r_+$ in that expression then is
\begin{equation}
\frac{d}{dr}\left(   \frac{r^4 \left(f_R g_R-f(r) g(r)\right) H(r)}{\left(-3 r^4+r_0^4\right) f(r) g(r)}\,w_{i}(r)+\left(\ldots\right) v_i(r)\right)\sim \frac{S_{i,5}(r)-C_{i,5}}{r^3 \left(3 r^4-r_0^4\right)^2 f(r)}\,.
\end{equation}
This would be a pole while the other terms are regular at $r=r_+$. Note in particular that the expression $S_{i,1}(r)/f(r)$ is finite as $r\rightarrow r_+$. Again, we may use this to fix an integration constant by setting the would-be residue to zero, making it a removable singularity. This time, we get
\begin{equation}
\label{bf:fixed_C5i}
\begin{aligned}
	C_{i,5}&=S_{i,5}(r_+)\\
	& =-\frac{3 \left(r_+^4+r_0^4\right) }{4 L^4 r_0 r_+ }\left(r_0\, q_{,i}-3 q\, r_{0,i}\right)\,,
\end{aligned}
\end{equation}
where as in the expression for $C_{i,2}$ we see that the familiar combination $r_+^4+r_0^4\propto\left(1+M\right)$ appears as a prefactor of $\left(r_0 \,q_{,i}-3 q\, r_{0,i}\right)$.

\subsection{Fixing \texorpdfstring{${C_{i,6}}$}{C6} by regularity at the cut-off}

For fixing ${{C_{i,6}}}$ we first trade $C_{i,1}$ and $C_{i,4}$ with $w_i(R)$ and $\mathbf{v}_i(R)$. This is of course not a physical condition imposed on them. It is only a slightly more convenient parametrization of these integration constants. We therefore evaluate \eqref{Integrated_Einstein_equation} and $\eqref{Relation_Between_V_and_VM}$ at $r=R$ and reinsert the expressions we get for $C_{i,1}$ and $C_{i,4}$ into \eqref{full_solution_w}. The Dirichlet condition $v_i(R)$ drops out and some of the integrals will essentially turn into definite integrals like
\begin{equation}
   \int \frac{1}{r^7 f(r)^2}dr\rightarrow \int_R^r \frac{1}{{r^\prime}^7 f(r^\prime)^2}dr^\prime\,.
\end{equation}
Now we fix ${{C_{i,6}}}$ by imposing that $w_i(r)$ should have no pole at $r=R$. This is clearly a sensible physical condition which preserves Minkowski space at the cut-off surface as can be seen from the perturbation ansatz \eqref{Vector_Perturbation_ansatz}. We therefore impose that the numerator in~\eqref{full_solution_w} after we replaced $C_{i,1}$ and $C_{i,4}$ should vanish at $r=R$. From this we get
\begin{equation}
\label{fixed_C6i}
\begin{split}
   C_{i,6}=\left(\frac{H_R^{1/4}  \left(-2 r_0^4 R^6 H_R+3 L^4 q^2+q^2 r_0^4\right)}{4 L^4 q \left(3 R^4-r_0^4\right) \sqrt{f_R g_R \left(L^4+r_0^4\right)}}\right)w_i(R)\\+\left(\frac{1}{2 R^4 H_R \left(3
   R^4-r_0^4\right)}\right)\mathbf{v}_i(R)+\int^R(\ldots)\,.
\end{split}
\end{equation}
If we reinsert this into \eqref{full_solution_w} the last term in \eqref{fixed_C6i} sets the remaining integrals to $\int_R^r (\ldots)dr^\prime$. So, in summary, fixing $C_{i,6}$ sets every single integral of \eqref{full_solution_w} to $\int_R^r (\ldots)dr^\prime$ and additionally contributes the terms explicitly spelled out in \eqref{fixed_C6i}, which are proportional to $w_i(R)$ and~$\mathbf{v}_i(R)$.

\subsection{Fluid frame choice and \texorpdfstring{$\mathbf{v}_i(R)$}{vm(R)}}

How do we fix the remaining integration constant, i.e., the linear combination of $w_i(R)$ and $\mathbf{v}_i(R)$ in \eqref{fixed_C6i} which is basically equivalent to $C_{i,1}-C_{i,4}$? For doing so we may carefully extract the limit of the solution \eqref{full_solution_w} for $r\rightarrow R$ using e.g.,
\begin{equation}
   \lim_{r\rightarrow R}\left(\frac{\int_R^r A(r^\prime)dr^\prime}{f_R g_R-f(r)g(r)}\right)=-\frac{A(R)}{g_R\,f'(R)+f_R\, g^\prime(R)}\,,
\end{equation}
and therefore see that the linear combination of $w_i(R)$ and $\mathbf{v}_i(R)$ given in \eqref{fixed_C6i} can essentially be expressed in terms of $w_i(R)$ alone. But as we will see next $w_i(R)$ is uniquely determined by fixing the fluid's frame ambiguity. So, in total, the Landau frame choice determines $w_i(R)$ and from using the explicit solution \eqref{full_solution_w}, we may read off $\mathbf{v}_i(R)$ or equivalently the linear combination $C_{i,1}-C_{i,4}$. This fixes the last integration constant we need for evaluating physical quantities.

\subsection{Landau frame choice}

The energy-momentum tensor which we have constructed is generically not in Landau frame. While the zeroth order is, generically the first order terms are such that they make it deviate from Landau frame unless we impose a specific condition on $w_i(R)$ to ensure it. Using \eqref{quasi_local_stress_energy_tensor}, the important component in the vector sector is
\allowdisplaybreaks
\begin{align}
T_{vi}^{(1)}={}&-\left(\frac{R^5 H_R^{5/4} }{\sqrt{g_R}}\right)\left(\frac{1}{2 \kappa_5^2}\right)\beta_{i,v}-\left(\frac{ H_R^{5/4} }{ R\, f_R \sqrt{g_R}}\right) \left(\frac{q}{2 \kappa_5^2}\right) q_{,i}\nonumber\\
&+\left(\frac{ H_R^{1/4}  }{R^5 f_R \sqrt{g_R} }\right)\left(4 L^8 R^2+L^4 \left(3 R^2 \left(r_0^4+R^4\right)-q^2\right)+2 r_0^4 R^6\right)\label{bf:landau_frame}\\
&\qquad\times\left(\frac{r_0^3}{2 \kappa_5^2 \left(2 L^4+r_0^4\right)}\right)r_{0,i}\nonumber\\
&+\left(\frac{ \sqrt{g_R}}{\kappa _5^2 R^{12}\sqrt{f_R} H_R}\right) \left(L^8 \left(2 r_0^4 R^4-3 q^2 R^2\right)+r_0^4 R^6 \left(2 R^6-q^2\right)\right.\nonumber\\ 
&\qquad\left. +L^4 \left(-2 q^4+q^2 \left(r_0^4 R^2-5 R^6\right)+4 r_0^4 R^8\right)\right)\,w_i(R)\nonumber\\
\stackrel{!}{=}{} & 0\,.\nonumber
\end{align}
We see that by choosing $w_i(R)$ such that this component vanishes, we take the fluid to be in the Landau frame. This fixes our last integration constant needed in \S\ref{sec:vector_sector}.

\end{appendix}

\bibliographystyle{JHEP}
\bibliography{References}

\providecommand{\href}[2]{#2}\begingroup\raggedright\begin{thebibliography}{10}

\bibitem{Damour:1978cg}
T.~Damour, {\it {Black Hole Eddy Currents}},  {\em Phys.Rev.} {\bf D18} (1978)
  3598--3604.

\bibitem{Price:1986yy}
R.~Price and K.~Thorne, {\it {Membrane Viewpoint on Black Holes: Properties and
  Evolution of the Stretched Horizon}},  {\em Phys.Rev.} {\bf D33} (1986)
  915--941.

\bibitem{Thorne:1986iy}
K.~S. Thorne, R.~Price, and D.~Macdonald, {\em Black Holes: The Membrane
  Paradigm}.
\newblock Yale University Press, New Haven, 1986.

\bibitem{damour1979quelques}
T.~Damour, {\em Quelques propri{\'e}t{\'e}s m{\'e}caniques,
  {\'e}lectromagn{\'e}tiques, thermodynamiques et quantiques des trous noirs}.
\newblock PhD thesis, Universit{\'e} Pierre et Marie Curie, Paris VI, 1979.

\bibitem{damour1982surface}
T.~Damour, {\it Surface effects in black hole physics},  in {\em Proceedings of
  the second Marcel Grossmann meeting on general relativity}, p.~587, North
  Holland, Amsterdam The Netherlands, 1982.

\bibitem{znajek1978electric}
R.~Znajek, {\it The electric and magnetic conductivity of a kerr hole},  {\em
  Monthly Notices of the Royal Astronomical Society} {\bf 185} (1978) 833--840.

\bibitem{Policastro:2001yc}
G.~Policastro, D.~Son, and A.~Starinets, {\it {The Shear viscosity of strongly
  coupled N=4 supersymmetric Yang-Mills plasma}},  {\em Phys.Rev.Lett.} {\bf
  87} (2001) 081601, [\href{http://xxx.lanl.gov/abs/hep-th/0104066}{{\tt
  hep-th/0104066}}].

\bibitem{Policastro:2002se}
G.~Policastro, D.~T. Son, and A.~O. Starinets, {\it {From AdS / CFT
  correspondence to hydrodynamics}},  {\em JHEP} {\bf 0209} (2002) 043,
  [\href{http://xxx.lanl.gov/abs/hep-th/0205052}{{\tt hep-th/0205052}}].

\bibitem{Bhattacharyya:2008jc}
S.~Bhattacharyya, V.~E. Hubeny, S.~Minwalla, and M.~Rangamani, {\it {Nonlinear
  Fluid Dynamics from Gravity}},  {\em JHEP} {\bf 0802} (2008) 045,
  [\href{http://xxx.lanl.gov/abs/0712.2456}{{\tt arXiv:0712.2456}}].

\bibitem{Rangamani:2009xk}
M.~Rangamani, {\it {Gravity and Hydrodynamics: Lectures on the fluid-gravity
  correspondence}},  {\em Class.Quant.Grav.} {\bf 26} (2009) 224003,
  [\href{http://xxx.lanl.gov/abs/0905.4352}{{\tt arXiv:0905.4352}}].

\bibitem{Hubeny:2011hd}
V.~E. Hubeny, S.~Minwalla, and M.~Rangamani, {\it {The fluid/gravity
  correspondence}},  \href{http://xxx.lanl.gov/abs/1107.5780}{{\tt
  arXiv:1107.5780}}.

\bibitem{Emparan:2009cs}
R.~Emparan, T.~Harmark, V.~Niarchos, and N.~A. Obers, {\it {World-Volume
  Effective Theory for Higher-Dimensional Black Holes}},  {\em Phys.Rev.Lett.}
  {\bf 102} (2009) 191301, [\href{http://xxx.lanl.gov/abs/0902.0427}{{\tt
  arXiv:0902.0427}}].

\bibitem{Emparan:2009at}
R.~Emparan, T.~Harmark, V.~Niarchos, and N.~A. Obers, {\it {Essentials of
  Blackfold Dynamics}},  {\em JHEP} {\bf 1003} (2010) 063,
  [\href{http://xxx.lanl.gov/abs/0910.1601}{{\tt arXiv:0910.1601}}].

\bibitem{Camps:2012hw}
J.~Camps and R.~Emparan, {\it {Derivation of the blackfold effective theory}},
  {\em JHEP} {\bf 1203} (2012) 038,
  [\href{http://xxx.lanl.gov/abs/1201.3506}{{\tt arXiv:1201.3506}}].

\bibitem{Emparan:2011br}
R.~Emparan, {\it {Blackfolds}},  \href{http://xxx.lanl.gov/abs/1106.2021}{{\tt
  arXiv:1106.2021}}.

\bibitem{Camps:2010br}
J.~Camps, R.~Emparan, and N.~Haddad, {\it {Black Brane Viscosity and the
  Gregory-Laflamme Instability}},  {\em JHEP} {\bf 1005} (2010) 042,
  [\href{http://xxx.lanl.gov/abs/1003.3636}{{\tt arXiv:1003.3636}}].

\bibitem{Bredberg:2011jq}
I.~Bredberg, C.~Keeler, V.~Lysov, and A.~Strominger, {\it {From Navier-Stokes
  To Einstein}},  {\em JHEP} {\bf 1207} (2012) 146,
  [\href{http://xxx.lanl.gov/abs/1101.2451}{{\tt arXiv:1101.2451}}].

\bibitem{Compere:2011dx}
G.~Compere, P.~McFadden, K.~Skenderis, and M.~Taylor, {\it {The Holographic
  fluid dual to vacuum Einstein gravity}},  {\em JHEP} {\bf 1107} (2011) 050,
  [\href{http://xxx.lanl.gov/abs/1103.3022}{{\tt arXiv:1103.3022}}].

\bibitem{Compere:2012mt}
G.~Compere, P.~McFadden, K.~Skenderis, and M.~Taylor, {\it {The relativistic
  fluid dual to vacuum Einstein gravity}},  {\em JHEP} {\bf 1203} (2012) 076,
  [\href{http://xxx.lanl.gov/abs/1201.2678}{{\tt arXiv:1201.2678}}].

\bibitem{Eling:2012ni}
C.~Eling, A.~Meyer, and Y.~Oz, {\it {The Relativistic Rindler Hydrodynamics}},
  {\em JHEP} {\bf 1205} (2012) 116,
  [\href{http://xxx.lanl.gov/abs/1201.2705}{{\tt arXiv:1201.2705}}].

\bibitem{Bhattacharyya:2008kq}
S.~Bhattacharyya, S.~Minwalla, and S.~R. Wadia, {\it {The Incompressible
  Non-Relativistic Navier-Stokes Equation from Gravity}},  {\em JHEP} {\bf
  0908} (2009) 059, [\href{http://xxx.lanl.gov/abs/0810.1545}{{\tt
  arXiv:0810.1545}}].

\bibitem{Emparan:2013ila}
R.~Emparan, V.~E. Hubeny, and M.~Rangamani, {\it {Effective hydrodynamics of
  black D3-branes}},  {\em JHEP} {\bf 1306} (2013) 035,
  [\href{http://xxx.lanl.gov/abs/1303.3563}{{\tt arXiv:1303.3563}}].

\bibitem{Brattan:2011my}
D.~Brattan, J.~Camps, R.~Loganayagam, and M.~Rangamani, {\it {CFT dual of the
  AdS Dirichlet problem : Fluid/Gravity on cut-off surfaces}},  {\em JHEP} {\bf
  1112} (2011) 090, [\href{http://xxx.lanl.gov/abs/1106.2577}{{\tt
  arXiv:1106.2577}}].

\bibitem{Emparan:2012be}
R.~Emparan and M.~Martinez, {\it {Black Branes in a Box: Hydrodynamics,
  Stability, and Criticality}},  {\em JHEP} {\bf 1207} (2012) 120,
  [\href{http://xxx.lanl.gov/abs/1205.5646}{{\tt arXiv:1205.5646}}].

\bibitem{Banerjee:2008th}
N.~Banerjee, J.~Bhattacharya, S.~Bhattacharyya, S.~Dutta, R.~Loganayagam,
  et~al., {\it {Hydrodynamics from charged black branes}},  {\em JHEP} {\bf
  1101} (2011) 094, [\href{http://xxx.lanl.gov/abs/0809.2596}{{\tt
  arXiv:0809.2596}}].

\bibitem{Erdmenger:2008rm}
J.~Erdmenger, M.~Haack, M.~Kaminski, and A.~Yarom, {\it {Fluid dynamics of
  R-charged black holes}},  {\em JHEP} {\bf 0901} (2009) 055,
  [\href{http://xxx.lanl.gov/abs/0809.2488}{{\tt arXiv:0809.2488}}].

\bibitem{Son:2009tf}
D.~T. Son and P.~Surowka, {\it {Hydrodynamics with Triangle Anomalies}},  {\em
  Phys.Rev.Lett.} {\bf 103} (2009) 191601,
  [\href{http://xxx.lanl.gov/abs/0906.5044}{{\tt arXiv:0906.5044}}].

\bibitem{Chamblin:1999tk}
A.~Chamblin, R.~Emparan, C.~V. Johnson, and R.~C. Myers, {\it {Charged AdS
  black holes and catastrophic holography}},  {\em Phys.Rev.} {\bf D60} (1999)
  064018, [\href{http://xxx.lanl.gov/abs/hep-th/9902170}{{\tt
  hep-th/9902170}}].

\bibitem{Cvetic:1999ne}
M.~Cvetic and S.~S. Gubser, {\it {Phases of R charged black holes, spinning
  branes and strongly coupled gauge theories}},  {\em JHEP} {\bf 9904} (1999)
  024, [\href{http://xxx.lanl.gov/abs/hep-th/9902195}{{\tt hep-th/9902195}}].

\bibitem{Harvey:1998bx}
J.~A. Harvey, R.~Minasian, and G.~W. Moore, {\it {NonAbelian tensor multiplet
  anomalies}},  {\em JHEP} {\bf 9809} (1998) 004,
  [\href{http://xxx.lanl.gov/abs/hep-th/9808060}{{\tt hep-th/9808060}}].

\bibitem{Emparan:2011hg}
R.~Emparan, T.~Harmark, V.~Niarchos, and N.~A. Obers, {\it {Blackfolds in
  Supergravity and String Theory}},  {\em JHEP} {\bf 1108} (2011) 154,
  [\href{http://xxx.lanl.gov/abs/1106.4428}{{\tt arXiv:1106.4428}}].

\bibitem{Maldacena:2008wh}
J.~Maldacena, D.~Martelli, and Y.~Tachikawa, {\it {Comments on string theory
  backgrounds with non-relativistic conformal symmetry}},  {\em JHEP} {\bf
  0810} (2008) 072, [\href{http://xxx.lanl.gov/abs/0807.1100}{{\tt
  arXiv:0807.1100}}].

\bibitem{Cvetic:1999xp}
M.~Cvetic, M.~Duff, P.~Hoxha, J.~T. Liu, H.~Lu, et~al., {\it {Embedding AdS
  black holes in ten-dimensions and eleven-dimensions}},  {\em Nucl.Phys.} {\bf
  B558} (1999) 96--126, [\href{http://xxx.lanl.gov/abs/hep-th/9903214}{{\tt
  hep-th/9903214}}].

\bibitem{Bhattacharyya:2008ji}
S.~Bhattacharyya, R.~Loganayagam, S.~Minwalla, S.~Nampuri, S.~P. Trivedi,
  et~al., {\it {Forced Fluid Dynamics from Gravity}},  {\em JHEP} {\bf 0902}
  (2009) 018, [\href{http://xxx.lanl.gov/abs/0806.0006}{{\tt
  arXiv:0806.0006}}].

\bibitem{Buchel:2003tz}
A.~Buchel and J.~T. Liu, {\it {Universality of the shear viscosity in
  supergravity}},  {\em Phys.Rev.Lett.} {\bf 93} (2004) 090602,
  [\href{http://xxx.lanl.gov/abs/hep-th/0311175}{{\tt hep-th/0311175}}].

\bibitem{Kovtun:2004de}
P.~Kovtun, D.~Son, and A.~Starinets, {\it {Viscosity in strongly interacting
  quantum field theories from black hole physics}},  {\em Phys.Rev.Lett.} {\bf
  94} (2005) 111601, [\href{http://xxx.lanl.gov/abs/hep-th/0405231}{{\tt
  hep-th/0405231}}].

\bibitem{Kim:1985ez}
H.~Kim, L.~Romans, and P.~van Nieuwenhuizen, {\it {The Mass Spectrum of Chiral
  N=2 D=10 Supergravity on S**5}},  {\em Phys.Rev.} {\bf D32} (1985) 389.

\bibitem{Kraus:1998hv}
P.~Kraus, F.~Larsen, and S.~P. Trivedi, {\it {The Coulomb branch of gauge
  theory from rotating branes}},  {\em JHEP} {\bf 9903} (1999) 003,
  [\href{http://xxx.lanl.gov/abs/hep-th/9811120}{{\tt hep-th/9811120}}].

\bibitem{Gibbons:1993sv}
G.~Gibbons and P.~Townsend, {\it {Vacuum interpolation in supergravity via
  super p-branes}},  {\em Phys.Rev.Lett.} {\bf 71} (1993) 3754--3757,
  [\href{http://xxx.lanl.gov/abs/hep-th/9307049}{{\tt hep-th/9307049}}].

\bibitem{Kraus:1999di}
P.~Kraus, F.~Larsen, and R.~Siebelink, {\it {The gravitational action in
  asymptotically AdS and flat space-times}},  {\em Nucl.Phys.} {\bf B563}
  (1999) 259--278, [\href{http://xxx.lanl.gov/abs/hep-th/9906127}{{\tt
  hep-th/9906127}}].

\bibitem{Brown:1992br}
J.~D. Brown and J.~York, James~W., {\it {Quasilocal energy and conserved
  charges derived from the gravitational action}},  {\em Phys.Rev.} {\bf D47}
  (1993) 1407--1419, [\href{http://xxx.lanl.gov/abs/gr-qc/9209012}{{\tt
  gr-qc/9209012}}].

\bibitem{Bai:2012ci}
X.~Bai, Y.-P. Hu, B.-H. Lee, and Y.-L. Zhang, {\it {Holographic Charged Fluid
  with Anomalous Current at Finite Cutoff Surface in Einstein-Maxwell
  Gravity}},  {\em JHEP} {\bf 1211} (2012) 054,
  [\href{http://xxx.lanl.gov/abs/1207.5309}{{\tt arXiv:1207.5309}}].

\bibitem{Wald:1984rg}
R.~M. Wald, {\em General Relativity}.
\newblock University Of Chicago Press, Chicago, 1984.

\bibitem{Torabian:2009qk}
M.~Torabian and H.-U. Yee, {\it {Holographic nonlinear hydrodynamics from
  AdS/CFT with multiple/non-Abelian symmetries}},  {\em JHEP} {\bf 0908} (2009)
  020, [\href{http://xxx.lanl.gov/abs/0903.4894}{{\tt arXiv:0903.4894}}].

\bibitem{Caldarelli:2012hy}
M.~M. Caldarelli, J.~Camps, B.~Gouteraux, and K.~Skenderis, {\it
  {AdS/Ricci-flat correspondence and the Gregory-Laflamme instability}},  {\em
  Phys.Rev.} {\bf D87} (2013), no.~6 061502,
  [\href{http://xxx.lanl.gov/abs/1211.2815}{{\tt arXiv:1211.2815}}].

\bibitem{Caldarelli:2013aaa}
M.~M. Caldarelli, J.~Camps, B.~Gouteraux, and K.~Skenderis, {\it
  {AdS/Ricci-flat correspondence}},  {\em JHEP} {\bf 1404} (2014) 071,
  [\href{http://xxx.lanl.gov/abs/1312.7874}{{\tt arXiv:1312.7874}}].

\bibitem{Russo:1998mm}
J.~G. Russo, {\it {New compactifications of supergravities and large N QCD}},
  {\em Nucl.Phys.} {\bf B543} (1999) 183--197,
  [\href{http://xxx.lanl.gov/abs/hep-th/9808117}{{\tt hep-th/9808117}}].

\bibitem{Russo:1998by}
J.~G. Russo and K.~Sfetsos, {\it {Rotating D3-branes and QCD in
  three-dimensions}},  {\em Adv.Theor.Math.Phys.} {\bf 3} (1999) 131--146,
  [\href{http://xxx.lanl.gov/abs/hep-th/9901056}{{\tt hep-th/9901056}}].

\bibitem{Cvetic:1996dt}
M.~Cvetic and D.~Youm, {\it {Near BPS saturated rotating electrically charged
  black holes as string states}},  {\em Nucl.Phys.} {\bf B477} (1996) 449--464,
  [\href{http://xxx.lanl.gov/abs/hep-th/9605051}{{\tt hep-th/9605051}}].

\bibitem{Behrndt:1998jd}
K.~Behrndt, M.~Cvetic, and W.~Sabra, {\it {Nonextreme black holes of
  five-dimensional N=2 AdS supergravity}},  {\em Nucl.Phys.} {\bf B553} (1999)
  317--332, [\href{http://xxx.lanl.gov/abs/hep-th/9810227}{{\tt
  hep-th/9810227}}].

\bibitem{Schwarz:1983qr}
J.~H. Schwarz, {\it {Covariant Field Equations of Chiral N=2 D=10
  Supergravity}},  {\em Nucl.Phys.} {\bf B226} (1983) 269.

\bibitem{Gunaydin:1984fk}
M.~Gunaydin and N.~Marcus, {\it {The Spectrum of the $s**5$ Compactification of
  the Chiral N=2, D=10 Supergravity and the Unitary Supermultiplets of U(2,
  2/4)}},  {\em Class.Quant.Grav.} {\bf 2} (1985) L11.

\bibitem{Pernici:1984xx}
M.~Pernici, K.~Pilch, and P.~van Nieuwenhuizen, {\it {Gauged Maximally Extended
  Supergravity in Seven-dimensions}},  {\em Phys.Lett.} {\bf B143} (1984) 103.

\bibitem{Gunaydin:1984qu}
M.~Gunaydin, L.~Romans, and N.~Warner, {\it {Gauged N=8 Supergravity in
  Five-Dimensions}},  {\em Phys.Lett.} {\bf B154} (1985) 268.

\bibitem{Myers:1986un}
R.~C. Myers and M.~Perry, {\it {Black Holes in Higher Dimensional
  Space-Times}},  {\em Annals Phys.} {\bf 172} (1986) 304.

\end{thebibliography}\endgroup
\end{document}